\documentclass[aps,prx,twocolumn,superscriptaddress,longbibliography]{revtex4-2}
\pdfoutput=1

\usepackage{amsmath}
\usepackage{amssymb}
\usepackage{amsfonts}
\usepackage{bm}
\usepackage{graphicx}
\usepackage{braket}
\usepackage{mathtools}

\usepackage[unicode=true,
 bookmarks=false,
 breaklinks=false,pdfborder={0 0 1},backref=false,colorlinks=true]
 {hyperref}
\hypersetup{
 linkcolor=blue,citecolor=blue,urlcolor=blue,linkcolor=blue,citecolor=blue,urlcolor=blue}
\usepackage{microtype}

\newcommand{\itp}{\affiliation{Institute for Theoretical Physics, University of Innsbruck, 6020 Innsbruck, Austria}}
\newcommand{\iqoqi}{\affiliation{Institute for Quantum Optics and Quantum Information of the Austrian Academy of Sciences,  6020 Innsbruck, Austria}}
\newcommand{\iisc}{\affiliation{Department of Instrumentation and Applied Physics, Indian Institute of Science,  Bengaluru, 560012, India}}
\newcommand{\iqi}{\affiliation{Institut f{\"u}r Quanteninformation GmbH,  6020 Innsbruck, Austria}}

\DeclareMathOperator{\tr}{\mathrm{Tr}}

\DeclareMathOperator{\Uen} {\mathcal{U}_{\mathrm{en}}}
\DeclareMathOperator{\Ude} {\mathcal{U}_{\mathrm{de}}}

\DeclareMathOperator{\psiin} {\psi_{\rm in}}

\usepackage{dcolumn}
\usepackage{mathtools}
\usepackage{verbatim}
\usepackage{bbold}

\newcommand{\abs}[1]{\lvert #1 \rvert} 
\newcommand{\ev}[1]{\langle #1 \rangle} 
\newcommand{\td}[1]{\tilde{#1}} 

\date{\today}

\begin{abstract}
We study optimal quantum sensing of multiple physical parameters using repeated measurements. In this scenario, the Fisher information framework sets the fundamental limits on sensing performance, yet the optimal states and corresponding measurements that attain these limits remain to be discovered. To address this, we extend the Fisher information approach with a second optimality requirement for a sensor to provide unambiguous estimation of unknown parameters. We propose a systematic method integrating Fisher information and Bayesian approaches to quantum metrology to identify the combination of input states and measurements that satisfies both optimality criteria. Specifically, we frame the optimal sensing problem as an optimization of an asymptotic Bayesian cost function that can be efficiently solved numerically and, in many cases, analytically. We refer to the resulting optimal sensor as a `quantum compass' solution, which serves as a direct multiparameter counterpart to the Greenberger-Horne-Zeilinger state-based interferometer, renowned for achieving the Heisenberg limit in single-parameter metrology. We provide exact quantum compass solutions for paradigmatic multiparameter problem of sensing two and three parameters using an SU(2) sensor. Our metrological cost function opens avenues for quantum variational techniques to design low-depth quantum circuits approaching the  optimal sensing performance in the many-repetition scenario. We demonstrate this by constructing simple quantum circuits that achieve the Heisenberg limit for vector field and 3D rotations estimation using a limited set of gates available on a trapped-ion platform. Our work introduces and optimizes sensors for a practical notion of optimality, keeping in mind the ultimate goal of quantum sensors to precisely estimate \emph{unknown} parameters.
\end{abstract}

\begin{document}

\title{Optimal Multiparameter Metrology: The Quantum Compass Solution}

\author{Denis V. Vasilyev}\iqoqi \iqi
\author{Athreya Shankar}\iisc
\author{Raphael Kaubruegger}\iqoqi \itp
\author{Peter Zoller}\iqoqi \itp

\maketitle

\section{Introduction}

Quantum sensing, a rapidly advancing field in quantum technologies, involves utilizing quantum systems in non-classical or entangled states to achieve estimation precision of physical parameters surpassing classical limits. It has become an established technology for single-parameter estimation, as seen in advancements such as gravitational wave detection~\cite{LIGO_2023} and progress in atomic clocks~\cite{Pedrozo-Penafiel:2020aa,robinson2022,Marciniak2022}. A current frontier of quantum metrology research is multiparameter estimation, where several parameters of a physical process are estimated simultaneously~\cite{Liu2020,Demkowicz2020}. Quantum-enhanced precision in estimating multiple parameters will benefit various fields, including imaging~\cite{Genovese_2016,Tsang2016,Rehacek2018}, spatially distributed quantum sensing networks~\cite{Urizar2013,Proctor2018,Sekatski2020,Hainzer2024,yang2023,Bringewatt2021}, aligning reference frames~\cite{Bartlett2007}, and the sensing of non-commuting rotations~\cite{Vaneph2013,Goldberg:2021,ferretti2023} and vector fields~\cite{Baumgratz2016,Gorecki2022,Kaubruegger2023}.

Our metrological task of interest is to estimate multiple parameters of a physical process through repeated interactions with a quantum system, followed by measurements. Quantum metrology addresses two key questions: What is the best precision for estimating parameters, and what constitutes the optimal quantum sensor, encompassing the input quantum state and measurement, achieving this precision? The Fisher information framework successfully addresses the first question in the many-repetition scenario~(see~\cite{Liu2020,Demkowicz2020} for recent reviews). Fundamental precision bounds are formulated in terms of the quantum Fisher information matrix~(QFIM), yielding the quantum Cram\'er-Rao bound~(CRB) on the estimation mean squared error~(MSE). In the multiparameter case, the bound is not necessarily tight and is further generalized to the Holevo CRB to account for the potential incompatibility of quantum measurements~\footnote{Incompatibility implies that the measurement of one parameter can impact the precision in measuring another parameter, rendering the construction of the optimal measurement a highly intricate task.}. Despite extensive research~\cite{Humphreys2013,Matsumoto:2002aa,Pezze2017,Yang2019,Len_2022,Miyazaki2022}, the question of the optimal multiparameter quantum sensor still lacks a satisfactory answer, as discussed below. Meanwhile, the emergence of programmable quantum sensors~\cite{Marciniak2022}, resulting from the fusion of quantum information concepts and quantum metrology, enables the experimental implementation of complex sensors capable of approaching the fundamental precision limits in the measurement of multiple parameters. This poses the challenge of explicitly identifying the optimal input states and corresponding measurements that yield optimal sensing, and finding ways to approximate these states and measurements on an experimental platform to achieve optimal sensing in practice.

We address this challenge by introducing a systematic method to identify the combination of the input state and measurement that forms the optimal quantum sensor, achieving fundamental precision bounds in the general setting of multiparameter estimation. In our method, we introduce a metrological cost function that is minimized by the optimal sensor and thus can be combined with the quantum variational approach~\cite{Koczor_2020,Kaubruegger2021,Kaubruegger2023,Marciniak2022} to construct quantum circuits that approximate the optimal sensor using a given set of gates available on an experimental platform.

A key insight is that the sole criterion of saturating the QFIM by a sensor at some point $\bm\phi_0$ in the parameter space does not guarantee the sensor's ability to distinguish parameters within a vicinity of that point. Consequently, there may exist a continuum of input states and measurements that saturate the QFIM~\cite{Pezze2017,Miyazaki2022}, but they differ in their ability to unambiguously identify parameters. Therefore, they are not all equally useful for the primary function of a sensor, which is to estimate unknown parameters.

To identify a unique optimal sensor that is practically relevant, we propose a second criterion for the sensor. This criterion requires maximizing the neighborhood of~$\bm\phi_0$ in the parameter space where parameters can be estimated \emph{unambiguously}. We observe that maximizing the domain of unambiguous estimation is closely linked to maximizing the information obtained from a single measurement shot. With this understanding, we approach the problem of optimizing the sensor performance at a given point~$\bm\phi_0$ using the Bayesian framework in the limit of asymptotically many measurement repetitions $K$. The Bayesian framework allows us to optimize post-measurement knowledge about parameters given their prior distribution, thereby directly maximizing information gain. This enables us to formulate a metrological cost function to be minimized over the sensor input states and measurements, which, unlike the Fisher information, is \emph{non-local}, i.e., it is taken to be the single-shot estimation MSE averaged over a prior distribution of parameter values with an infinitesimal width~$\sim 1/K$, centered at~$\bm\phi_0$.

The Bayesian cost function, in the asymptotic limit~$\smash{K\to\infty}$, is independent of the prior distribution of the parameters and is linked to the Fisher information matrix, according to the Bernstein-von Mises theorem~\cite{DasGupta2008, Kleijn2012, Li2018}. The non-local nature of the cost function leads to a Taylor expansion terms in powers of $1/K$, which define higher-order information matrices similar to those introduced by Bhattacharyya~\cite{Bhattacharyya1947,gessner2023}. These higher-order terms create increasingly non-local optimization problems that gradually eliminate the degeneracy of the space of Fisher-optimal sensor solutions. As a result, the optimal sensor that emerges is unique, performs optimally at the given parameter value~$\bm\phi_0$, maximizes single-shot information gain, and has the largest known neighborhood of unambiguous estimation around that value. 

Surprisingly, the asymptotic Bayesian cost can be analytically minimized, as demonstrated by solving the paradigmatic multiparameter problem of sensing two and three parameters using an SU(2) sensor. The analytical solution is facilitated by the efficient numerical minimization of the Bayesian cost, that offers insights into the symmetry properties of the optimal solution. This efficiency is enabled by the multi-convex nature of the Bayesian cost minimization problem with respect to the state, measurement, and estimator degrees of freedom~\cite{Kaubruegger2023}. Furthermore, the numerical method is versatile, not reliant on specific symmetry classes~\cite{Miyazaki2022} or intuition for initial state guesses~\cite{Baumgratz2016}. This generality allows for the approximate identification of the optimal sensor, considering practical constraints like limiting measurements to the projective class. The numerical method allows us to identify input states that minimize the quantum (Holevo) CRB, defining the Heisenberg limit (HL) in situations where a lack of symmetry hinders intuitive guessing, as demonstrated in SU(2) sensors with a bias field.

All identified optimal sensor solutions saturate the multiparameter HL and share a distinct feature with the well-known Greenberger-Horne-Zeilinger (GHZ) state-based interferometer, renowned for achieving the HL in single-parameter metrology. The GHZ-state interferometer yields only two possible outcomes in a \emph{single} measurement repetition, revealing the sign but not the magnitude of the parameter. Similarly, our optimal sensors exhibit single-shot estimators evenly distributed on a hypersphere in the parameter space, indicating merely directions toward the true phase in each measurement. Hence, we term such optimal sensor solutions as \emph{quantum compass} (QC) solutions. The directional nature of single-shot estimators suggests that the QC solution can be seen as a multiparameter counterpart to the GHZ-state-based interferometer. These explicit QC solutions allow us to study the limits of sensing performance away from the optimization point~$\bm\phi_0$ and conjecture a relationship between ultimate precision and locality properties of the corresponding estimation in the many-repetitions regime.

Finally, we have framed the optimal sensing problem as an optimization of the asymptotic Bayesian cost function. This paves the way for the use of quantum variational techniques in designing quantum circuits for sensing in many-measurement scenario. These circuit-based sensors can realize near-optimal multiparameter sensing, even with a restricted set of quantum resources available on an experimental platform. We demonstrate this by constructing simple quantum circuits that achieve the HL for three-parameter estimation, such as vector magnetic field or 3D rotation sensing. The low-depth circuits are designed to be readily implementable on a trapped-ion platform~\cite{Leibfried2005, Monz2011} and use only global rotations and entangling one-axis twisting operations for state preparation and generating optimal measurements.

This paper is organized as follows: Sec.~\ref{sec:estimation_theory} summarizes quantum estimation theory for multiple parameters in the many-measurement scenario. Our main results are in Sec.~\ref{sec:optimal_multiparameter}, where we formulate the problem of finding the optimal sensor as an optimization of an asymptotic Bayesian cost function. We expand this cost using Taylor series to connect it to Fisher and higher-order information matrices, enabling analytical solutions. In Secs.~\ref{sec:two-parameter_QC} and~\ref{sec:three-parameter_QC}, we apply this approach to find exact QC solutions for sensing two and three parameters using an SU(2) sensor. In Sec.~\ref{sec:VQC}, we combine our cost function with quantum variational techniques to find simple quantum circuits for vector field and 3D rotations sensing in the many-measurement scenario. The paper concludes in Sec.~\ref{sec:conclusion}.

\section{Quantum Multiparameter Estimation Theory}
\label{sec:estimation_theory}

We consider a quantum system initialized in state~$\ket{\psi_\mathrm{in}}$ as a sensor to estimate multiple classical parameters, represented by a vector of $d$~unknown phases $\bm\phi = \{\phi_1,\ldots,\phi_d\}$. This vector is imprinted onto the input state $\ket{\psi_\mathrm{in}}$ through a parameter-dependent dynamical evolution, such as a unitary operation $U(\bm{\phi})$. Subsequently, these phases are estimated from a measurement conducted on the evolved state $\ket{\psi_{\bm{\phi}}} = U(\bm{\phi})\ket{\psi_\text{in}}$. The most general measurement process is described by a positive operator-valued measure (POVM), which is a set $\{M_\mu\}$ of positive Hermitian operators, $M_\mu \succeq 0$, such that $\sum_\mu M_\mu = \mathbb{1}$. Given a specific vector of phases $\bm\phi$, the measurement outcomes, labeled~$\mu$, occur with probabilities
\begin{equation}
    p(\mu | \bm\phi) = \tr\{M_{\mu}\ket{\psi_{\bm{\phi}}} \bra{\psi_{\bm{\phi}}}\}.
    \label{eq:cond_prob}
\end{equation}

We focus on scenarios where the signal being measured remains essentially constant across multiple measurement repetitions, enabling repeated probing by the sensor. Alternatively, independent measurements can be conducted in parallel on several identically prepared systems. In these cases, the outcomes of $K$ independent and identical measurements $\bm{\mu}=\{\mu_1,\ldots,\mu_K\}$ are characterized by the conditional probability, which is expressed as the product of single-shot probabilities defined in Eq.~\eqref{eq:cond_prob}:
\begin{equation}    
	p(\bm\mu | \bm\phi) = \prod_{k=1}^K p(\mu_k | \bm\phi).
    \label{eq:likelihood}
\end{equation}
The probability Eq.~\eqref{eq:likelihood} is often referred to as the \emph{likelihood}.

Finally, the estimation of the phases $\bm\phi$ is performed using an estimator function that maps the measurement outcomes $\bm\mu$ to a vector of estimates $\bm\xi_{\bm\mu}$ of the true phases. The quality of estimation is quantified by the mean squared error (MSE), which averages the squared differences between the estimates and the true values over all potential measurement outcomes:
\begin{equation}
    \mathrm{MSE}(\bm{\phi}) = \sum_{\bm{\mu}} (\bm{\phi}-\bm{\xi}_{\bm{\mu}})\cdot (\bm{\phi}-\bm{\xi}_{\bm{\mu}})\,p(\bm\mu | \bm\phi).
    \label{eq:mse}
\end{equation}
The MSE, Eq.~\eqref{eq:mse}, represents the figure of merit for the sensing performance. The goal of quantum metrology is to find the input state $\ket{\psi_\mathrm{in}}$, the measurement $\{M_\mu\}$, and the estimators $\bm{\xi}_{\bm{\mu}}$ that minimize the estimation error under relevant constraints imposed by a measurement task. 

In this work, we focus on \emph{local estimation} of parameters in the scenario of many measurement repetitions. Specifically, the task is to achieve maximum sensitivity to slight variations in parameters around a known a priori value~$\bm\phi_0$, approximately determined, for instance, through preliminary coarse measurements. Lower bounds on MSE in this context are determined using the Fisher information approach.

\subsection*{Fisher Information Framework}
\label{subsec:FI_framework}

Within the Fisher information framework, unbiasedness constraints are imposed on estimation strategies to derive lower bounds on estimation error. An unbiased estimator $\bm{\xi}_{\bm{\mu}}^{\rm u}$ converges to the true phase value on average across all measurement realizations~\footnote{Specifically, a locally unbiased estimator satisfies $\sum_{\bm{\mu}} \bm{\xi}^{\rm u}_{\bm{\mu}}p(\bm\mu | \bm\phi)=\bm{\phi}$ and $\sum_{\bm{\mu}} \bm{\xi}^{\rm u}_{\bm{\mu}}\nabla p(\bm\mu | \bm\phi)=\mathbb{1}$}. For a given true phase value $\bm\phi$, the MSE of such estimators is lower bounded by the following chain of inequalities~\cite{Liu2020,Demkowicz2020}:
\begin{equation}
\label{eq:QBounds_chain}
K\times\mathrm{MSE}(\bm{\phi})\ge \Delta_{\rm CRB} \ge \Delta_{\rm QCRB} \ge \Delta_{\rm HL}.
\end{equation}

The first inequality, known as the Cram\'er-Rao bound~(CRB), sets the lower bound on the estimation error for an arbitrary unbiased estimator. The CRB reads $\Delta_{\rm CRB}\equiv\mathrm{Tr}\mathcal{F}^{-1}$, where
\begin{equation}
    \mathcal{F}=\sum_{\mu}\frac{\bm{\nabla}p(\mu | \bm\phi)\bm{\nabla}^{T}p(\mu | \bm\phi)}{p(\mu | \bm\phi)}.
    \label{eq:FIM}
\end{equation}
is the Fisher information matrix~(FIM) which depends on the input state $\ket{\psi_\mathrm{in}}$, the measurements $\{M_\mu\}$, and the true phase vector value $\bm{\phi}$. While there is no guarantee that estimators saturating the CRB exist for an arbitrary number of measurements, with a large number of measurements $K\gg1$, there is at least one such estimator: the maximum likelihood estimator (MLE). The MLE is defined as
\begin{equation}    
	\bm{\xi}_{\bm{\mu}}^{\mathrm{ML}} \equiv \mathrm{arg}\;\mathrm{max}_{\bm{\phi}\in\Omega^*}\; p(\bm\mu | \bm\phi).
	\label{eq:MLE}
\end{equation}
The MLE is normally distributed, saturates the CRB and thus, in the limit $K\to\infty$, it is as good as or superior to any other estimator. In practice, the likelihood function may have multiple local maxima, making the identification of a unique global maximum challenging and requiring global optimization. To ensure an unambiguous and efficient evaluation of a unique $\bm{\xi}_{\bm{\mu}}^{\mathrm{ML}}$ through local maximization of $p(\bm\mu | \bm\phi)$, the parameter space domain $\Omega^*$ is chosen to encompass only a single maximum of the likelihood.

The second inequality in Eq.~\eqref{eq:QBounds_chain} is the quantum Cram\'er-Rao bound (QCRB), which sets the fundamental quantum limit on sensing performance for a given input state $\ket{\psi_\mathrm{in}}$. The QCRB is expressed as $\Delta_{\rm QCRB} \equiv \mathrm{Tr}\mathcal{F}_Q^{-1}$, where $\mathcal{F}_Q$ is the quantum Fisher information matrix (QFIM) obtained by maximizing the FIM over all possible POVMs, $\smash{\mathcal{F}_Q=\max_{\{M_{\bm\mu}\}}\mathcal{F}}$. The optimization result can be expressed in terms of the symmetric logarithmic derivative~(SLD)~\cite{Paris2009,Liu2020}, such that for pure states and unitary evolution, the SLD and the QFIM are given by~\cite{Fujiwara1995}:
\begin{align}
\label{eq:sld}
L_i &= 2\left(\ket{\partial_{\phi_i}\psi_{\bm{\phi}}}\bra{\psi_{\bm{\phi}}} + \ket{\psi_{\bm{\phi}}}\bra{\partial_{\phi_i}\psi_{\bm{\phi}}}\right)\\
\label{eq:QFIM}
[\mathcal{F}_Q]_{ij} &= \frac12\bra{\psi_{\bm\phi}}L_iL_j+L_jL_i\ket{\psi_{\bm\phi}}.
\end{align}
Thus, the QFIM depends on the input state $\ket{\psi_\mathrm{in}}$ and the true phase vector $\bm{\phi}$. Note that the saturability of the QCRB is a nuanced issue in the multiparameter case, leading to the formulation of a universally tight bound, the Holevo~CRB~(HCRB), as further discussed in the Appendix~\ref{app:QCRB_saturability}.

Finally, the last inequality in Eq.~\eqref{eq:QBounds_chain} is the Heisenberg limit (HL) characterizing the maximum phase sensitivity allowed by quantum mechanics. The HL is derived by further minimizing the QCRB over all input states and depends solely on the phase value of interest, $\bm\phi$.

The QCRB and the HL define fundamental quantum limits, setting the benchmark performance level for optimal quantum sensors. However, as we discuss below, reaching the limits at a particular parameter value alone is not sufficient to qualify a sensor as optimal; to be practically relevant, an optimal sensor must also guarantee unambiguous estimation in a neighborhood of this value. In other words, the domain $\Omega^*$ should be maximized around the phase value of interest $\bm\phi$.

\section{Optimal Multiparameter Sensing}
\label{sec:optimal_multiparameter}
In this section, we establish criteria for an optimal quantum sensor for local estimation and develop a systematic method to determine the combination of input state $\ket{\psi_\text{in}}$ and measurement $\{M_\mu\}$ that satisfy these criteria. Initially, we demonstrate that merely minimizing the CRB is inadequate for defining an optimal sensor that guarantees unambiguous estimation. Therefore, in the subsequent subsection, we introduce the requirement of unambiguous estimation to formulate the optimality criteria. Consequently, developing a cost function that encapsulates the optimal sensor meeting these criteria necessitates extending beyond the Fisher information. To address this, in the following subsection, we approach the many-measurement estimation problem within the Bayesian framework. We identify an asymptotic Bayesian cost as a pertinent cost function. Minimizing this function with respect to the input state and measurement results in a unique sensor that achieves fundamental quantum limits, maximizes single-shot information gain, and extends the range of unambiguous estimation~$\Omega^*$.

\subsection{Issue of Estimation Unambiguity}
\label{subsubsec:unambiguity}
In many measurement scenario, a common approach to optimal sensing involves using the CRB as a cost function to optimize input states and measurements~\cite{Koczor_2020,Meyer:2021aa,Ma2021,le2023variational}. The resulting state and measurement, combined with an appropriate estimator (e.g., the MLE), constitute a quantum sensor capable of reaching the QCRB with a sufficiently large number of measurements $K$. However, multiple combinations of input states and measurements can yield the maximum FIM. The issue lies in the fact that these various `optimal' sensors can significantly differ in their ability to unambiguously identify parameters, thus making them all not equally useful in practice. This aspect of the sensor optimization problem is evident already in the single-parameter case.

\begin{figure}[t] 
   \centering
   \includegraphics[]{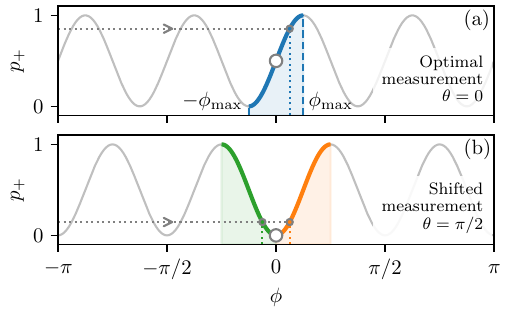}
   \caption{The range of unambiguous phase estimation for GHZ state-based interferometers consisting of $N=4$ atoms. (a) The optimal measurement basis, $\ket{\psi_{\pm}}_{\theta=0}$, places the phase of interest $\phi_0=0$~(white circle) in the middle of the invertible estimation region. The gray line represents the probability of projecting the initial GHZ state into the $\ket{\psi_{+}}$ state. The blue curve highlights a single fringe that enables the unambiguous mapping (indicated by dotted lines) of the observed probability $p_+$ to phases within the interval $\smash{\phi\in (-\phi_{\rm max},\phi_{\rm max})}$, marked by vertical dashed blue lines.
   (b) A shifted measurement basis, $\ket{\psi_{\pm}}_{\theta=\pi/2}$, places the phase $\phi_0=0$ (white circle) at the boundary between two fringes (green and orange curves). This measurement does not allow unambiguous mapping as each observed probability corresponds to two possible phase values, shown by the green and orange dotted lines.}
   \label{fig:1D_GHZ}
\end{figure}

\emph{Single-parameter estimation example}. Let us consider the estimation of a single phase $\phi$ in the vicinity of $\phi_0=0$. The phase is applied to an input state of $N$ spin-$1/2$ atoms using the unitary evolution $U(\phi)=e^{-iJ_z\phi}$, where $J_z=\sum_i\sigma_i^z/2$ represents the collective spin $z$-projection, and $\sigma^z_i$ stands for the Pauli spin operators of individual atoms. In this case the optimal sensor consists of the GHZ state $\ket{\psi_\mathrm{in}}=(\ket{\downarrow}^{\otimes N} + \ket{\uparrow}^{\otimes N})/\sqrt{2}$ as the optimal input state and projective measurements on the two bases $\ket{\psi_\pm}_{\theta}=(e^{-i\theta/2}\ket{\downarrow}^{\otimes N} \pm ie^{i\theta/2}\ket{\uparrow}^{\otimes N})/\sqrt{2}$ as the optimal measurement with a free parameter $\theta$. The conditional probabilities to project onto $\ket{\psi_\pm}_{\theta}$ states read $p_\pm(\phi)=[1\mp\sin(N\phi-\theta)]/2$. The FIM and QFIM of the sensor are maximal and independent of the measurement basis parameter~$\theta$, $\mathcal{F}=\mathcal{F}_Q=N^2$. However, its ability to unambiguously identify phases does depend on $\theta$ as illustrated in the Fig.~\ref{fig:1D_GHZ}. In panel (a), the optimal sensor with $\theta=0$ shows the broadest unambiguous estimation range around the point $\phi_0=0$ located at the center of the interference fringe. In contrast, in panel (b), the sensor with $\theta=\pi/2$ fails to uniquely identify any phase other than $\phi=0$. This makes it unsuitable for the primary function of a sensor, which is to estimate an unknown parameter.

Thus, the choice of input state and measurement can significantly impact the capability of a sensor to uniquely identify phases, even when the FIM remains at its maximum. We aim for sensors capable of identifying a broad range of parameter values with maximum sensitivity, corresponding to Fig.~\ref{fig:1D_GHZ}(a) in the example above. Therefore, we formulate the optimality criteria as follows.

\subsection{Optimality criteria}

We define an optimal quantum sensor for local estimation around a given phase vector $\bm\phi_0$ based on two criteria:
\begin{enumerate}
    \item[(i).] Saturation of the QCRB (HCRB) with a given input state or the HL in general, at the point $\bm\phi_0$.
    \item[(ii).] Assuming (i) is met, the sensor exhibits the widest domain of unambiguous estimation~$\Omega^*$ centered around the point~$\bm\phi_0$.
\end{enumerate}
Drawing inspiration from the quantum variational approach~\cite{Koczor_2020,Kaubruegger2021,Kaubruegger2023}, our aim is to formulate a cost function to be minimized over the sensor's input states and measurements, resulting in a solution that satisfies the two criteria of optimality.

Given that the FIM of the sensor is insensitive to criterion~(ii), directly formulating such a cost function would require incorporating the convexity properties of the associated likelihood to assess the domain $\Omega^*$. Such a cost function would be extremely cumbersome to calculate and optimize. Instead, we consider a cost function beyond the Fisher information and approach the many-repetition estimation scenario from the perspective of the Bayesian framework. We demonstrate that, in the limit of an asymptotically large number of measurements $K$, the Bayesian approach yields a cost function amenable to numeric and analytic optimizations, providing optimal sensor solutions satisfying criterion (i), while criterion (ii) is replaced by a physically related requirement of maximizing the information gained about the signal in a single measurement shot.

As a side note, the local estimation optimality criteria contrast with those in \emph{global estimation} scenarios, where the goal is to estimate parameters within a \emph{predefined} range of possible values. Global estimation is relevant in scenarios like monitoring a time-dependent (stochastic) signal, where only a single measurement attempt is available to estimate current parameter values~\cite{Sekatski2017,Macieszczak2014,Chabuda2016,Kaubruegger2021,Marciniak2022,Kaubruegger2023}. In such cases, it may be advantageous to trade off some estimation MSE for the ability to unambiguously estimate parameters in a larger range~\cite{Gorecki2020,Gorecki2022}.

\subsection{Local Estimation Optimality in Bayesian Framework}
\label{subsec:Bayesian_framework}

In the Bayesian approach, the vector of phases $\bm{\phi}$ is treated as a continuous random variable characterized by a prior probability distribution $\mathcal{P}(\bm{\phi})$ representing our initial guess or knowledge about the true phases. By applying Bayes' theorem, we can combine this prior knowledge with the outcomes ${\bm\mu}$ of $K$ measurements characterized by the corresponding likelihood $p(\bm\mu|\bm\phi)$. This allows us to obtain the posterior probability density function after~$K$ measurements,
\begin{equation}
p(\bm\phi|{\bm\mu}) = \frac{p(\bm\mu|\bm\phi)\mathcal{P}(\bm\phi)}{p({\bm\mu})},
\label{eq:posterior_K}
\end{equation}
where $p(\bm{\mu}) = \int_{\Omega} d\bm{\phi}\, p(\bm\mu | \bm\phi)\mathcal{P}(\bm{\phi})$ is the probability to observe the measurement outcomes $\bm{\mu}$.

The posterior probability distribution, Eq.~\eqref{eq:posterior_K}, represents our knowledge about the phases after $K$ measurements and allows us to obtain an estimation $\bm\xi_{\bm\mu}$ of the true phase vector value $\bm\phi_0$ as the conditional expectation of $\bm\phi$ given the known observed values $\bm\mu$. The variance of the posterior, averaged over possible measurement outcomes for a given phase vector $\bm\phi_0$, establishes a confidence interval around the estimator, quantifying the lack of information about the true phase vector. Therefore, to maximize the information gained by the measurement in the Bayesian framework, we adopt the average posterior variance as a cost function to minimize~\cite{Li2018}:
\begin{equation}
\mathcal{C} = \sum_{\bm\mu} \int d\bm\phi\, (\bm\phi-\bm\xi_{\bm\mu})^2\,p(\bm\phi|{\bm\mu})p(\bm\mu | \bm\phi_0),
\label{eq:cost_Bayes_K}
\end{equation}
where we use shorthand notation for the scalar product, $(\bm\phi-\bm\xi_{\bm\mu})^2 \equiv (\bm\phi-\bm\xi_{\bm\mu})\cdot (\bm\phi-\bm\xi_{\bm\mu})$.

To link the Bayesian estimation performance, represented by the cost $\mathcal{C}$, to criterion (i) formulated within the Fisher information approach, we consider the limit of a large number of measurements $K\gg1$. In this limit, as per the Bernstein-von Mises theorem \cite{DasGupta2008, Kleijn2012, pezze2014, Li2018}, the posterior $p(\bm\phi|{\bm\mu})$, Eq.~\eqref{eq:posterior_K}, converges to a normal distribution $p(\bm\phi|{\bm\mu}) \to\mathcal{P}_K(\bm\phi-\bm\phi_0)$:
\begin{equation}
\mathcal{P}_K(\bm\phi-\bm\phi_0)\equiv \frac{\exp\left[-\frac{K}2(\bm\phi-\bm\phi_0)\cdot \mathcal{F}\cdot (\bm\phi-\bm\phi_0)\right]}{\sqrt{(2\pi)^d/\mathrm{det}(K\mathcal{F})}},
\label{eq:BvM_K}
\end{equation}
centered at the true phase value, $\bm\phi_0$, and with the covariance matrix being the inverse of the FIM, $\mathcal{F}^{-1}$, evaluated at~$\bm\phi_0$. Consequently, the variance of asymptotic Bayesian estimation is given by the CRB: $K\mathcal{C}\geq\mathrm{Tr}\mathcal{F}^{-1}$~\cite{pezze2014}. Hence, a sensor that minimizes the Bayesian cost $\mathcal{C}$ in the asymptotic limit minimizes the CRB, thus reaching the QCRB or the HCRB and meeting criterion~(i).

However, as discussed above, sensors that achieve the QCRB can exhibit significant differences in their domains of unambiguous estimation~$\Omega^*$, criterion~(ii). The state and measurement that minimize the Bayesian cost Eq.~\eqref{eq:cost_Bayes_K} are notable for inherently maximizing the information gained by the measurement. In the asymptotic limit $K\gg1$, among sensors with identical FIM, the sensor minimizing the cost $\mathcal{C}$ stands out by requiring the fewest measurement repetitions for the posterior to converge to the asymptotic normal distribution $\mathcal{P}_K(\bm\phi-\bm\phi_0)$. In other words, it requires the least number of measurement repetitions~$K'$ to converge to the CRB. The explicit sensor solutions provided below illustrate that this property is closely linked to maximizing the domain of unambiguous estimation $\Omega^*$, criterion~(ii).

Intuitively, this connection arises because the posterior~$p(\bm\phi|{\bm\mu})$ can only adopt the Gaussian shape $\mathcal{P}_K(\bm\phi-\bm\phi_0)$ if the width of the Gaussian is smaller than the distance between the optimization point $\bm\phi_0$ and the nearest boundary point $\bm\phi_B$ of the domain $\Omega^*$. This imposes a lower bound on the number of measurements $K'$ required to approach the asymptotic posterior, $\mathrm{var}(\mathcal{P}_K) = \mathrm{Tr}\mathcal{F}^{-1}/K'\lesssim(\bm\phi_0-\bm\phi_B)^2$. Since the FIM is upper bounded by the QFIM, the only way for a sensor to maximize the information gained per measurement (minimize $K'$) is to push the boundary of the domain as far away from the optimization point $\bm\phi_0$ as possible.

The relationship between the convergence rate to the CRB and the proximity to the boundaries of the domain~$\Omega^*$ can also be observed within the frequentist paradigm by considering generalizations of the CRB, such as the one introduced by Bhattacharyya~\cite{Bhattacharyya1947}, involving higher-order information matrices beyond the FIM. We illustrate this using the example of the single-parameter GHZ-state interferometer in Appendix~\ref{app:BhB}.

In summary, we have demonstrated that in the limit of a large number of measurements, the sensor optimal with respect to the Bayesian cost $\mathcal{C}$ satisfies criterion~(i), and we have provided motivation for it to also meet criterion~(ii). However, solving the problem of minimizing the cost~$\mathcal{C}$ is extremely challenging due to the highly non-linear nature of the complete likelihood $p(\bm\mu|\bm\phi)$ with respect to the input state $\ket{\psi_\text{in}}$ and measurement $\{M_\mu\}$.

\subsection{New Cost Function. Quantum Compass Solution}
\label{subsec:new_cost}

We seek a scalable method for identifying optimal sensors with large numbers $N$ of qubits (atoms) and/or estimating a large number $d$ of parameters. This requires a cost function like $\mathcal{C}$, maximizing information gain per measurement and expressed in terms of the single-shot conditional probability $p(\mu | \bm\phi)$, in order to facilitate the optimization process, as discussed below.

To achieve this, we introduce an extra measurement repetition with an outcome $\mu$, besides the $K$ measurements with outcomes $\bm\mu$ discussed earlier. By leveraging the $\delta$-like properties of the asymptotic posterior $\mathcal{P}_K$, we demonstrate in Appendix~\ref{app:Cost_derivation} that the Bayesian cost $\mathcal{C}$ for $K+1$ measurements converges to
\begin{equation}
\Xi_K=\sum_{\mu}\int d\bm\phi\,(\bm\phi-\bm\zeta_{\mu})^2 p(\mu | \bm\phi)\mathcal{P}_K(\bm\phi-\bm\phi_0).
\label{eq:cost_Bayes_single_shot}
\end{equation}
While the asymptotic cost $\Xi_K$, Eq.~\eqref{eq:cost_Bayes_single_shot}, corresponds to the limit of the Bayesian cost $\mathcal{C}$, Eq.~\eqref{eq:cost_Bayes_K}, and defines the precision of estimation with $K+1$ measurements of a fixed true phase~$\bm\phi_0$, it is distinctly formulated in terms of the {single-shot} conditional probability~$p(\mu | \bm\phi)$. Thus, it can be understood as the average posterior variance for a \emph{single} measurement of a \emph{random} true phase $\bm\phi$, distributed according to the narrow asymptotic prior $\mathcal{P}_K(\bm\phi-\bm\phi_0)$. To emphasize the single-shot nature of the asymptotic cost $\Xi_K$, we alter the notation for the corresponding single-shot estimator to~$\bm\zeta_{\mu}$.

Due to its single-shot formulation, the new cost~$\Xi_K$ has the important property of being multi-convex with respect to the state, measurement, and estimator degrees of freedom, making it amenable to efficient numerical optimization across these degrees of freedom, as demonstrated in Ref.~\cite{Kaubruegger2023} (see also~\cite{bavaresco2023}). In Appendix~\ref{app:num_id}, we provide details of an iterative algorithm that minimizes the metrological cost~$\Xi_K$, starting from a random state and measurement and progressively increasing the number of measurements~$K$ towards the asymptotic limit where the optimal solution meets the optimality criteria~(i) and~(ii).

Formally the optimization problem is formulated as follows
\begin{equation}
\begin{aligned}
\ket{\psi_\text{in}^*},\;\{M_\mu^*\} &= \lim_{K\to\infty}\left(\ket{\psi_\text{in}}_K,\;\{M_\mu\}_K\right),\\
\ket{\psi_\text{in}}_K,\;\{M_\mu\}_K &= \mathrm{arg}\;\mathrm{min}_{\ket{\psi_\text{in}},\{M_\mu\},\bm\zeta_{\mu}}\;\Xi_K.
\end{aligned}
\label{eq:QC_solutions}
\end{equation}
Subsequently, we explore sensor solutions $\ket{\psi_\text{in}^*},\;\{M_\mu^*\}$. These sensor solutions in combination with the MLE, Eq.~\eqref{eq:MLE}, achieve the minimum QCRB (HCRB), thus reaching the HL. Simultaneously, they maximize the information gain from a single measurement and expand the domain~$\Omega^*$. As discussed in the Introduction, we refer to these sensor solutions as QC solutions due to their single-shot estimator properties. Further details are provided in Sections~\ref{sec:two-parameter_QC} and \ref{sec:three-parameter_QC}.

\begin{figure}[t] 
   \centering
   \includegraphics[]{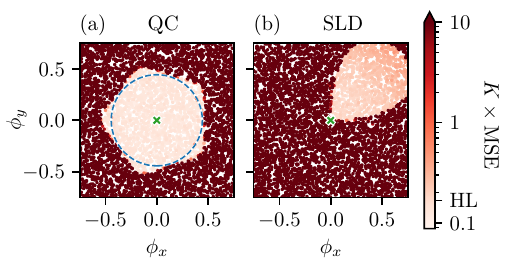}
   \caption{Simulated two-parameter sensing experiment using MLE and sensors comprising $N=4$ qubits. Two sensors, the 2D QC solution~(a) and an SLD-based sensor~(b), are optimized to saturate the HL at $\bm\phi_0=\{0,0\}$~(green cross). Each point represents a randomly chosen true parameter vector $\bm\phi=\{\phi_x,\phi_y\}$, with the color indicating the MSE (scaled by number of measurements $K$) of the MLE. The darker-shade points occupy regions where the MLE converges to incorrect local maxima. The blue dashed circle delimits the region $|\bm\phi|<\phi_\mathrm{max}\approx 0.443$, where the QC solution unambiguously estimates phase vectors along \emph{any} direction. On the other hand, the SLD-based sensor only estimates points unambiguously within a single quadrant, pre-selected by the choice of the initial guess for the MLE (see Appendix~\ref{app:mle}).}
   \label{fig:QC_vs_SLD}
\end{figure}

As a preview of the upcoming examples, Fig.~\ref{fig:QC_vs_SLD}(a) shows the performance of the QC solution for $N=4$ qubits, in estimating two parameters $\bm\phi=\{\phi_x,\phi_y\}$ encoded by the unitary $U(\bm\phi)=\exp[-i(\phi_x J_x+\phi_y J_y)]$ (see Sec.~\ref{sec:two-parameter_QC} for details). The figure presents the results of a simulated maximum likelihood estimation experiment, with the QC solution optimized at $\bm\phi_0=\{0,0\}$ (green cross). The simulation procedure is described in Appendix~\ref{app:mle}. Each point represents a randomly chosen true phase vector and its color indicates the MSE of the MLE, Eq.~\eqref{eq:MLE}, in shades of red. Points with low estimation errors (light red) belong to the domain of unambiguous estimation~$\Omega^*$, while points outside this region exhibit high errors (dark red) due to estimators converging to incorrect local maxima of the likelihood function. For comparison, panel~(b) represents the performance of a sensor employing a measurement based on the SLD, a standard construction for demonstrating QCRB saturability~\cite{Demkowicz2020}.

Unlike the SLD-based sensor, which saturates the QCRB only at the optimization point~$\bm\phi_0$ (green cross), the QC achieves the QCRB across the entire domain of unambiguous estimation, covering any phase vector with an absolute value up to $\phi_{\rm max}\approx0.443$ (blue dashed circle). Importantly, similar to the GHZ-state-based interferometer with $\theta=\pi/2$ [see Fig.~\ref{fig:1D_GHZ}(b)], the 2D SLD-based sensor fails to unambiguously estimate any phase vector except the optimization point~$\bm\phi_0$. This point is located at the intersection of the boundaries of four estimation domains, one of which is highlighted in Fig.~\ref{fig:QC_vs_SLD}(b) due to the choice of the initial guess for the MLE (see Appendix~\ref{app:mle}). In other words, for any true phase vector in the vicinity of~$\bm\phi_0$, the measurement outcomes of the SLD-based sensor are compatible with four estimates. In contrast, the unambiguous estimation domain $\Omega^*$ of the 2D QC solution is maximized around the optimization point $\bm\phi_0$, similar to the optimal GHZ-state-based interferometer ($\theta=0$) shown in Fig.~\ref{fig:1D_GHZ}(a). This general feature suggests that the QC solution can be viewed as a multiparameter counterpart of the GHZ-state-based interferometer.

\subsubsection*{Asymptotic Bayesian Cost: Analytical Solution}
\label{subsec:analytic_quantum_compass}

Surprisingly, the QC solution, i.e. the input state~$\ket{\psi_\text{in}^*}$ and measurement~$\{M_\mu^*\}$, Eq.~\eqref{eq:QC_solutions}, that minimize the cost~$\Xi_K$, Eq.~\eqref{eq:cost_Bayes_single_shot}, in the limit $K\to\infty$ can be analytically determined. This involves expanding the metrological cost in powers of $1/K$, thereby revealing its connection to the Fisher and higher-order information matrices.

We begin by noting that for a fixed input state $\ket{\psi_\mathrm{in}}$ and measurements $\{M_\mu\}$, the estimators minimizing the cost~$\Xi_K$ are known analytically and are given by 
\begin{eqnarray}    
\bm{\zeta}_\mu^* = \int d \bm{\phi} \, \bm{\phi} \, p(\bm{\phi}|\mu), 
\label{eq:MMSEE}
\end{eqnarray}
as a consequence of the cost $\Xi_K$ being quadratic with respect to the estimators. Here $p(\bm{\phi}|\mu) = p(\mu | \bm\phi)\mathcal{P}(\bm{\phi})/p(\mu)$ is the posterior probability distribution of the true phase vector $\bm{\phi}$ given that the outcome~$\mu$ is measured, and $p(\mu) = \int d \bm{\phi}\, p(\mu | \bm\phi)\mathcal{P}(\bm{\phi})$ is the probability of observing outcome $\mu$.

We substitute the optimal Bayesian estimators~\eqref{eq:MMSEE} into the
cost function $\Xi_K$. As a result, the Bayesian cost function is expressed in the following form, depending solely on the sensor's input state and measurement:
\begin{equation}
\Xi_K=\mathrm{var}(\mathcal{P}_K)-\sum_{\mu}\frac{\bm{\varrho}_{\mu}\cdot\bm{\varrho}_{\mu}}{\varrho_{\mu}}.
\label{eq:Cost}
\end{equation}
Here $\mathrm{var}(\mathcal{P}_K)=\int d\bm{\phi}\,(\bm{\phi}-\bm{\phi}_0)^2\,\mathcal{P}_K(\bm{\phi}-\bm{\phi}_0)$
is the asymptotic prior phase variance and we define scalar and vector terms
\begin{equation}
\begin{aligned}
\varrho_{\mu} & =\int d\bm{\phi}\,p(\mu | \bm\phi)\mathcal{P}_K(\bm{\phi}-\bm{\phi}_0),\\
\bm{\varrho}_{\mu} & =\int d\bm{\phi}\,(\bm{\phi}-\bm{\phi}_0)\,p(\mu | \bm\phi)\mathcal{P}_K(\bm{\phi}-\bm{\phi}_0).
\end{aligned}
\label{eq:scalar_vector_terms}
\end{equation}
Note that the expression for the cost $\Xi_K$ in the form of Eq.~\eqref{eq:Cost} resembles the generating function in the Bhattacharyya~(see App.~\ref{app:BhB}) and other generalized bounds~\cite{gessner2023}. Similar to the generating function, we can expand the metrological cost in powers of $1/K$ since we are interested in minimizing it in the limit $K\to\infty$. 

To simplify, we present here the expansion using the asymptotic prior~$\mathcal{P}_K$ defined by an isotropic FIM, $\mathcal{F}=F\mathbb{1}$, which is relevant to most of the sensors we examine in the following sections. We also rescale the number of measurements $K$ to absorb~$F$, such that $\mathcal{P}_K(\bm{\phi}-\bm{\phi}_0)\propto \exp[-(\bm{\phi}-\bm{\phi}_0)\cdot(\bm{\phi}-\bm{\phi}_0)K/2]$. The Taylor expansion of the asymptotic cost reads:
\begin{equation}
\Xi_K=\mathrm{var}(\mathcal{P}_K)-\frac1{K^2}\left\{{C}^{(1)} + \frac{{C}^{(2)}}{K} + \frac{{C}^{(3)}}{K^{2}}+\ldots\right\},
\label{eq:Cost_expansion}
\end{equation}
for details see Appendix~\ref{app:asymptotic_cost}.

In general, the coefficients $C^{(\ell)}$ depend on the covariance matrix of the asymptotic normal prior $\mathcal{P}_K(\bm{\phi}-\bm{\phi}_0)$, Eq.~\eqref{eq:BvM_K}. As the covariance matrix is the inverse of the FIM of the sensor, we should use the maximum FIM (often known) to define the metrological cost. By minimizing this cost, we can determine the input state and measurement of the unique sensor that not only maximizes the single-shot information gain and expands the domain~$\Omega^*$, but also saturates the maximum FIM. In cases where the maximum FIM is unknown, it can be determined concurrently with the optimal sensor solution by iteratively optimizing $\Xi_K$ and adjusting the FIM for the prior, as will be elaborated in Ref.~\cite{Shankar202x}.

In the following, we study the QC solutions for two- and three-parameter estimation using a sensor which encodes the parameters via SU(2) rotations. In these cases, the maximum FIM at the most sensitive point $\bm\phi_0=\vec0$ is known and isotropic. The coefficients $C^{(\ell)}$ for the corresponding isotropic~\footnote{In the general case of an anisotropic maximum FIM, $\mathcal{F}$, optimization of the sensor can be carried out in a new coordinate system $\phi_i'=\sum_j[\mathcal{F}^{1/2}]_{ij}\phi_j$, where the FIM becomes the identity matrix. This leads to an isotropic normal prior, and the MSE is determined with the inverse FIM, $\mathcal{F}^{-1}$, as a weight matrix.} normal prior are provided in Appendix~\ref{app:asymptotic_cost}. The leading term is given by the FIM, $C^{(1)}={\rm Tr}\,\mathcal{F}$, while the higher-order terms are defined by corresponding higher-order information matrices~$\mathcal{I}^{(\ell)}$ involving higher-order derivatives of conditional probabilities $p(\mu | \bm\phi)$, similar to the generalized information coefficients in the Bhattacharyya bound.

Minimizing the cost $\Xi_K$ in the limit as $K\to\infty$ corresponds to the successive maximization of the Taylor expansion terms $C^{(\ell)}$ in Eq.~\eqref{eq:Cost_expansion}. Our approach eliminates the degeneracy among sensors saturating the QCRB or the HL, as captured by the leading term $C^{(1)}$ determined by the FIM. This is achieved by optimizing as many of the higher-order terms $C^{(\ell>1)}$ as necessary to uniquely identify a sensor with the highest single-shot information gain. To illustrate this we consider the aforementioned single-parameter sensor.

\emph{Single-parameter estimation example}. The Bayesian metrological cost $\Xi_K$ for the GHZ-state based interferometer comprises the following Taylor terms:
\begin{align}
C^{(1)} &= \mathcal{F} = N^2,\\
C^{(2)} &= -\frac{N^4}{\cos^2\theta}.
\end{align}
As expected, while $C^{(1)}$ remains invariant with respect to the measurement basis phase $\theta$, the second-order information term~$C^{(2)}$ has a maximum at~$\theta=0$ (see Appendix~\ref{app:asymptotic_cost}). This corresponds to the optimal sensor depicted in Fig.~\ref{fig:1D_GHZ}(a). Hence, this single-parameter GHZ state interferometer belongs to the family of QC solutions.

Minimizing the cost Eq.~\eqref{eq:Cost_expansion} analytically involves using the method of Lagrange multipliers to incorporate the POVM constraints. This results in a set of polynomial equations for variables parametrizing the state and measurement. In general, there may be many local minima and thus several solutions for the set of equations. However, in practice, it is straightforward to numerically identify the correct solution by leveraging the multi-convexity of the metrological cost, Eq.~\eqref{eq:cost_Bayes_single_shot}, as described in Appendix~\ref{app:num_id}. The obtained numerical solution can be used to identify the relevant symmetries of the solution, efficiently parameterize the state and measurement, and either solve the resulting simplified problem with arbitrary precision or even identify the solution analytically as we present below for SU(2) sensors. 

To conclude this section, we emphasize our main finding---the metrological cost $\Xi_K$, Eq.~\eqref{eq:cost_Bayes_single_shot}. Optimizing this cost for an asymptotic number of measurement repetitions ($K\to\infty$) establishes a systematic method, Eq.~\eqref{eq:QC_solutions}, for identifying the optimal sensor for local estimation, referred to as the QC solution. Analytical solutions are attainable for sufficiently symmetric cases, while the numerical method (see Appendix~\ref{app:num_id}) enables the approximate identification of the optimal sensor, accounting for practical constraints like projective measurements. The numerical method also facilitates obtaining QC solutions with an anisotropic maximum FIM. In the subsequent sections of the paper, we will apply the proposed technique to two- and three-parameter SU(2) sensors, and discuss the remarkable properties of the resulting QC solutions. Finally, we demonstrate how the metrological cost $\Xi_K$, in combination with variational metrology techniques, allows for the design of simple quantum circuit-based sensors that approximate the theoretically optimal performance of the QC solutions.

\section{Two-parameter SU(2) QC solutions}
\label{sec:two-parameter_QC}

In this and the next section, we apply our method to identify QC solutions for paradigmatic multiparameter sensing problems. In this section, our focus is on estimating two phases, $\bm\phi=\{\phi_x,\phi_y\}$, encoded by the SU(2) unitary $U(\bm\phi) = \exp[-i(\phi_x J_x+\phi_y J_y)]$. Here, $J_{x,y,z}$ are collective spin operators obeying the standard commutation relation $[J_x,J_y]=iJ_z$. The physical system underlying this can be an ensemble of $N$ spin-$1/2$ atoms or a qudit with spin $J=N/2$. This sensor is applicable for tasks such as direction reference alignment~\cite{Bagan2001,Kolenderski2008,Bartlett2007} or detecting two-dimensional (2D) magnetic or electric fields~\cite{Vaneph2013,Baumgratz2016,Kaubruegger2023}.

The section is organized as follows. First, we present analytical two-parameter QC solutions for even $N$, which are optimized for estimating phases around the point $\bm\phi_0=\{0,0\}$. Then, we discuss the sensing performance of the QC solution in view of the local estimation optimality criteria discussed in Sec.~\ref{sec:optimal_multiparameter}. Subsequently, we consider the odd~$N$ QC solutions, which exhibit a unique feature of multiparameter quantum metrology --- the incompatibility of optimal measurements. Finally, by applying our method at points away from the origin, we demonstrate how it can be used to determine the Heisenberg limit (HL) for two-parameter sensing at any point $\bm\phi$.

\subsection{2D QC solution for Even Number of Atoms}
\label{subsec:2D_QC_even}

\begin{figure}[t] 
   \centering
   \includegraphics[width=2.5in]{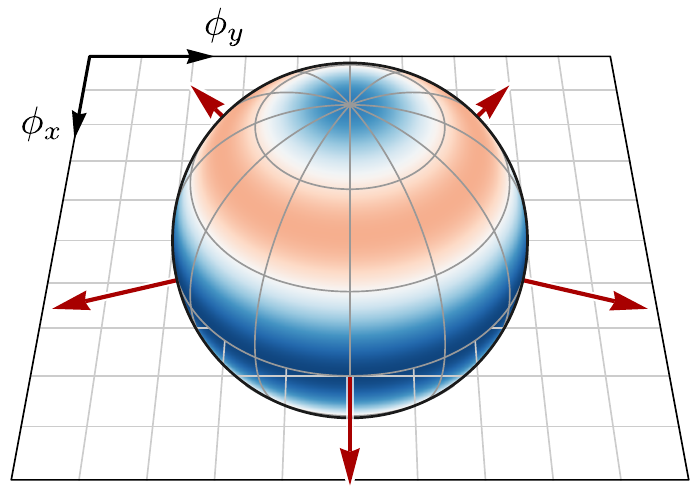} 
   \caption{Two-parameter QC solution for $N=4$ qubits. The Wigner function representing the input state is shown, along with the directions of the single-shot estimators~$\bm{\zeta}^*_{\mu}$ (red arrows), each corresponding to one of the $N+1=5$ measurement outcomes $\mu$.}
   \label{fig:2D_QC_Wigner}
\end{figure}

We employ the numerical method outlined in Sec.~\ref{subsec:new_cost} and App.~\ref{app:num_id} to identify, within the symmetric subspace~\footnote{Reference~\cite{Kaubruegger2023} demonstrated that the optimal solution for a Bayesian SU(2) sensor with a narrows prior distribution, $K^{-1}\lesssim 1/N$, resides in the symmetric subspace of the entire $N$-qubit Hilbert space. This finding aligns with previous research that identified input states belonging to the symmetric subspace and capable of saturating the QCRB for two- or three-parameter SU(2) sensors~\cite{Kolenderski2008}.}, the state, measurement, and corresponding Bayesian estimators, Eq.~\eqref{eq:MMSEE}, that are optimized for estimating phase values around $\bm\phi_0=\{0,0\}$ in the limit $K\to\infty$. Figure~\ref{fig:2D_QC_Wigner} displays the Wigner function of the optimal state, along with the single-shot estimators represented as a set of directions in the parameter space. Such a visualization offers a clear way to observe and understand the symmetry properties of a numerical sensor solution. Specifically, for two parameters, the numerically observed rotational symmetry in the estimators and input state leads to the formulation of the following ansatz for the optimal measurement. For an even number of atoms $N$, the optimal input state $\ket{\psi_{\rm in}^{\rm 2D}}$ and the measurement basis $\ket{\mu}$ of the 2D QC solution are given by:
\begin{align}
\label{eq:2D_QC_state}
\ket{\psi_{\rm in}^{\rm 2D}} &= \ket{J,0},\\
\label{eq:2D_QC_measurement}
\ket{\mu} &= e^{-i(\varphi_{\mu} -\pi) J_z}\ket{f}.
\end{align}
Here $J=N/2$ and $\ket{J,m}$ denote basis composed of the eigenstates of the collective spin operator $J_z$, such that $J_z\ket{J,m}=m\ket{J,m}$.

The state in Eq.~\eqref{eq:2D_QC_state} was previously identified as the one that sets the ultimate limit in two-parameter sensing~\cite{Vaneph2013}.
The corresponding measurement basis, Eq.~\eqref{eq:2D_QC_measurement}, that achieves this limit is defined by a fiducial state~$\ket{f}$, which is rotated by an angle $\varphi_{\mu} = 2\pi \mu/(2J+1)$, corresponding to a measurement outcome $\mu=-J,\ldots,J$, with the projector $M_{\mu}=\ket{\mu}\bra{\mu}$. Strong symmetry constraints in the 2D QC allow for the direct guess of the analytical form of the fiducial state from the numerical solution:
\begin{equation}
\ket{f} = \frac{1}{\sqrt{2J+1}}\sum_{m=-J}^Je^{i\frac{\pi}2|m|}\ket{J,m}.
\label{eq:2D_fiducial_state}
\end{equation}

The optimal single-shot Bayesian estimators Eq.~\eqref{eq:MMSEE} for the 2D QC solution take the form of a ring $\bm{\zeta}^*_{\mu}=\{{\zeta}^*_{x},{\zeta}^*_{y}\}=\{R\cos\varphi_{\mu},R\sin\varphi_{\mu}\}$, where the angles $\varphi_{\mu}$ offer evenly distributed estimated directions in the two-parameter space (see Fig.~\ref{fig:2D_QC_Wigner}). The radius of this ring is given by $R=2\sqrt{J(J+1)}K^{-1} + O(K^{-2})$. As all estimators are positioned on a ring, a single measurement with the optimal sensor yields only directional information in the parameter space. However, the MLE, Eq.~\eqref{eq:MLE}, which combines outcomes from multiple measurement samples, reveals the full information about the true phase vector~$\bm\phi$. This feature is consistent across all optimal sensor solutions that we subsequently examine, leading us to term these limiting sensor solutions as multiparameter quantum compasses. 

As an aside, we initially coined the term `quantum compass' in Ref.~\cite{Kaubruegger2023} to describe a sensor optimized within the Bayesian framework for single-shot sensing tasks, which are characterized by prior distributions with small but finite widths. This contrasts with the many-repetitions scenario that we explore in this study, where quantum compass solutions saturate the Fisher information limits. The quantum compass solutions from Ref.~\cite{Kaubruegger2023} can serve as a starting point for the numerical method outlined in Appendix~\ref{app:num_id} to find the QC solutions that we are interested in this study.

\subsubsection{Sensing Performance of the 2D QC solution}

In order to assess the performance of the 2D QC solution in relation to criteria (i) and (ii), we examine the MSE. We begin by establishing the corresponding bounds on the MSE using the FIM and QFIM. Subsequently, we analyze the ML estimator that can achieve these bounds.

The QC input state $\ket{\psi_{\rm in}^{\rm 2D}}$, \eqref{eq:2D_QC_state}, for even number of atoms $N$ is \emph{quasiclassical} i.e., the commutator of the SLDs, Eq.~\eqref{eq:sld}, has zero expectation value with respect to the state. This indicates the absence of measurement incompatibility between the two parameters, allowing for the saturation of the QFIM with an optimal measurement (refer to Appendix~\ref{app:QCRB_saturability}). Furthermore, the input state and the optimal measurement basis~$\ket{\mu}$ exhibit antiunitary symmetry, which denotes an invariance under the simultaneous operations of $\pi$-rotation around the $y$-axis and complex conjugation~\cite{Miyazaki2022}. This implies that the corresponding sensor saturates the QFIM of the input state for all values of the phase vector. Indeed, using Eqs.~\eqref{eq:FIM} and \eqref{eq:QFIM} together with Eqs.~\eqref{eq:2D_QC_state} and \eqref{eq:2D_QC_measurement} we obtain the limit on the MSE of the 2D QC solution:
\begin{equation}
\label{eq:2D_QC_QCRB}
K\times\mathrm{MSE}(\bm{\phi})\ge \tr{\mathcal{F}^{-1}} = \tr{\mathcal{F}_Q^{-1}} = \frac{1+(\mathrm{sinc}|\bm\phi|)^{-2}}{2J(J+1)}.
\end{equation}
Equation~\eqref{eq:2D_QC_QCRB} explicitly demonstrates that the state and measurement of the 2D QC solution meet criterion~(i) as they saturate the QFIM, which in turn corresponds to the HL at $\bm\phi_0$~\cite{Vaneph2013}.

\begin{figure}[t] 
   \centering
   \includegraphics[]{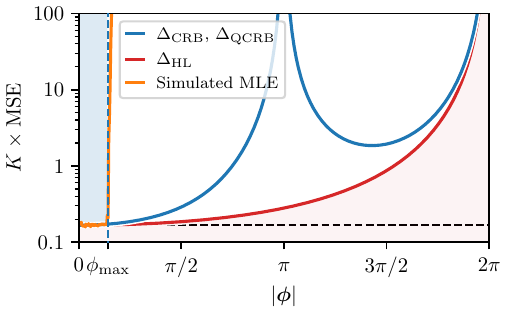} 
   \caption{Sensing performance of the 2D QC solution for $N=4$ qubits. The blue solid line shows the CRB ($\Delta_\text{CRB}$) for the 2D QC solution as a function of $|\bm\phi|$, which coincides with the QCRB ($\Delta_\text{QCRB}$) for all parameter values. The orange solid line shows the MSE (scaled by number of measurements $K$) averaged over direction, obtained from a simulated maximum likelihood experiment. The vertical dashed line delimits the region (blue) where the MLE achieves the $\Delta_\text{QCRB}$. The red solid line shows the HL. The region (red) below this line is inaccessible to any $4$-qubit sensor. The QC solution saturates the HL only at $\bm\phi=0$, where the sensor has the highest sensitivity, indicated by the horizontal dashed line.
   }
   \label{fig:2D_MSE}
\end{figure}

Figure~\ref{fig:2D_MSE} shows the sensing performance of the 2D quantum compass comprising $N=4$ atoms. The blue curve represents the precision limit given by Eq.~\eqref{eq:2D_QC_QCRB} as a function of the absolute value of the phase vector. Any quantum sensor optimized to achieve maximum precision at $\bm\phi_0=\{0,0\}$ cannot surpass the limit set by the blue curve. The black dashed line indicates the HL at $\bm\phi_0=\{0,0\}$ and $N=4$, $\Delta_{\rm HL} = 1/[J(J+1)]=1/6$~\cite{Vaneph2013}. The performance of the quantum compass closely approaches the black dashed line only for phase values around zero. Furthermore, the MSE diverges when the absolute values of the phase vector are multiples of $\pi$. The divergence at even multiples of~$\pi$ is a fundamental property of multiparameter SU(2) sensors, see Appendix~\ref{app:QCRB_divergence}. However, the divergence at odd multiples of~$\pi$ arises due to an additional symmetry of the input state~$\ket{J,0}$, which is invariant under $\pi$-rotations around any axis in the $xy$-plane. Consequently, it is possible to find a different quantum compass solution that surpasses the blue curve by optimizing the Bayesian cost at a shifted phase vector $\bm\phi=\bm\phi_0$, e.g., with $|\bm\phi_0|=\pi$, as discussed below in Sec.~\ref{subsec:HL_in_2D}. However, such a sensor will not excel at measuring small phase values and, as a result, will not approach the HL at $\bm\phi_0=\{0,0\}$ (dashed line).

Finally, as discussed in Sec.~\ref{subsec:FI_framework}, when the measurement data consists of a large number $K$ of independent samples $\bm\mu=\{\mu_1,\ldots,\mu_K\}$ the CRB given by Eq.~\eqref{eq:2D_QC_QCRB}, can be achieved using the MLE method outlined in Eq.~\eqref{eq:MLE}. In Figure~\ref{fig:2D_MSE}, the orange line represents the MSE scaled by the number of measurements $K$ for the 2D QC solution employing the MLE technique. Specifically, we simulate the estimation experiment using $K=10^5$ measurement repetitions for 5000 randomly selected true phase vectors within a circle $|\phi|<1$. The circle is divided into 50 rings of equal thickness. The MSE of the sample points belonging to each ring is averaged and presented in the figure as a function of the corresponding ring radius $|\phi|$. The details of the simulated maximum likelihood estimation are given in Appendix~\ref{app:mle}.

The figure shows that the performance of the 2D QC solution combined with the MLE (orange line) achieves the QCRB (blue line) for all phases up to a certain threshold $|\phi|<\phi_{\rm max}$. However, for phase vectors with absolute values exceeding this threshold, the estimation error grows dramatically beyond the scale of the figure due to MLE convergence to an incorrect local maximum. The limit $\phi_{\rm max}$ for the 2D QC solution with $N=4$ atoms is approximately $0.443$, which is indicated by the vertical dashed line in the figure. As a result, we observe that although the MLE achieves the CRB, it is only possible within a certain domain $\Omega^*$ of the parameter space where phases can be unambiguously estimated, as discussed earlier. In the context of criterion~(ii), we now examine the properties of the domain $\Omega^*$ for the 2D QC solution more closely to provide arguments suggesting that the domain is maximized around the optimization point $\bm\phi_0$.

\begin{figure}[t] 
   \centering
   \includegraphics[]{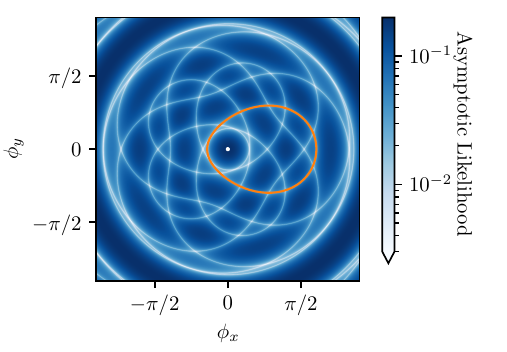}
   \caption{Asymptotic likelihood of the 2D QC solution for $N=4$ qubits. Taking the true parameter $\bm\phi_0=\{0,0\}$ (central white dot), Eq.~(\ref{eq:rescaled_likelihood}) is plotted as a function of $\phi_x,\phi_y$ in parameter space. Domains of large likelihoods (dark blue) are demarcated by white curves of zero likelihood. The central domain contains a local maximum of the likelihood at $\bm\phi_0$. The orange curve indicates the zeros of the likelihood that occur due to a vanishing probability for a particular outcome, $p(\mu=0|\bm\phi)=0$.
   }
   \label{fig:2D_QC_Likelihood}
\end{figure}

\subsubsection{Domain of unambiguous estimation for the 2D QC solution}
\label{subsec:Domain_2D}

The Fisher information, being estimator-agnostic, does not reveal properties of estimators such as unambiguity. Hence, we delve into the properties of a specific asymptotically optimal estimator, the MLE, Eq.~\eqref{eq:MLE}, to elucidate criterion~(ii) for the 2D QC solutions.

The MLE approach entails searching for the maximum of the likelihood $p(\bm\mu|\bm\phi)$ which typically exhibits a complex structure with multiple local maxima. This results in domains where phases can be unambiguously estimated through a straightforward local optimization strategy. These domains of unambiguous estimation can be seen as the multidimensional counterparts of the interference fringes observed in single-parameter estimation scenarios~(see Fig.~\ref{fig:1D_GHZ}).

To gain insight into the underlying causes of the estimation ambiguity we explore the properties of the likelihood function of the QC solution. Specifically, we examine $p(\bm\mu|\bm\phi)$ in the limit $K\to\infty$, where the MLE, Eq.~\eqref{eq:MLE}, becomes optimal and the frequencies of different measurement outcomes can be approximated by their corresponding probabilities, such that the rescaled likelihood for a true phase $\bm\phi_0$ reads:
\begin{equation}
p(\bm\mu|\bm\phi)^{1/K}\to\prod_{\mu=-J}^J p(\mu|\bm\phi)^{p(\mu | \bm\phi_0)}.
\label{eq:rescaled_likelihood}
\end{equation}
Note that the asymptotic likelihood, Eq.~\eqref{eq:rescaled_likelihood}, is independent of a specific realization of measurement outcomes~$\bm\mu$, thus, representing universal properties of the sensor.

In Fig.~\ref{fig:2D_QC_Likelihood} we show the asymptotic likelihood of the 2D QC solution for $N=4$ atoms as a function of the phases $\phi_x$ and $\phi_y$. The true vector of phases is $\smash{\bm\phi_0=\{0,0\}}$ (white dot). The color gradient, from white (low likelihood) to dark blue (high likelihood), illustrates the likelihood distribution. The figure reveals 2D counterparts of the 1D interference fringes observed in the GHZ-state interferometer, manifested as dark blue high-likelihood regions separated by white zero-likelihood curves. Each dark blue domain’s peak corresponds to an estimate of $\bm\phi_0$. The correct estimate of $\bm\phi_0$ is confined to the central (approximately) circular dark blue domain, where the likelihood peaks precisely at $\{0,0\}$. This domain is also apparent in the simulated estimation experiment data presented in Fig.~\ref{fig:QC_vs_SLD}(a).

To clarify the criterion~(ii), we need to precisely determine the radius of the domain $\Omega^*$ for the 2D QC solutions. Notably, the expression for the asymptotic likelihood, Eq.~\eqref{eq:rescaled_likelihood}, indicates that if any probability $p(\mu|\bm\phi)$ is zero for a specific $\bm\phi'$, then $\bm\phi'$ is a root of the likelihood function. The likelihood is a positive function, thus continuous lines of roots form barriers between regions where the likelihood may have a single maximum (be convex), as shown in Fig.~\ref{fig:2D_QC_Likelihood}. This is further exemplified by the single-parameter GHZ-state interferometer, which demonstrates how zeros in the probabilities can lead to estimation ambiguity, as depicted in Fig.~\ref{fig:1D_GHZ}(b). Therefore, instead of examining the likelihood, we can inspect the roots of individual probabilities $p(\mu|\bm\phi)$, which simplifies the analysis, as presented in Appendix~\ref{app:domains}.

\begin{figure}[t] 
   \centering
   \includegraphics[]{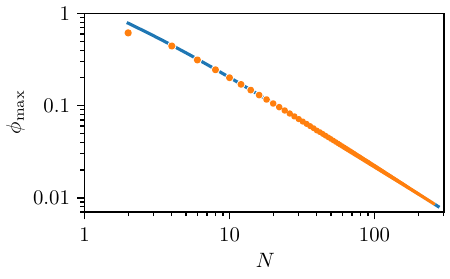}
   \caption{Radius of the domain of unambiguous estimation~$\Omega^*$ for the 2D QC solution as a function of the number of qubits~$N$. The orange dots correspond to exact numerical values, while the blue line represents the asymptotic behavior of $\phi_{\rm max}$ as given by Eq.~\eqref{eq:phi_max_asymptotic}.}
   \label{fig:2D_phi_max}
\end{figure}

We numerically calculate the domain radii for the QC comprising various numbers of atoms $N=2,\ldots,256$ and plot them as orange dots in Fig.~\ref{fig:2D_phi_max}. One can show that, in the asymptotic limit, $N\gg1$, the domain radius is given by
\begin{equation}
\phi_{\rm max} = \frac{\gamma_0}{\sqrt{J(J+1)}},
\label{eq:phi_max_asymptotic}
\end{equation}
where $J=N/2$ and $\gamma_0\approx 1.10836$ is the first root of the equation $1 - \gamma \{J_0(\gamma) + \frac{\pi}{2} [\pmb{H}_0(\gamma) J_1(\gamma)-\pmb{H}_1(\gamma) J_0(\gamma)]\} = 0$, with $\pmb{H}_{n}(\gamma)$ being the Struve function and $J_{n}(\gamma)$ the Bessel function of the first kind. The asymptotic $\phi_{\rm max}$ given by Eq.~\eqref{eq:phi_max_asymptotic} is represented by the blue line in Fig.~\ref{fig:2D_phi_max}.

\subsubsection{Relation between the maximum domain size and ultimate sensing precision}
\label{subsec:Domain_HL_relation}

The expression Eq.~\eqref{eq:phi_max_asymptotic} indicates an interesting connection between the asymptotic domain radius $\phi_{\rm max}$ and the HL. In our case with two parameters, the HL is defined by the maximal QFIM $\mathcal{F}_Q=F\mathbb{1}_2$, where the diagonal elements are $F=2J(J+1)$. Remarkably, the asymptotic normal distribution~$\mathcal{P}_K$, Eq.~\eqref{eq:BvM_K}, defined by the QFIM appears to be entirely contained within the domain~$\Omega^*$ of the 2D QC solution when the number of measurements is as low as~$K\sim1$. This is because the variance of~$\mathcal{P}_K$ along the direction towards the nearest domain boundary is of the order of the squared distance to the boundary: $\phi_{\rm max}^2/\mathrm{var}(\mathcal{P}_K)=KF\phi_{\rm max}^2 = 2K\gamma_0^2\approx2.4K$.
This suggests that, on average, the 2D QC solution can achieve the CRB with a number of measurement repetitions on the order of 1. This behavior mirrors that of the GHZ-state interferometer, where the convergence to the CRB with $K\simeq3$ repetitions is precisely indicated by the Bhattacharyya bound, as discussed in Appendix~\ref{app:BhB}.

These findings reveal that both the optimal single-parameter and two-parameter sensors exhibit domains of unambiguous estimation that are \emph{just about} large enough to allow for rapid attainment of the CRB with $K\sim1$.
Based on this, we conjecture a universal relation between the ultimate precision of a quantum sensor and the maximum size of its domain of unambiguous estimation:
\begin{equation}
\phi_{\rm max}=\frac{r}{\sqrt{F}},
\label{eq:general_domain_size}
\end{equation}
where $F$ is the maximal eigenvalue of the FIM defining the HL, such that for a general $d$-parameter estimation it can be estimated as $F\approx d/\Delta_{\rm HL}$. The coefficient $r$ is of order~1. For the single-parameter GHZ-state interferometer, we have $\phi_{\rm max} = \pi/(2N)$ and $F=N^2$, yielding $r_{\rm 1D}=\pi/2\approx1.5707$. Similarly, for the two-parameter QC solution, we obtain a similar value: $r_{\rm 2D}=\sqrt2 \gamma_0\approx 1.5675$. In the following, we demonstrate that a 3D QC solution designed to estimate three parameters encoded by all three generators of the SU(2) group also exhibits a domain~$\Omega^*$ characterized by the coefficient of a similar magnitude, $r_{\rm 3D}\approx1.5$ (see Sec.~\ref{subsec:Domain_3D}).

In summary, the domain of unambiguous estimation~$\Omega^*$ of the 2D QC solutions exhibits a symmetric circular shape (or tending toward a spherical shape in the case of 3D QC solutions, see Sec.~\ref{subsec:Domain_3D}), centered around the optimization point~$\bm\phi_0$, with a radius $\phi_{\rm max}$ scaling similar to the GHZ-state interferometer, as captured by Eq.~\eqref{eq:general_domain_size}. This suggests that the QC solutions maximize the volume of the domain~$\Omega^*$, criterion (ii), in line with our expectations for a sensor maximizing single-shot information gain, as discussed in Sec.~\ref{subsec:Bayesian_framework} and App.~\ref{app:BhB}. Furthermore, the proposed relation between ultimate precision and domain size, Eq.~\eqref{eq:general_domain_size}, allows us to precisely define what `local’ means in the \emph{local estimation} paradigm of quantum metrology.

\subsection{2D QC Solutions for Odd Number of Atoms}
We now turn our attention to optimal two-parameter sensing with an odd number of atoms $N$. This scenario is notable as it exhibits features arising from the incompatibility of optimal measurements, which is a unique hallmark of multiparameter quantum metrology. This incompatibility makes it impossible to saturate the QCRB. In this case, the achievable sensing performance is described by the HCRB. We demonstrate that our approach allows for the identification of explicit optimal solutions in such a general multiparameter scenario.

For an odd number of atoms $N$ with the corresponding half-integer collective spin $J=N/2$, the numerical solution suggests $z$ rotation symmetry for the state and measurement, similar to the case of an even number of atoms. Thus, the projective measurement basis is defined by rotating a fiducial state around the $z$ axis according to~Eq.~\eqref{eq:2D_QC_measurement}. The corresponding input and fiducial states of the QC, optimized for $\bm\phi_0=\{0,0\}$, read:
\begin{align}
\label{eq:2D_QC_state_odd_N}
\ket{\psi_{\rm in(odd)}^{\rm 2D}} &= \ket{J,\tfrac12},\\
\label{eq:2D_fiducial_state_odd_N}
\ket{f_{\rm odd}} &= \frac{1}{\sqrt{2J+1}}\sum_{m=-J}^Je^{i\frac{\pi}2|m-1/2|}\ket{J,m}.
\end{align}

In this case the input state Eq.~\eqref{eq:2D_QC_state_odd_N} is not \emph{quasiclassical} as the commutator of generators $[J_x,J_y]$ has a non-zero expectation value (see App.~\ref{app:QCRB_saturability}). As a result, the QCRB, $\Delta_{\rm QCRB}=1/[J(J+1)-1/4]$~\cite{Vaneph2013}, is not saturable for any choice of measurement. Instead, the QC measurement defined by the fiducial state~Eq.~\eqref{eq:2D_fiducial_state_odd_N} saturates the HCRB. The CRB of the 2D QC solution, calculated at $\bm\phi=\{0,0\}$ using Eq.~\eqref{eq:FIM}, and the HCRB corresponding to the state~$\ket{\psi_{\rm in(odd)}^{\rm 2D}}$~, which can be analytically evaluated for pure states as outlined in Ref.~\cite{Matsumoto:2002aa}, read:
\begin{equation}
\Delta_{\rm CRB}=\Delta_{\rm HCRB} = \frac4{\left[J+\frac12+\sqrt{(J-\frac12)(J+\frac32)}\right]^2}.
\end{equation}
Again, this equality demonstrates that the 2D QC solution meets criterion (i) by saturating the HCRB, which defines the two-parameter HL in this setting~\footnote{This is the minimal HCRB across eigenstates of the collective $J_z$~\cite{Matsumoto:2002aa}, which exhibit isotropic sensitivity due to their rotational invariance around the $z$-axis, mirroring the symmetry of the sensing problem.}. Note that the state Eq.~\eqref{eq:2D_QC_state_odd_N} differs from the one proposed as the optimal for odd~$N$ in Ref.~\cite{Vaneph2013}.

The domain $\Omega^*$ for an odd $N$ is as wide as that of a QC solution with a similarly sized even $N$, thereby meeting criterion (ii). However, unlike the 2D QC solution with an even $N$ (Sec.~\ref{subsec:2D_QC_even}), the odd $N$ solution lacks additional symmetries. Consequently, the CRB for the 2D QC with odd $N$ fails to saturate the phase-dependent fundamental limit set by the HCRB everywhere except at the optimization point $\bm\phi_0$. We will further explore a similar behavior of the CRB for the three-parameter QC solution with $N=4$ qubits, which lacks antiunitary symmetry (see Sec.~\ref{subsec:3D_QC_N4}).

\subsection{Heisenberg Limit in Two-Parameter SU(2) Interferometry}
\label{subsec:HL_in_2D}

The numerical approach described in Appendix~\ref{app:num_id} does not rely on any assumptions regarding the symmetries of the input state and measurement of the sensor. This flexibility enables us to determine the multiparameter HL in scenarios where a lack of symmetry hinders intuitive guessing of an optimal state that minimizes the QCRB or HCRB. Here, we demonstrate this capability by investigating the HL in two-parameter SU(2) interferometry.

Unlike the single-parameter case, where the HL is independent of the phase value to be estimated, in multiparameter metrology, the HL varies significantly depending on the specific phase value of interest. Our goal, therefore, is to calculate the two-parameter HL, considering a bias phase vector $\bm\phi_0$ as the reference point for sensor optimization. In other words, we are interested in the optimal estimation of small deviations $\delta\bm\phi=\bm\phi-\bm\phi_0$ from the bias phase vector. The corresponding phase encoding Hamiltonian reads $H = |\bm\phi_0|J_\parallel + \delta\bm\phi\cdot\bm J$, where $J_\parallel=\bm J\cdot\bm\phi_0/|\bm\phi_0|$ represents the phase shift generator along~$\bm\phi_0$. Thus, the bias phase vector introduces an anisotropy in the parameter space, leading to a distinction between parallel and perpendicular directions to $\bm\phi_0$ which are associated with the generator $J_\parallel$ and the remaining generators that do not commute with~$J_\parallel$, respectively. As a result, the maximum FIM permitted by quantum mechanics exhibits this distinction in the form of anisotropic sensitivity to the parameters along and perpendicular to the direction of~$\bm\phi_0$. In our approach to finding the optimal sensor in the form of the QC, we address this anisotropy of the target FIM by introducing anisotropic prior distributions.

Specifically, we consider a sensor composed of $N=4$ atoms and optimize the Bayesian cost $\Xi_K$ using infinitesimally narrow asymptotic prior distributions $\mathcal{P}_K$ centered at $\bm\phi_0=\{\phi_0,0\}$ (we choose coordinates with $x$-axis along~$\bm\phi_0$). These prior distributions exhibit anisotropy in directions parallel and perpendicular to $\bm\phi_0$. 
For each $\bm\phi_0$, we vary the prior anisotropy and find the point where the sensor minimizing the cost $\Xi_K$ has an FIM with a matching anisotropy. At this special value of anisotropy, minimizing the cost $\Xi_K$ is equivalent to minimizing the CRB, thus determining the HL. Using this technique, we identify the following ansatz for the optimal input state in the $\ket{J,m}$ basis: 
\begin{align}
\label{eq:psi_HL_2D}
\ket{\psi_{\rm in}^{\rm 2D}}_{\bm\phi_0} &= \exp\left(i\frac{\phi_0}2 J_{\parallel}\right)\ket{\lambda},\\
\ket{\lambda} &= \frac{1}{\sqrt{1+\lambda^2}}\left[\ket{2,0} - \frac{\lambda}{\sqrt{2}}(\ket{2,-2} + \ket{2,2})\right],
\end{align}
where $J_{\parallel} = J_x$ and $\lambda\in[0,\infty)$.

The state given by Eq.~\eqref{eq:psi_HL_2D} interpolates between the $m=0$ state, Eq.~\eqref{eq:2D_QC_state}, which is optimal for two-parameter sensing at $\bm\phi_0=\{0,0\}$, and the GHZ state which is optimal at $\bm\phi_0=\{2\pi,0\}$ where only one component can be estimated (see Appendix~\ref{app:QCRB_divergence}). Since the state is quasiclassical, the QCRB provides an achievable lower bound. By using the QFIM given by Eq.~\eqref{eq:QFIM}, we can calculate the QCRB for the state described by Eq.~\eqref{eq:psi_HL_2D}:
\begin{equation}
\label{eq:QCRB_HL_2D}
\Delta_{\rm QCRB} = \frac{1}{4}
   \left\{\frac{1+\lambda ^2}{\big[(\sqrt{3}+\lambda)
   \,\mathrm{sinc}\frac{\phi_0}{2}\big]^2}+
   \frac{1+\lambda ^2}{(\sqrt{3}-\lambda)^2}\right\}
\end{equation}
The HL for estimating two parameters using a sensor with $N=4$ atoms is obtained by minimizing the QCRB, Eq.~\eqref{eq:QCRB_HL_2D}, at $\bm\phi_0$ with respect to the parameter~$\lambda$, $\Delta_{\rm HL}=\min_\lambda\Delta_{\rm QCRB}$. Although an analytical solution can be found, it requires identifying a root of a fourth-order polynomial, which lacks a simple expression. 

The resulting HL is presented as the red curve in Fig.~\ref{fig:2D_MSE} where the horizontal axis is associated with the bias phase vector magnitude. The region of sensing performance below the red curve is forbidden by quantum physics for a $4$-atom sensor. Unlike in the single-parameter metrology, the HL can not be achieved by a single sensor. Instead, the red curve should be understood as an envelope of MSE curves of various sensors optimized at different $\bm\phi_0$.
The detailed properties of the QC optimized at various bias phases $\bm\phi_0$ and for different $N$ including the study of the corresponding domains of unambiguous estimation will be considered in~Ref.~\cite{Shankar202x}.

\section{Three-parameter SU(2) QC solutions}
\label{sec:three-parameter_QC}

In this section we discuss quantum compass solutions for the general SU(2) sensor estimating three parameters $\bm\phi=\{\phi_x,\phi_y, \phi_z\}$ encoded by the SU(2) unitary $U(\bm\phi) = \exp[-i(\phi_x J_x+\phi_y J_y+\phi_z J_z)]$ employing all three generators of the group. We remind that $J_{x,y,z}$ are collective spin operators obeying the standard commutation relation $[J_x,J_y]=iJ_z$ describing a physical system represented by an ensemble of $N$ spin-$1/2$ atoms or a qudit of spin $J=N/2$. The three-parameter sensor enables the Cartesian frame alignment~\cite{Kolenderski2008,Bartlett2007}, rotation sensing~\cite{Goldberg:2021}, and the simultaneous detection of three components of magnetic or electric fields~\cite{Baumgratz2016,Kaubruegger2023}.

Similar to the two-parameter case, we aim to find the optimal solution for a three-parameter SU(2) sensor within the permutationally symmetric subspace of the entire $N$-qubit Hilbert space. In this subspace, pure states can be elegantly visualized using the Majorana representation~\cite{Bjork2015,Giraud2010}, which we find to be useful for understanding state, measurement, and estimator symmetries of a 3D~QC.

\subsection{Majorana Representation}

The Majorana representation is a one-to-one mapping between particle permutation symmetric states and a set of points on the unit sphere, known as the Majorana constellation. This representation is closely related to the Husimi $Q$ function. 

To obtain the Majorana representation of a pure spin-$J$ state $\ket{\Psi} = \sum_{m=-J}^J \Psi_m \ket{J,m}$, we evaluate its overlap with a coherent spin state (CSS) $\ket{\theta,\varphi}$, where $\theta$ and $\varphi$ are the polar and azimuthal angles that define the average spin direction of the CSS. The overlap is proportional to a polynomial
\begin{eqnarray}
    G(z) = \sum_{m=-J}^J \binom{2J}{J+m}^{1/2} \Psi_m z^{J+m}.
\end{eqnarray}
Here, $z=\cot(\theta/2)e^{i\varphi}$ represents the stereographic projection of the point $(\theta,\varphi)$ on the unit sphere onto the complex plane. The Majorana constellation is defined by the $2J$ complex roots $\{z_k^*\}$ of the polynomial $G(z)$, which are mapped via inverse stereographic projection onto the unit sphere, $\{z_k^*\}\to \{(\theta_k^*,\phi_k^*)\}$. If the degree of $G(z)$ is less than $2J$, the set of roots is appropriately supplemented by points at infinity, which are then mapped to the North Pole of the unit sphere. From the viewpoint of phase space representations, the Majorana constellation is the set of points where the Husimi $Q$ function for $\rho=\ket{\Psi}\bra{\Psi}$ vanishes. Lastly, the Majorana constellation undergoes rigid body rotation under the action of unitaries from the SU(2) group on the associated quantum state, providing an intuitive understanding of the symmetries of 3D QC solutions, as discussed below.

\subsection{3D Quantum Compass solution for $N=4$}
\label{subsec:3D_QC_N4}
\begin{figure}[t] 
   \centering
   \includegraphics[width=2in]{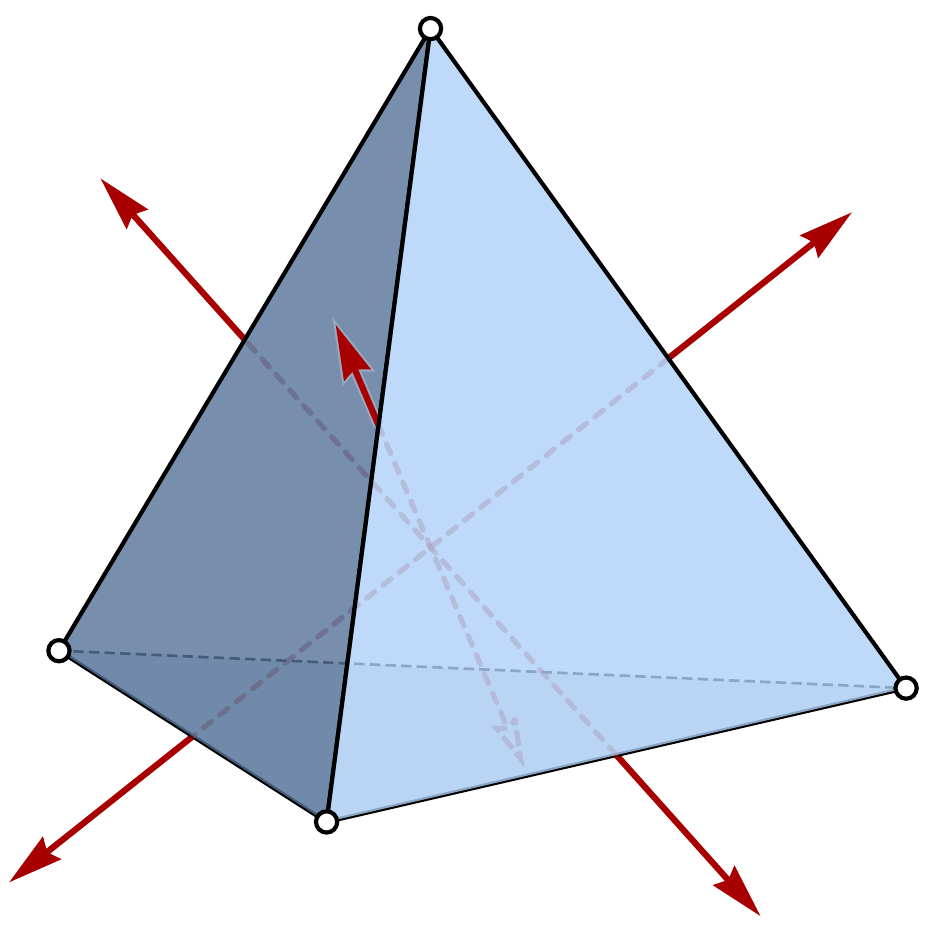} 
   \caption{Three-parameter QC solution for $N=4$ qubits. The Majorana constellation representing the input state (blue tetrahedron) is shown, along with the directions of the single-shot estimators (red arrows), each corresponding to one of the~$6$ measurement outcomes $\mu$ of the optimal POVM given in Eq.~(\ref{eq:POVM_n4_3D}).
   }
   \label{fig:Majorana_N4}
\end{figure}

While the 2D QC solution can be universally described for any~$N$ using one family of input states and measurements, the 3D QC solution is unique for each $N$. Hence, we start with the $N=4$ qubit QC solution, which is the smallest sensor capable of isotropic sensing performance; in other words, it exhibits an isotropic QFIM. Notably, this 3D QC solution lacks antiunitary symmetry, illustrating the generality of our approach in identifying optimal quantum sensors.

Using the numerical method (see App.~\ref{app:num_id}) to find the approximate QC solution optimized for $\bm\phi_0=\{0,0,0\}$ we arrive at an input state parametrized in $\ket{J,m}$ basis as 
\begin{equation}
\ket{\psi_{\rm in}^{\rm 3D}}_{N=4} = \sqrt{1-x^2}\ket{2,1} + x\ket{2,-2}.
\label{eq:psi_3D_N4}
\end{equation}
The optimal state is given by $x=1/\sqrt{3}$, corresponding to the unique \emph{quasiclassical} state that maximizes the QFIM in this parametrization. The Majorana constellation of the state forms a regular tetrahedron, the most symmetric arrangement of 4 points under 3D rotations, as depicted in the Fig.~\ref{fig:Majorana_N4}. The Majorana representation offers a geometric interpretation of antiunitary symmetry in states, observed through Majorana constellations consisting of point pairs on opposite sides of the unit sphere. Consequently, the regular tetrahedron, representing the $N=4$ QC input state, does not possess antiunitary symmetry. Recently, this state was experimentally realized using optical photons~\cite{ferretti2023}.

The optimal measurement is a POVM, $M_j=\ket{\mu_j}\bra{\mu_j}$, $j=1,\ldots,6$ with the corresponding single-shot estimators, Eq.~\eqref{eq:MMSEE}, shown in the Fig.~\ref{fig:Majorana_N4} with red arrows. The evident tetrahedral symmetry, with each estimator corresponding to a Majorana constellation edge, suggests the ansatz for the POVM formed using a single fiducial state $\ket{\eta}$:
\begin{equation}
\begin{aligned}
\ket{\mu_1}&=\ket{\eta},\\
\ket{\mu_2}&=e^{-i\frac{2\pi}{3}J_z}\ket{\mu_1},\\
\ket{\mu_3}&=e^{i\frac{2\pi}{3}J_z}\ket{\mu_1}\\
\ket{\mu_4}&=e^{-i\pi J_y} e^{-i\frac{\pi}{2\sqrt3}(\sqrt2J_x-J_z)} \ket{\eta},\\
\ket{\mu_5}&=e^{-i\frac{2\pi}{3}J_z}\ket{\mu_4}, \\
\ket{\mu_6}&=e^{i\frac{2\pi}{3}J_z}\ket{\mu_4}
\end{aligned}
\label{eq:POVM_n4_3D}
\end{equation}
The fiducial state can be found in an analytical form:
\begin{equation}
\ket{\eta} = \frac1{\sqrt6}\Big\{1,-i,\frac{\sqrt2+i}{\sqrt{3}},i,1\Big\}.
\label{eq:3D_eta_n4}
\end{equation}
The details of the analytical optimization are presented in Appendix~\ref{app:3D_QC_solutions}.

\subsubsection{Sensing Performance of the 3D Quantum Compass}
\begin{figure}[t] 
   \centering
   \includegraphics[]{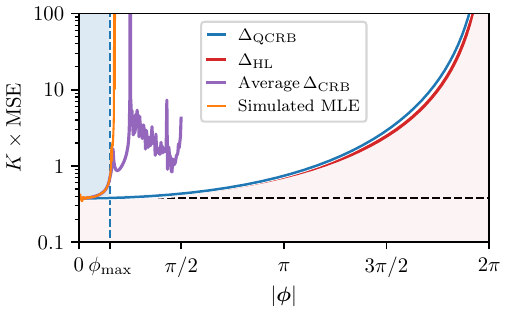} 
   \caption{
   Sensing performance of the 3D QC solution for $N=4$ qubits. The purple solid line shows the CRB ($\Delta_\text{CRB}$) for the 3D QC solution, averaged over direction, as a function of $|\bm\phi|$. Unlike the 2D QC solution, the CRB coincides with the QCRB ($\Delta_\text{QCRB}$, blue solid line) only at $\bm\phi=0$. The orange solid line shows the MSE (scaled by number of measurements~$K$) averaged over direction (see Appendix~\ref{app:mle}). The vertical dashed line delimits the region (blue) where the MLE achieves the CRB. The red solid line shows the HL. The region (red) below this line is inaccessible to any $4$-qubit sensor. The QC solution saturates the HL only at $\bm\phi=0$, where the sensor has the highest sensitivity, indicated by the horizontal dashed line.
   }
   \label{fig:3D_MSE}
\end{figure}

Similar to the previous section on two-parameter sensing, we evaluate the 3D QC's performance in the context of criteria (i) and (ii) by analyzing the FIM, the QFIM, the HL, and conducting numerical simulations of the estimation experiment using the MLE. Using the 3D QC input state Eq.~\eqref{eq:psi_3D_N4} and measurement we calculate the FIM, Eq.~\eqref{eq:FIM}, to obtain the CRB, $\Delta_{\rm CRB}=\tr{\mathcal{F}^{-1}}$. The CRB is non-isotropic with respect to the direction of $\bm\phi$, thus we average it over the $4\pi$ solid angle and show as a function of $|\bm\phi|$ in Fig.~\ref{fig:3D_MSE} with the violet curve.

The QFIM is calculated analytically and provides the QCRB matching the calculation in Ref.~\cite{Baumgratz2016} (for $J=2$):
\begin{equation}
\Delta_{\rm QCRB}=\tr{\mathcal{F}_Q^{-1}} = \frac{3\left[1+2\,\mathrm{sinc}(|\bm\phi|/2)^{-2}\right]} {4J(J+1)}
\label{eq:QCRB_3D}
\end{equation}
The QCRB, Eq.~\eqref{eq:QCRB_3D}, is shown in Fig.~\ref{fig:3D_MSE} with the blue curve. Because the input state~\eqref{eq:psi_3D_N4} is \emph{quasiclassical}, the QCRB can be saturated with the optimal measurement, as shown in Fig.~\ref{fig:3D_MSE}, but only at $|\bm\phi|=0$ due to the absence of antiunitary symmetry. It is also straightforward to show analytically that $\Delta_{\rm CRB}=\Delta_{\rm QCRB}$ at $|\bm\phi|=0$ using Eq.~\eqref{eq:FIM} for the 3D QC state and measurement.

The QCRB can be compared to the phase-dependent HL, which reads (for integer $J>1$)
\begin{equation}
\Delta_{\rm HL} = \frac{\left[1+2\left|\mathrm{sinc}(|\bm\phi|/2)\right|^{-1}\right]^2} {4J(J+1)}.
\label{eq:3D_HL}
\end{equation}
The HL matches the QCRB and the CRB of the 3D QC solution at $|\bm\phi|=0$, thus, criterion~(i) is satisfied for $N=4$ (and larger even $N$, see below).
For additional details on how the phase-dependent HL is derived, refer to the Appendix~\ref{app:HL_in_3D}. The HL is represented as the red curve in Fig.~\ref{fig:3D_MSE}, delineating the region of sensing performance forbidden by quantum physics (light red shaded area). Compared to the two-parameter case, the gap between the QCRB and the HL in three-parameter sensing is less pronounced. The HL for $N=4$ at the optimization point $\bm\phi_0$ reads $\Delta_{\rm HL} = 9/[4J(J+1)]=3/8$~\cite{Kolenderski2008,Baumgratz2016} and is indicated with the horizontal dashed line.

When assessing the sensor's performance via the CRB (violet line), a notable deviation from the QCRB becomes evident around $\phi_{\rm max} = \arcsin(\sqrt{2/3})/2\approx0.4777$ (vertical dashed line). This deviation is associated with the root of the conditional probabilities closest to the optimization point $\bm\phi_0$. In contrast to antiunitary symmetric cases, these conditional probability roots are isolated and do not form a hypersurface in the parameter space. Thus, the set of roots can specify the radius but not the full shape of the domain of unambiguous estimation.

The MLE performance, representing the actual sensing precision achieved in a numerical simulation (see App.~\ref{app:mle}), is presented with an orange line. Similar to the 2D QC solution, we simulate the estimation experiment for $20000$ randomly selected true phase vectors within a sphere of $|\bm\phi|<0.84$. The sphere is divided into $111$ shells of equal thickness and the MSE of the sample points within each shell is averaged and plotted as a function of the shell radius $|\bm\phi|$. The curve demonstrates that the MLE achieves the CRB of the 3D QC solution for all parameters within a sphere $|\bm\phi|<\phi_\mathrm{max}$, beyond which the shell-averaged estimation error grows sharply since the MLE converges to incorrect local maxima for some sample points. The simulation results thus indicate that the roots closest to $\bm\phi_0$ determine the radius $\phi_\mathrm{max}$ of the domain of unambiguous estimation. The corresponding criterion~(ii) is discussed in Section~\ref{subsec:Domain_3D} below.

\subsection{3D QC solutions for different $N$}
\label{subsec:3D_QC_gallery}

\begin{table}[t]
\centering
\begin{tabular}{c|c|c|c}
\textbf{J}   & \textbf{Constellation} & \textbf{State} & \textbf{Domain $\Omega^*$}  \\ \hline

3 
& 
\begin{minipage}{1in}
\includegraphics[width=1in]{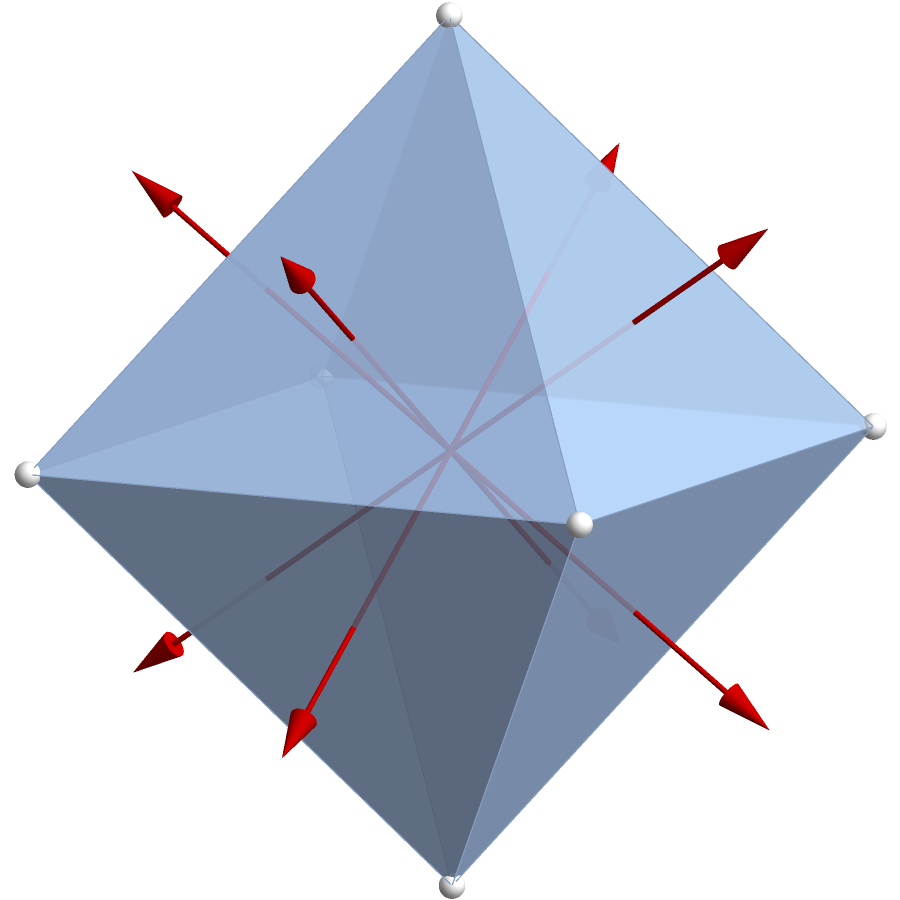} 
\end{minipage}
& 
$\Psi_{\pm1}=\frac1{\sqrt2}$
&
\begin{minipage}{1in}
\includegraphics[width=0.8in]{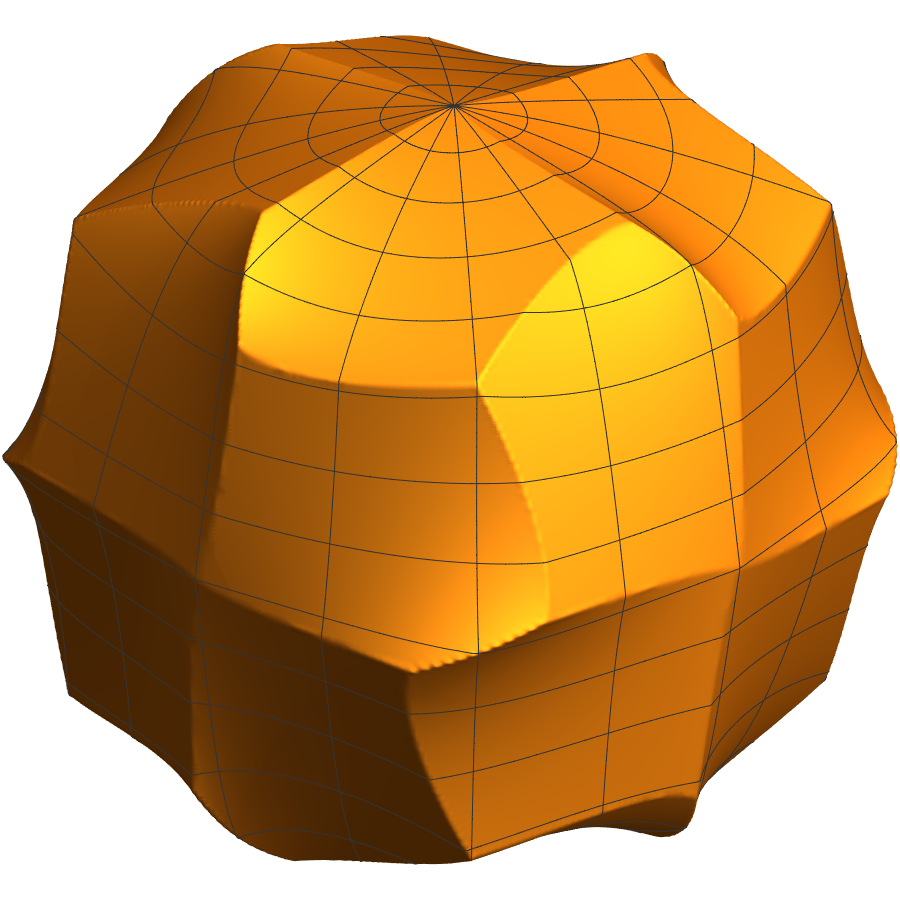} \\
$\phi_{\rm max} \approx 0.384$
\end{minipage}\\ \hline
4 & 
\begin{minipage}{1in}
\includegraphics[width=1in]{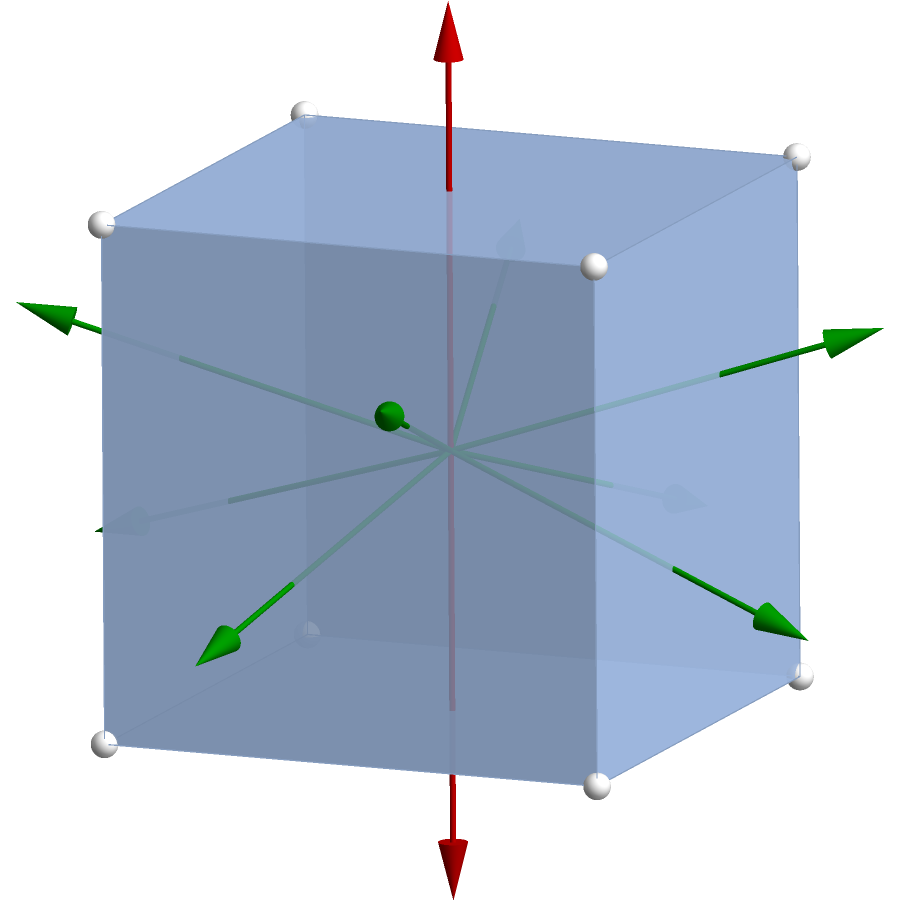} 
\end{minipage}
&
\begin{minipage}{2cm}
$\Psi_{0}=\sqrt\frac7{12}$\\
$\Psi_{\pm4}=\sqrt\frac5{24}$
\end{minipage}
&
\begin{minipage}{1in}
\includegraphics[width=0.8in]{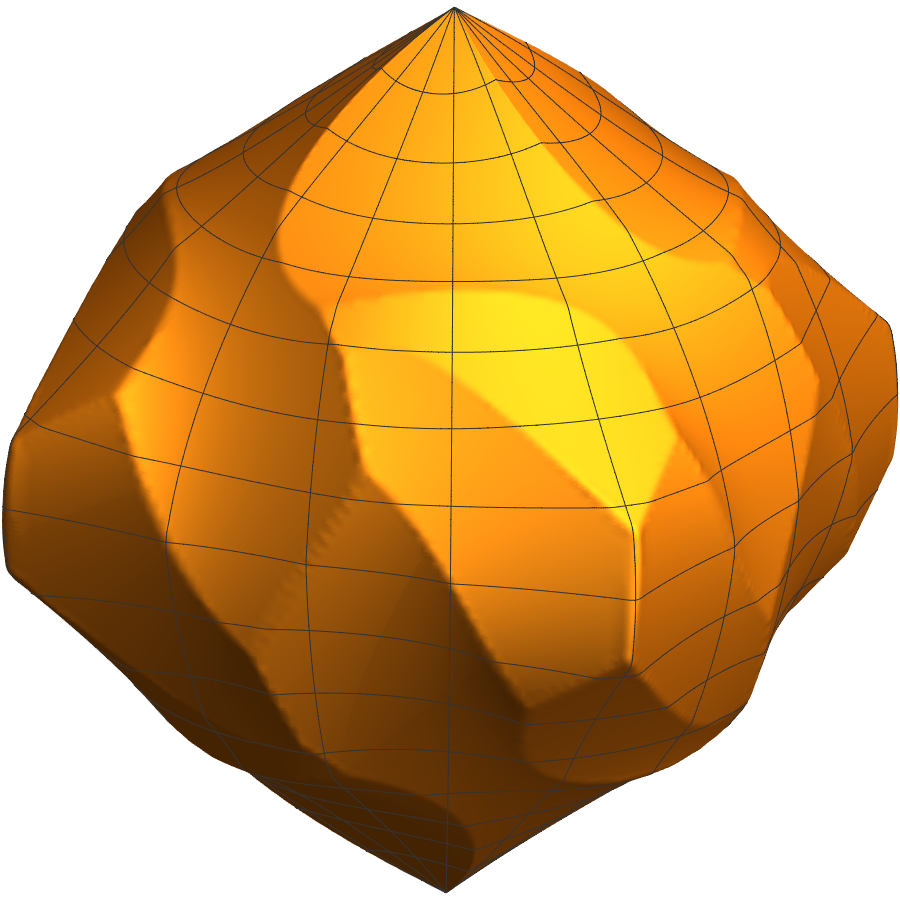} \\
$\phi_{\rm max} \approx 0.295$
\end{minipage}\\ \hline
6 & 
\begin{minipage}{1in}
\includegraphics[width=1in]{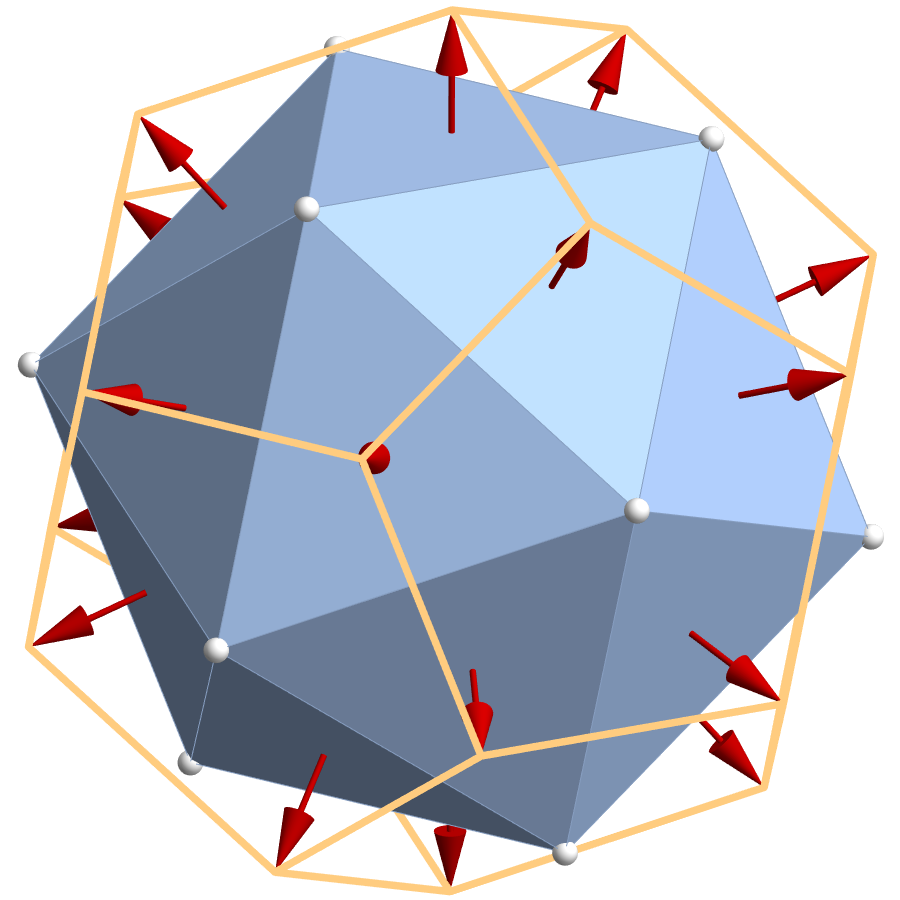} 
\end{minipage}
&
\begin{minipage}{2cm}
$\Psi_{0}=\frac{\sqrt{11}}{5}$\\
$\Psi_{\pm5}=\pm\frac{\sqrt7}{5}$
\end{minipage}
&
\begin{minipage}{1in}
\includegraphics[width=0.8in]{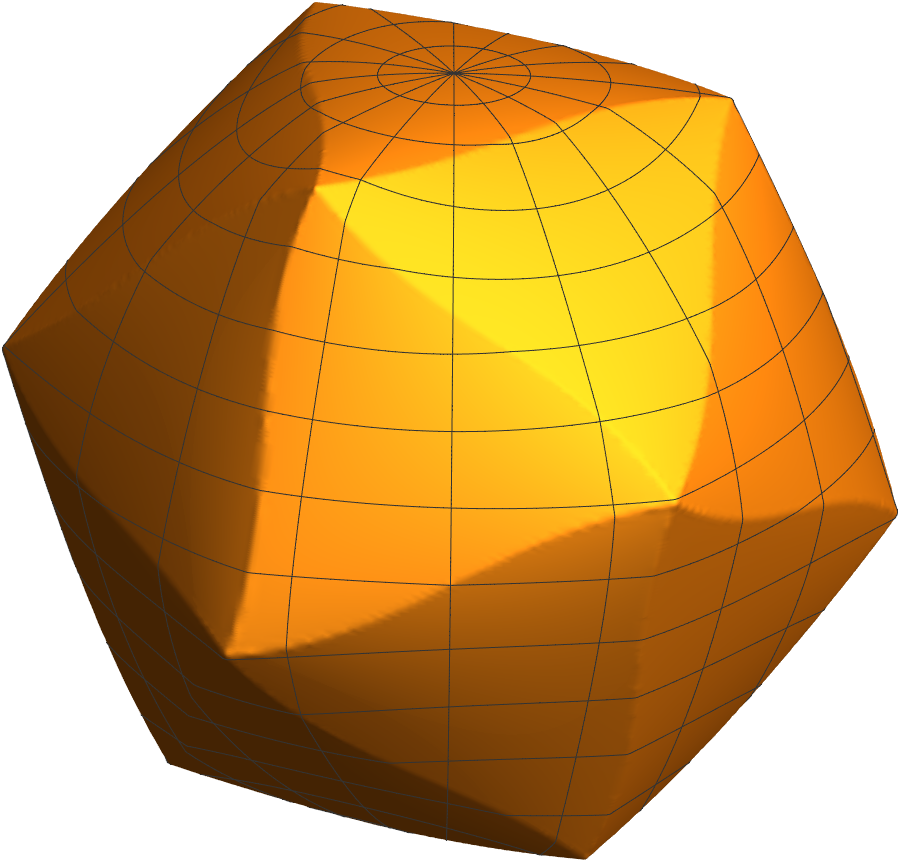} \\
$\phi_{\rm max} \approx 0.203$
\end{minipage}\\ \hline
16 & 
\begin{minipage}{1in}
\includegraphics[width=1in]{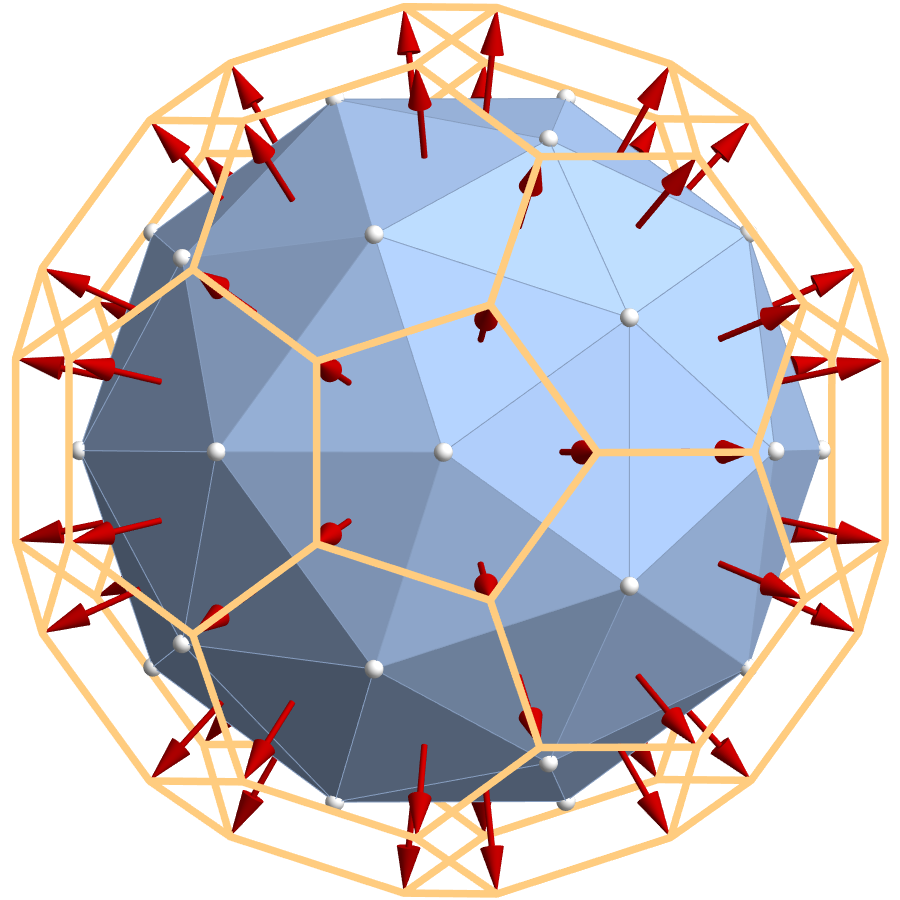} 
\end{minipage}
&
see App.~\ref{app:3D_QC_solutions}\\ \hline

\end{tabular}
\caption{Gallery of 3D QC solutions for different sensor sizes $J=N/2$ (first column). The second column depicts the Majorana constellation of the input state (blue polyhedra). The arrows indicate the directions of the single-shot estimators, each corresponding to one of the outcomes of the optimal POVMs (presented in Appendix~\ref{app:3D_QC_solutions}). For $J=4$, the estimators are classified into two groups (red, green) based on the underlying fiducial states parametrizing the POVM. For $J=6,16$, a polyhedron (orange lines) with vertices along the estimator directions is additionally drawn to highlight the relative orientation of estimators with respect to the input state. The third column lists the non-zero elements of the input state. The last column depicts the shape of the domain of unambiguous estimation, along with the corresponding radius (minimum distance to origin).
}
\label{tab:gt}
\end{table}

We now explore 3D QC solutions for various atom numbers, specifically even values such as $N=6,\,8,\,12,\,32$. ($N=3$ is considered separately in Sec.~\ref{subsec:circuit-based_sensors} and App.~\ref{subapp:3D_QCS_N3}) This choice streamlines the identification process due to the isotropic sensing performance of the QC solutions, leading to isotropic asymptotic prior distributions and symmetric antiunitary invariant optimal solutions. In these cases, the achievable MSE is described by the QCRB, Eq.~\eqref{eq:QCRB_3D}, and the domain of unambiguous estimation is easily assessed through the zeros of measurement probability amplitudes (see Sec.~\ref{subsec:Domain_2D}). Table~\ref{tab:gt} summarizes the optimal states, their Majorana constellations, single-shot estimates, and the domains of unambiguous estimation.

$N=6$: The optimal state Majorana constellation is a regular octahedron, a platonic solid with $6$ vertices and $8$ faces. There are $N_e=8$ estimators, forming a cube which is dual to the octahedron. Leveraging octahedral symmetry, we can parameterize the optimal measurement with a single fiducial state $\ket\eta$, which also possesses antiunitary symmetry. For the exact expression for $\ket\eta$, refer to Appendix~\ref{app:3D_QC_solutions}. 

$N=8$: The optimal state is a cube, with $N_e=10$ estimators divided into two groups. One pair of estimators passes through the centers of two opposite faces, while the remaining $8$ pass near the centroids of the triangles that make up the side faces. Consequently, the optimal measurements are parameterized by two fiducial states, $\ket\eta_1$ and $\ket\eta_2$, which define the rest of the measurement projectors through suitable rotations that preserve the Majorana constellation and associated estimator vectors (see Appendix~\ref{app:3D_QC_solutions}). 

$N=12$: Similar to $N=6$, the $N=12$ sensor also exhibits a high degree of symmetry. The optimal state is a platonic solid, a regular icosahedron with $12$ vertices and $20$ faces. The $N_e=20$ estimators form a regular dodecahedron, the dual to the icosahedron. A single antiunitary invariant fiducial state $\ket\eta$ defines all the measurement projections through rotations within the icosahedral rotation group (see Appendix~\ref{app:3D_QC_solutions}).

Observing the optimal solutions for various values of~$N$, we find that the 3D QC solution tends to align estimators with the faces of the Majorana constellation of the input state. Consequently, the Majorana constellations of 3D QC solution not only distribute vertices evenly around the unit sphere but also tend to maximize the number of faces in the constellation. It can be shown that a polyhedron with $N$ vertices can have at most $2N-4$ faces. The aforementioned exceptionally symmetric QC solutions for $N=6$ and $N=12$ saturate this upper bound, having $\smash{N_e=2N-4}$, with the state constellation and estimator configuration being geometrically dual. Another implication of this observation is that highly symmetric constellations with a low count of faces are unlikely to be optimal for sensing. For instance, it is straightforward to demonstrate using our numerical method that an input state represented by a regular dodecahedron for $N=20$, as discussed in Ref.~\cite{Kolenderski2008}, can be outperformed by less symmetric constellations with more faces.

$N=32$: Using this observation we conjecture the optimal state for $N=32$ to be a pentakis dodecahedron with $32$ vertices and $60$ faces. The $2N-4=60$ estimator directions pass through the face centers, forming a truncated icosahedron, which is also the geometry of the fullerene~$C_{60}$ molecule. We numerically confirmed the predicted configuration of estimators for the input state.

The Majorana representation naturally connects the optimization of 3D QC states to the problem of distributing $N$ points on the unit sphere in a `most symmetric' manner. This problem has a rich history with various solutions depending on the specific cost function being optimized~\cite{Conway1996,Saff1997}. Furthermore, tables of spherical codes~\cite{Sloane} are useful in defining initial guesses for quantum states during the numerical search for 3D QC solutions. The 3D QC states are also related to prior research that sought to identify extremal quantum states using various criteria. This earlier work identified platonic solids~\cite{Kolenderski2008} and introduced state families like the Kings of quantumness~\cite{Bjork2015} and Queens of Quantum~\cite{Giraud2010}.

\subsection{Domain of unambiguous estimation: POVM vs Projective Measurement}
\label{subsec:Domain_3D}

Whereas 2D QC solutions are realized within the projective measurement (PM) class (Sec.~\ref{sec:two-parameter_QC}), distinct 3D QC solutions exist for both the general POVM class measurements and the PM. This allows us to explore the trade-off in sensing performance, in relation to criteria (i) and (ii), between 3D QC solutions using the general POVM and the more experimentally accessible PM.

Specifically, we numerically identify optimal PMs for the 3D QC solutions for $N=6$ and $12$ qubits, featuring highly symmetric optimal input states. The high degree of symmetry of these states makes them optimal for both the POVM and PM solutions (typically, constraints on the measurement class may influence the corresponding optimal state). The resulting optimal PMs in these cases, along with the input states, are antiunitary invariant, i.e. the CRB saturates the QCRB for all phase values (see Sec.~\ref{sec:two-parameter_QC})~\cite{Miyazaki2022}. Consequently, the MSE achievable with the MLE for an asymptotic number of measurements~$K$ remains identical between the PM-based and POVM-based 3D QC solutions, hence, both fulfill criterion (i).

\begin{figure}[t] 
   \centering
   \includegraphics[]{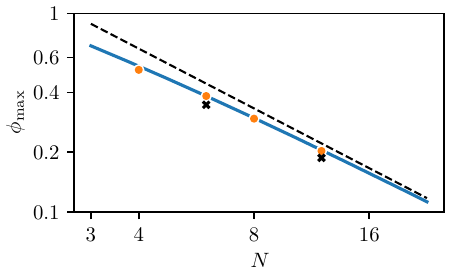} 
   \caption{Radius of the domain of unambiguous estimation for the 3D QCS as a function of the number of qubits $N$. The orange dots correspond to exact numerical values of the POVM-based QCS, while the black crosses represent the domain radius of the PM-based QCS. The solid blue line represents the empirical asymptotic scaling $\phi_{\rm max}\sim 1.33/\sqrt{J(J+1)}$, and the dashed black line shows the $1/J$ scaling for comparison.}
   \label{fig:3D_QC_domain_radius}
\end{figure}

However, the PM and POVM-based QC solutions demonstrate distinct domains of unambiguous estimation~$\Omega^*$, criterion (ii). The shapes and radii of the domains for the POVM-based QC solutions are obtained following Sec.~\ref{subsec:Domain_2D} and presented in Table~\ref{tab:gt}. In Figure~\ref{fig:3D_QC_domain_radius}, the radii of the POVM QC are plotted as a function of the system size with orange dots, while the black crosses show the corresponding radii of the PM QC solutions: $\phi_{\rm max}\approx 0.347(0.187)$ for $N=6(12)$. Clearly, the PM QC solution can unambiguously estimate phase vectors in a smaller range of absolute values compared to the POVM solution. This difference is also reflected in the smaller value of the metrological cost $\Xi_K$ Eq.~\eqref{eq:cost_Bayes_single_shot} for the POVM QC solution at a given number of measurement samples $K$, indicating higher information gain per single measurement and, consequently, faster convergence of the estimation error to the CRB with an increase in~$K$. Note that in terms of the cost expansion coefficients, Eq.~\eqref{eq:Cost_expansion}, this implies that the higher-order terms $C^{(\ell>1)}$ are larger for the POVM-based QC solution, while the Fisher information-based terms $C^{(1)}$ are identical between the PM and POVM solutions.

Finally, from Fig. \ref{fig:3D_QC_domain_radius}, we extract the empirical scaling of the domain radius as a function of the system size. We obtain the scaling $\phi_{\rm max}\sim 1.33/\sqrt{J(J+1)}$, corresponding to the coefficient $r_{\rm 3D}\approx 1.5$ in the proposed relation between the domain radius and the estimation uncertainty, Eq.~\eqref{eq:general_domain_size}, for 3D QC solution. The value of this coefficient is close to those of the 2D QC solution and the GHZ state interferometer (discussed earlier in Sec.\ref{sec:two-parameter_QC}), reinforcing the universality of the relation Eq.~\eqref{eq:general_domain_size} between the ultimate quantum precision of a sensor and the size of the corresponding domain of unambiguous estimation.

\section{Variational Quantum Circuits for Three-Parameter SU(2) Sensors}
\label{sec:VQC}

In the previous two sections, we demonstrated how the optimization method outlined in Sec.~\ref{sec:optimal_multiparameter} can determine the theoretical limits on sensing performance, yielding the QC solutions for two and three-parameter estimation with SU(2) sensors. In this section, we illustrate how the optimal sensor performance can be approximated using a restricted set of gates available on an experimental platform. Specifically, we consider estimation of three parameters $\bm\phi=\{\phi_x,\phi_y, \phi_z\}$ encoded by the SU(2) unitary $U(\bm\phi) = \exp[-i(\phi_x J_x+\phi_y J_y+\phi_z J_z)]$ (see Sec.~\ref{sec:three-parameter_QC}). We employ the asymptotic Bayesian cost introduced in Sec.~\ref{sec:optimal_multiparameter} as a metrological cost for variational optimization~\cite{Koczor_2020,Kaubruegger2021,Kaubruegger2023,Marciniak2022} of quantum circuit-based sensors constructed from the native gates of a given experimental setup. By utilizing these gates, we construct circuits that closely approach the QC solution for three-parameter sensing with $N=3$ qubits and, for an arbitrary number of qubits, achieve the HL while sacrificing some volume of the domain~$\Omega^*$.

\subsection{Quantum circuit model of a quantum sensor}
\label{sec:CircuitModel}
We consider the quantum circuit model of a sensor, where the initial state 
\begin{equation}
    \ket{\psiin} = \Uen \, \ket{\psi_{0}}
\end{equation}
is prepared by an entangling unitary $\Uen$ from a product state, e.g., $\ket{\psi_0} = \ket{\downarrow}^{\otimes N}$. The decoding unitary $\Ude$ transforms the measurement basis $\ket\mu$ provided by a physical observable in the experiment, resulting in effective measurement projectors
\begin{equation}
    \Pi_{\mu} = \Ude^{\dagger} \ket{\mu}\bra{\mu} \Ude.
\end{equation}
The corresponding conditional probability of obtaining an outcome $\mu$ given a phase $\bm\phi$ is $p(\mu|\bm{\phi}) = |\bra{\mu}\Ude U(\bm\phi)\Uen\ket{\psi_0}|^2$.

\subsection{Resource gates and variational ansatz}
We consider a set of native gates that includes global rotations $\mathcal{R}_{\ell}(\vartheta)=e^{-i\vartheta J_{\ell}}$ with $\ell=x,y,z$ and an entangling gate in the form of a one-axis twisting (OAT) operation $\mathcal{T}_z(\vartheta) = e^{-i\vartheta J_{z}^2}$. This choice is motivated by the feasibility of implementing programmable quantum sensors on a trapped-ion platform, where OAT is natively realized as a M\o{}lmer-S\o{}rensen gate \cite{Leibfried2005, Monz2011}, and spin rotations are performed via laser or microwave driving.

We construct the entangling and decoding unitaries following Refs.~\cite{Kaubruegger2021,Kaubruegger2023}, using a sequence of the native gates, $\mathcal{R}_{\ell}$ and $\mathcal{T}_{z}$, parametrized by a corresponding set of interaction angles $\bm\vartheta=\{\vartheta_1,\vartheta_2,\ldots\}$. By evaluating the corresponding conditional probability $p(\mu|\bm{\phi}) = |\bra{\mu}\Ude(\bm\vartheta_{\rm de}) U(\bm\phi)\Uen(\bm\vartheta_{\rm en})\ket{\psi_0}|^2$, we compute the asymptotic Bayesian cost Eq.~\eqref{eq:cost_Bayes_single_shot} as a function of the interaction angles $\Xi_K(\bm\vartheta_{\rm en}, \bm\vartheta_{\rm de})$ using an isotropic distribution $\mathcal{P}_K$ centered at $\bm\phi_0=\{0,0,0\}$. Minimizing the cost with respect to $\bm\vartheta_{\rm en}$ and $\bm\vartheta_{\rm de}$ for $K\gg1$ (as discussed in Sec.~\ref{subsec:new_cost}) yields quantum circuits approximating the sensing performance of the QC solutions. We adjust the number of gates in the entangling and decoding unitaries to strike a balance between circuit depth and sensor performance. The simplest solution is presented below.

\subsection{Quantum circuits for three-parameter SU(2) sensors}
\label{subsec:circuit-based_sensors}

We find that a sensor saturating the HL for three-parameter estimation can be made using only one entangling gate in each $\Uen$ and $\Ude$. The sensor input state is the GHZ state, prepared from $\ket{\psi_0}=\ket{J,J}$ with $J=N/2>1$ using the entangling unitary:
\begin{equation}
\label{eq:Uen_N=3}
\Uen = \mathcal{R}_\ell\left(\frac{\pi}2\right)\mathcal{T}_z\left(\frac{\pi}2\right) \mathcal{R}_x\left(\frac{\pi}2\right).
\end{equation}
The axis of the last rotation is $\ell=x(y)$ for even (odd) number of qubits $N$, ensuring alignment of the prepared GHZ state along the $z$-axis.

The corresponding measurement is the projective measurement of collective $J_z$ (projection onto $\ket\mu=\ket{J,\mu}$) combined with the decoding unitary:
\begin{equation}
\label{eq:Ude_N=3}
\Ude = \mathcal{R}_y\left(\frac{\pi}2\right) \mathcal{T}_z\left[\frac{\pi}{2(N-1)}\right].
\end{equation}

The performance at the optimization point $\bm\phi_0$ of the quantum sensor defined by the entangling and decoding unitaries Eqs.~\eqref{eq:Uen_N=3} and \eqref{eq:Ude_N=3} is given by the diagonal FIM~($N>2$)
\begin{equation}
\mathcal{F} = \left(\begin{matrix} 
      N & 0 & 0 \\
      0 & N & 0 \\
      0 & 0 & N^2\\
   \end{matrix}
   \right)
 \label{eq:FIM_circuit_sensor}
\end{equation}
The FIM is equal to the QFIM of the GHZ state generated by $\Uen$, thus the sensor saturates the QCRB. The sensor exhibits anisotropic sensitivity, with the highest sensitivity to phase changes along the $z$-axis.

\begin{figure}[t] 
   \centering
   (a)
   \includegraphics[width=1.35in]{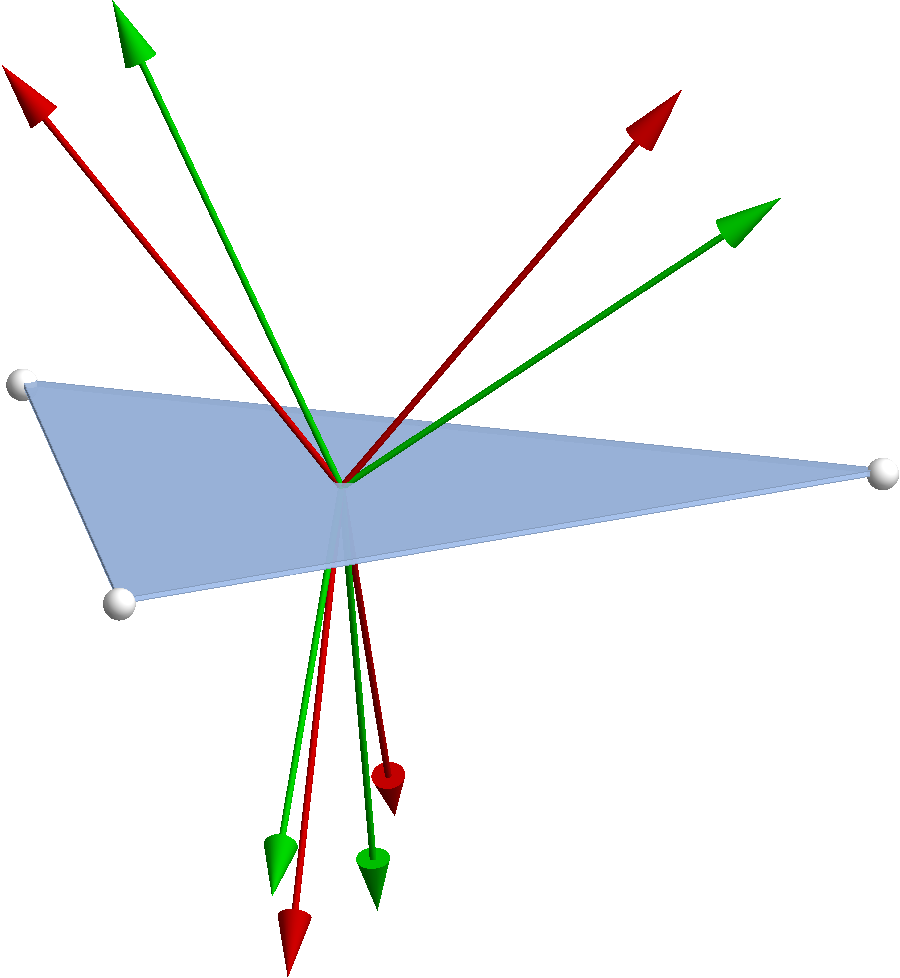}
   \hspace{0.2cm}
   \includegraphics[width=1.5in]{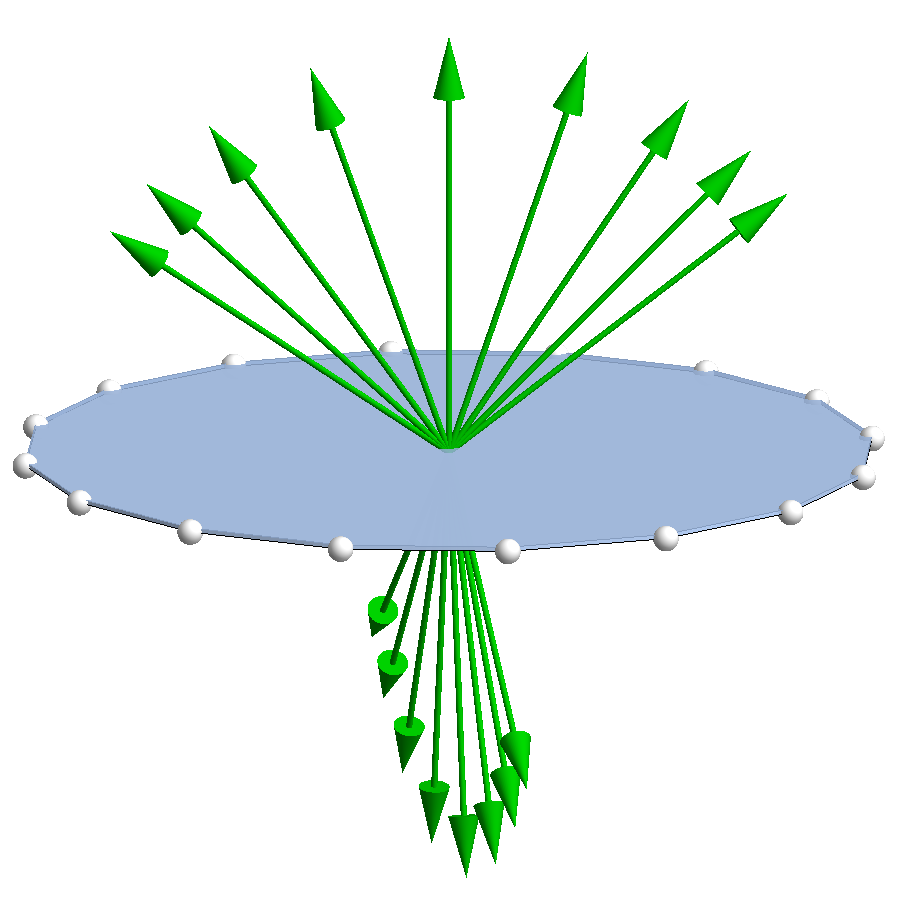} 
   (b)
   \caption{Circuit-based sensors for $N=3$ (a) and $N=16$ qubits (b). Majorana constellations illustrating the GHZ input states (blue polygons) are shown alongside the directions of single-shot estimators for the circuit-based sensors (green arrows). Additionally, the (a) panel displays estimators for the 3D QC solution (red arrows) for comparison.
   }
   \label{fig:Majorana_N3_16}
\end{figure}

Remarkably, for $N=3$, the FIM~\eqref{eq:FIM_circuit_sensor} reaches the quantum limit, which can be confirmed by minimizing the QCRB (maximizing the QFIM) across all states in the 4-dimensional Hilbert space. Consequently, the sensor attains the fundamental precision limit at $\bm\phi_0$ for a fixed input state (altering the state's orientation is discussed below), yielding $\Delta_\mathrm{CRB}=\Delta_\mathrm{HL}=7/9$ and, thus, meeting criterion~(i). This performance can be compared to the classical limit where each of the three qubits independently measures different phase components. The SQL variance is $\Delta_{\rm SQL}=9/N=3$, resulting in a roughly~$3.86$ improvement factor of the entangled sensor defined by $\Uen$ and $\Ude$ over the classical limit.

Moreover, the measurement generated by $\Ude$~\eqref{eq:Ude_N=3} for $N=3$ is nearly optimal, leading to the size of the domain $\Omega^*$ of the circuit-based sensor closely matching that of the QC solution, thereby approaching criterion~(ii). This is indicated by the metrological cost expansion coefficient $C^{(2)}=-169$ [see Eq.~\eqref{eq:Cost_expansion}] for the circuit-based sensor, slightly below the optimal $C^{(2)}=-157$ for the 3D QC solution with $N=3$ (refer to Appendix~\ref{subapp:3D_QCS_N3}). Figure~\ref{fig:Majorana_N3_16} demonstrates the near optimality of the $\Ude$-based measurement for $N=3$ qubits in the (a) panel, where single-shot estimators of the circuit-based sensor (green arrows) are only slightly shifted away from the estimators of the QC solution, which also employs projective measurement. Note that, due to anisotropy of the optimal sensitivity, the 3D QC for $N=3$ has another solution based on POVM with similar sensing performance, as discussed in Appendix~\ref{subapp:3D_QCS_N3}.

The anisotropic sensitivity of circuit-based sensors, Eq.~\eqref{eq:FIM_circuit_sensor}, can be mitigated by partitioning the $K$ measurement repetitions into three groups and rotating the GHZ state and the measurement basis between these groups, in order to align the maximum sensitivity direction with the $x$, $y$, or $z$ axes. This is achieved by adding extra rotation gates to $\Uen$ and $\Ude$. As the quantum Fisher information is additive for uncorrelated measurements~\cite{pezze2014}, the FIM of the sequence of measurements becomes isotropic, given by $\mathcal{\widetilde F} = \frac{N(N+2)}{3}\mathbb{1}_3$, achieving the HL, $\Delta_{\rm CRB}=\Delta_{\rm HL} = 9/[4J(J+1)]$~\cite{Kolenderski2008,Baumgratz2016}. Hence, by modifying the state and measurement directions, the circuit-based sensor can meet criterion~(i) for a general~$N$.

In terms of criterion (ii) for a general $N$, the domain~$\Omega^*$ radius for the circuit-based sensors is upper bounded by $\phi_{\rm max} = \pi/(2N)$ as the GHZ state is invariant under rotations by $2\pi/N$ around the axis of high sensitivity [see Fig.~\ref{fig:1D_GHZ}(a)]. Comparing this with the domain radius of the 3D QC solutions presented in Sec.~\ref{subsec:Domain_3D}, we estimate that the volume of $\Omega^*$ for the circuit-based sensor, Eqs.~\eqref{eq:Uen_N=3} and \eqref{eq:Ude_N=3}, is approximately five times smaller than that of the theoretically optimal sensor. A more detailed analysis is necessary to assess the domain $\Omega^*$ and sensing performance within it because the circuit-based sensors, defined by the unitaries $\Uen$ and $\Ude$, are not antiunitary symmetric. Therefore, these sensors saturate the QCRB of the GHZ state only at $\bm\phi=\bm\phi_0$, with performance degrading away from this point, similar to the behavior of the 3D QC solution for $N=4$ shown in Fig.~\ref{fig:3D_MSE}.

To highlight the differences between the QC solutions and the circuit-based sensors for a general $N$, we present the state and estimators of the circuit-based sensor defined by Eqs.~\eqref{eq:Uen_N=3} and \eqref{eq:Ude_N=3} for~$N=16$ in Fig.~\ref{fig:Majorana_N3_16}(b). Notably, the single-shot estimators of the circuit-based sensor are not as uniformly distributed as those of the 3D QC solutions.

In summary, our method for finding optimal sensors in the many-measurement scenario, combined with variational quantum metrology techniques, enables the construction of simple quantum circuits that reach the HL and allow us to estimate parameter values within a range comparable to that of the theoretically optimal sensor. More advanced circuit-based sensors, which include more gates, can be developed to achieve isotropic sensitivity with fixed states and broaden the domain $\Omega^*$. The exact QC solutions can facilitate the variational optimization by offering a reference in terms of relevant symmetries for the variational circuits ansatz.

\section{Conclusion and Outlook}
\label{sec:conclusion}

In this study, we introduced a systematic method for identifying optimal quantum sensors capable of achieving fundamental precision bounds in multiparameter estimation scenarios, as defined within the Fisher information framework. Our key insight is to extend the Fisher information approach with the requirement for the sensor to provide unambiguous estimation in a neighborhood of the parameter space surrounding the point of interest~$\bm\phi_0$. We show that this requirement is naturally accommodated by the Bayesian approach in the limit of asymptotically large number of measurement repetitions.

By integrating Fisher information and Bayesian approaches, we establish explicit optimal sensor solutions---referred to as QC solutions. These solutions not only attain fundamental precision bounds but also expand the domain of unambiguous estimation and enhance the information gain per single measurement repetition, leading to a faster reduction of the estimation error towards the CRB. As an illustration, we applied our approach to find exact QC solutions for two- and three-parameter estimation using SU(2)-quantum interferometers. These explicit optimal sensor solutions allow us to explore the limits of sensing performance away from the optimization point~$\bm\phi_0$ providing insights into how `local' the optimal local estimation strategies are in many-repetition scenario. The QC solutions explicitly reveal the interplay between fundamental precision limits and the range of unknown parameter values that can be estimated unambiguously with such precision, as summarized in Eq.~\eqref{eq:general_domain_size}.

Furthermore, we demonstrated that our method, which is based on optimizing the asymptotic Bayesian cost function, naturally integrates with the quantum variational approach for programmable quantum sensors. As an illustrative example, we present the design of simple quantum circuit-based sensors, which can be readily implemented on a trapped-ion platform and exhibit the ultimate sensitivity allowed by quantum physics for detecting 3D magnetic fields or 3D rotations. This opens the door for applying quantum variation techniques to design multiparameter sensors that approach optimal performance in many-repetition scenario	 across various sensing platforms~\cite{Leroux2010,Riedel2010, Gross2010,Chalopin2018,Zhou2020,Zheng2023}.

Our work sets the stage to explore several intriguing avenues. Initially, we applied our method to sensing problems with ultimate precision limits defining isotropic sensitivities. Yet, the approach is general and applicable to a variety of metrological scenarios. One such scenario involves anisotropic sensitivity, which we have touched upon in relation to the phase-dependent HL for SU(2) sensors and three-parameter sensing with $N=3$ qubits. Future investigations will delve deeper into QC solutions exhibiting anisotropic sensitivity~\cite{Shankar202x}. Moreover, our method naturally extends to sensing scenarios involving non-unitary quantum evolution, such as quantum thermometry~\cite{Mehboudi_2019,Rubio2021}. Another potential application is in multiparameter sensing with nuisance parameters, such as measuring a spatially distributed field in the presence of correlated noise~\cite{hamann2023}. This could enable the use of atoms or ions as sensors to test the interplay between quantum physics and general relativity~\cite{Mieling2022,Pikovski2015}. The redshift measurement using an entangled network of such sensors is affected by laser noise acting as a nuisance parameter.

A promising future research area is quantum metrology in the finite-repetition regime, where only a few measurement samples are available~\cite{Rubio_2018,meyer2023quantum}. In this regime, the inherent benefits of QC solutions become apparent, as they maximize the information gain from individual measurements. When combined with Bayesian estimation using non-informative priors across the maximized domain of unambiguous estimation, QC solutions can offer near-optimal sensing performance or serve as a starting point for optimizing sensors for specific numbers of measurement repetitions.

QC solutions can find application in global estimation scenarios, where multiple unknown parameters are estimated within a predefined range of possible values. Inspired by the single-parameter estimation protocol employing a cascade of GHZ states~\cite{Kessler2014}, one can utilize a cascade of QC solutions with increasing sensitivity (increasing number of qubits). By employing the quantum phase estimation protocol, this measurement scheme allows for parameter estimation within a given range. 
While it is expected to perform slightly worse (by a constant factor) than the optimal quantum sensor for global estimation as defined in Ref.~\cite{Kaubruegger2023}, it still enables achieving the Heisenberg scaling of sensitivity with the number of employed qubits~$N$. Importantly, approaching the fundamental precision limit for global estimation of multiple parameters may require an extensive number of entangling gates~\cite{Kaubruegger2023}, whereas the sensing performance of QC solutions can be approximated using simple circuit-based sensors, which are straightforward to scale up for preparing cascades of such sensors.

Finally, our approach to optimal multiparameter sensing in the many-measurement scenario identifies higher-order information matrices that generalize the FIM as a crucial ingredient for defining optimal sensors in the form of QC solutions. Recently, Ref.~\cite{gessner2023} revealed generalizations of the QFIM that allow for establishing tighter bounds in single-parameter estimation with mixed states and a finite number of measurements. Our study opens a new direction for theoretical investigation into the properties of these higher-order information matrices within the context of multiparameter metrology. Specific questions include understanding quantum limits associated with these higher-order information matrices and exploring their behavior in cases of anisotropic sensitivity.

In summary, our work not only advances the theoretical understanding of quantum measurements but also establishes QC solutions as a solid basis for benchmarking future multiparameter sensing experiments and guiding the design of next-generation quantum sensors.

\begin{acknowledgments} 
AS thanks Samarth Hawaldar for helpful discussions. This research is supported by the Austrian Science Fund (FWF) [10.55776/COE1] and the European Union -- NextGenerationEU, by the European Union’s Horizon 2020 research and innovation program under Programmable Atomic Large-Scale Quantum Simulation (PASQuanS) 2, by the Institut f{\"u}r Quanteninformation GmbH, and by is supported by the US Air Force Office of Scientific Research (AFOSR) via IOE Grant No.~FA9550-19-1-7044 LASCEM. Innsbruck theory is a member of the National Science Foundation (NSF) Quantum Leap Challenge Institute Q-Sense. PZ acknowledges funding from the Austrian Science Foundation (FWF, P 32597 N). AS acknowledges the support of a C. V. Raman Post-Doctoral Fellowship, Indian Institute of Science (IISc).
\end{acknowledgments}

\newpage

\appendix

\section{Quantum Cramer Rao bound in multiparameter metrology}
\label{app:QCRB_saturability}

In this appendix, we first discuss the issue of saturability of the QCRB for the simultaneous estimation of multiple parameters. Subsequently, we describe the construction of the sensor used in Fig.~\ref{fig:QC_vs_SLD}(b). Finally, we explain why the QCRB diverges for $\abs{\bm{\phi}}\to2n\pi$, $n=1,2,\ldots$ in the multiparameter case. 

\subsection{Saturability of QCRB}

As discussed in Sec.~\ref{subsec:FI_framework}, the MSE for estimating a $d$-dimensional phase vector $\bm{\phi}$ is lower bounded as 
\begin{eqnarray}
    K \times \mathrm{MSE}(\bm{\phi})\geq \Delta_\mathrm{CRB} \geq \Delta_\mathrm{QCRB}\equiv\tr\mathcal{F}_Q^{-1},
\end{eqnarray}
where $\mathcal{F}_Q$ is the QFIM, Eq.~(\ref{eq:QFIM}). The first inequality can be saturated by using an optimal estimator, e.g., the MLE for $K\gg 1$. Here, we summarize relevant results concerning the saturability of the second inequality, following the discussion in Ref.~\cite{Demkowicz2020}. The saturability of this bound, i.e. attaining $\Delta_\mathrm{CRB}=\Delta_\mathrm{QCRB}$, is not guaranteed in general. For a general family of mixed states $\{\rho_{\boldsymbol{\phi}}\}$ parameterized by $\bm\phi$, a necessary and sufficient condition for saturating this bound is
\begin{eqnarray}
     \tr\{\rho_{\boldsymbol{\phi}}[L_i,L_j]\} =0, \forall \; i,j,
    \label{eqn:weak_commutative}
\end{eqnarray}
i.e., the expectation values of the commutators of all pairs of SLDs [Eq.~(\ref{eq:sld})] with respect to the state $\rho_{\boldsymbol{\phi}}$ vanish~\cite{ragy2016compatibility}.  A pure state model (pure input state and unitary parameter encoding) for which this condition is satisfied at a particular value $\bm{\phi}$ is said to be \emph{quasiclassical} at $\bm{\phi}$~\cite{Miyazaki2022}. A nonvanishing expectation value for any pair of SLDs is a signature of measurement incompatibility; it indicates that there is no single measurement that simultaneously saturates the precision given by the QFIM for all the parameters.

Equation~(\ref{eqn:weak_commutative}) can be further simplified in the case of a pure input state $\ket{\psiin}$ and unitary parameter encoding~\cite{Baumgratz2016}. In particular, for $SU(2)$ sensors, Eq.~(\ref{eqn:weak_commutative}) is equivalent to the condition 
\begin{eqnarray}    \ev{[A_i(\boldsymbol{\phi}),A_j(\boldsymbol{\phi})]}_{\ket{\psiin}} = 0, \forall \; i,j,
\end{eqnarray}
where $A_i(\boldsymbol{\phi}),i=x,y,z$ are defined in Eq.~(\ref{eqn:ai_exp}). Hence, for $\ket{\psiin}$ such that $\ev{J_i}_{\ket{\psiin}}=0$ for $i=x,y,z$, the evolved state $\ket{\psi_{\boldsymbol{\phi}}}$ is quasiclassical for all values of $\boldsymbol{\phi}$, in the case of both two-parameter and three-parameter sensing. 

For models that do not satisfy Eq.~(\ref{eqn:weak_commutative}), a tight lower bound to $\Delta_\mathrm{CRB}$ is instead given by the Holevo-Cramer-Rao bound (HCRB)~\cite{holevo2011probabilistic,Demkowicz2020}. The HCRB can be formulated as the minimization problem:
\begin{equation}
    \Delta_\mathrm{HCRB} = \min_{\boldsymbol{X},V} (\tr V | V\succeq Z[\boldsymbol{X}], \ev{ \{L_i,X_j\}}_{\ket{\psi_{\boldsymbol{\phi}}}} = 2\delta_{ij}).
    \label{eq:hcrb_formulation}
\end{equation}
Here, $\boldsymbol{X}$ represents a vector of $d$ Hermitian matrices acting on the system's Hilbert space and $Z[\boldsymbol{X}]$ is a $d\times d$ matrix with elements $Z_{ij} = \ev{X_i X_j}_{\ket{\psi_{\boldsymbol{\phi}}}}$. The constraint on the anticommutators arises from the unbiasedness condition on the estimators. Furthermore, $V$ is a $d\times d$ real symmetric matrix; by introducing $V$ and minimizing its trace subject to $V\succeq Z[\boldsymbol{X}]$, the HCRB formulation accounts for the general case where the SLDs may not satisfy Eq.~(\ref{eqn:weak_commutative}), which results in a non-zero matrix  $\mathrm{Im}\{Z[\boldsymbol{X}]\}$. In contrast, the QCRB bound is obtained by simply discarding $\mathrm{Im}\{Z[\boldsymbol{X}]\}$ and is equivalent to the minimization: 
\begin{equation}
    \Delta_\mathrm{QCRB} = \min_{\boldsymbol{X}} (\tr Z[\boldsymbol{X}] | \ev{\{L_i,X_j\}}_{\ket{\psi_{\boldsymbol{\phi}}}} = 2\delta_{ij}).
    \label{eq:qcrb_formulation}
\end{equation}
The minimum value of this bound can be shown to occur when $\boldsymbol{X}=\mathcal{F}_Q^{-1}\boldsymbol{L}$, where  $\boldsymbol{L}$ denotes the vector of SLDs. The corresponding minimum value is the familiar result $\Delta_\mathrm{QCRB}=\tr\mathcal{F}_Q^{-1}$. Finally, it can be shown that $\Delta_\mathrm{HCRB}$ is at most a factor of $2$ larger than $\Delta_\mathrm{QCRB}$~\cite{Demkowicz2020,carollo2019onquantumness,tsang2020quantum}.

\subsection{Construction of SLD sensor}

Here, we summarize the construction of the measurement projectors based on the SLDs $L_i$~\cite{Baumgratz2016}, which is used for the simulation of the maximum likelihood estimation in Fig.~\ref{fig:QC_vs_SLD}(b). The eigenspaces of the set of operators $\{L_i\}$ are spanned by the $d+1$ vectors 
\begin{eqnarray}
    \ket{\xi_0} &=& \ket{\psi_{\bm{\phi}}}, \nonumber\\
    \ket{\xi_i} &=& \ket{\partial_{\phi_i}\psi_{\bm{\phi}}}, \; k=1,\ldots,d. 
\end{eqnarray}
These vectors are linearly independent and hence Gram-Schmidt orthogonalization can be used to find an orthogonal basis in this $d+1$ dimensional subspace, with the first vector taken to be $\ket{\xi_0}$. A POVM consisting of $d+2$ elements is thus obtained by taking the $d+1$ rank 1 projectors on to these orthogonal vectors and one more element that is chosen to satisfy the completeness requirement. 

\subsubsection{Evaluation of $\ket{\partial_{\phi_i}\psi_{\bm{\phi}}}$}

Additionally, for unitary evolution under $U(\bm{\phi})= \exp(-i\bm{\phi}\cdot \bm{J})$, the differential state $\ket{\partial_{\phi_i}\psi_{\bm{\phi}}}$ can be evaluated in terms of $\ket{\psiin},U(\bm\phi)$ and the generators $J_i,i=x,y,z$, as 
\begin{eqnarray}
    \ket{\partial_{\phi_i}\psi_{\bm{\phi}}} = -i U(\bm{\phi}) A_i(\bm{\phi}) \ket{\psi_\mathrm{in}},
    \label{eqn:dpsi}
\end{eqnarray}
where the operators $A_i$ are given by~\cite{Baumgratz2016} 
\begin{eqnarray}
    A_i (\bm{\phi}) = \int_0^1 d\alpha e^{i\alpha \bm{\phi}\cdot \bm{J} } J_i e^{-i\alpha \bm{\phi}\cdot \bm{J} }. 
\end{eqnarray}
Writing $\bm{\phi} = \abs{\bm{\phi}}(n_x\hat{x}+n_y\hat{y}+n_z\hat{z})$, these can be straightforwardly evaluated to yield 
\begin{align}
    A_i (\bm{\phi}) &= \left[ n_i^2 + \mathrm{sinc}(\abs{\bm{\phi}})(1-n_i^2) \right] J_i \nonumber\\
    &+ \left[n_i n_j [1-\mathrm{sinc}(\abs{\bm{\phi}})] - n_k\frac{\abs{\bm{\phi}}}{2} \mathrm{sinc}^2\left(\abs{\bm{\phi}}/2\right)\right] J_j \nonumber\\
    &+ \left[n_k n_i [1-\mathrm{sinc}(\abs{\bm{\phi}})] + n_j \frac{\abs{\bm{\phi}}}{2} \mathrm{sinc}^2\left(\abs{\bm{\phi}}/2\right)\right] J_k,
    \label{eqn:ai_exp}
\end{align}
where the indices $i,j,k$ must be permuted in cyclic order over $x,y,z$ to obtain $A_x,A_y,A_z$. 

\subsection{QCRB for $\abs{\bm{\phi}}=2n\pi$}
\label{app:QCRB_divergence}

Here, we show that in the multiparameter case with SU(2) phase encoding, $\Delta_{\mathrm{QCRB}}\to\infty$ for $\abs{\bm{\phi}}\to2n\pi$, $n=1,2,\ldots$. Without loss of generality, we consider the QCRB when the phase vector is along the $x$ direction, i.e. $\bm{\phi}=2\pi\hat{x}$. From Eq.~(\ref{eqn:ai_exp}), we get 
\begin{eqnarray}
    A_x = J_x,\; A_y = A_z = 0,
\end{eqnarray}
which, using Eq.~(\ref{eqn:dpsi}), implies that $\ket{\partial_{\phi_y}\psi_{\bm{\phi}}}=\ket{\partial_{\phi_z}\psi_{\bm{\phi}}}=0$. This result shows that the state is unchanged to first order under small rotations perpendicular to $\bm{\phi}$, and hence the sensor cannot distinguish between states in a neighborhood of any point with $\abs{\bm{\phi}}=2n\pi$. Accordingly, the QFIM acquires zero eigenvalues along the perpendicular directions, resulting in $\Delta_{\mathrm{QCRB}}\to\infty$.

In contrast, if the phases were instead encoded by commuting generators, say, $\{K_i\}$ such that $[K_i,K_j]=0\;\forall\;i,j$, then the corresponding operators $A_i(\bm{\phi})=K_i$ independent of $\bm\phi$ and hence $\ket{\partial_{\phi_i}\psi_{\bm\phi}}\neq 0$ for any $\bm\phi$. As a result, there is in general no divergence in $\Delta_{\mathrm{QCRB}}$ at $\abs{\bm{\phi}}=2n\pi$ in the case of commuting generators. Moreover, both $\ket{\psi_{\bm\phi}}$ \emph{and} $\{\ket{\partial_{\phi_i}\psi_{\bm\phi}}\}$ are $2\pi$ periodic in this case, making the QFIM--- and thus the QCRB--- also $2\pi$ periodic. Hence, it appears that non-commutativity of generators is a necessary condition for the observed divergence.

\section{Bhattacharyya Bounds}
\label{app:BhB}
In this Appendix, we illustrate the connection between the maximization of the domain of unambiguous estimation, criterion~(ii), and the maximization of single-shot information gain within the frequentist paradigm. To this end, we consider one of the bounds which extends beyond Fisher information. Bhattacharyya~\cite{Bhattacharyya1947} presented a family of bounds that are tighter than the CRB and account for the unbiasedness of the estimator over a progressively wider region surrounding the true phase value by adding constraints on the higher-order derivatives of the likelihood~$p(\bm\mu | \bm\phi)$.

Here we consider a single-parameter version of the Bhattacharyya bounds (BhB) for application to the GHZ-state sensor example discussed in Sec.~\ref{subsubsec:unambiguity} of the main text. To establish the BhB, we introduce generalized information coefficients using the generating function:
\begin{equation}
\chi(h,k) = \sum_{\bm\mu}\frac{p(\bm\mu | \phi+h)p(\bm\mu | \phi+k)}{p(\bm\mu | \phi)}
\label{eq:BhB_genfunc}
\end{equation}
The generalized information coefficients $\mathcal{I}_{ij}$ forming a matrix $\mathcal{I}$ are obtained as the coefficients of $h^ik^j/(i!j!)$ in the Taylor expansion of $\chi(h,k)$ in powers of $h$ and $k$ up to order $\kappa$, such that $i,j\leq\kappa$. The BhB of order $\kappa$ reads
\begin{equation}
\mathrm{MSE}({\phi})\ge \Delta_{\rm BhB} \equiv (\mathcal{I}^{-1})_{11}.
\label{eq:BhB}
\end{equation}
Note that the number of measurements $K$ is implicit in the BhB~\eqref{eq:BhB} as the generalized information coefficients are defined using the complete likelihood $p(\bm\mu | \phi)$. The CRB is recovered as the BhB of order $\kappa=1$.

\begin{figure}[t] 
   \centering
   \includegraphics[]{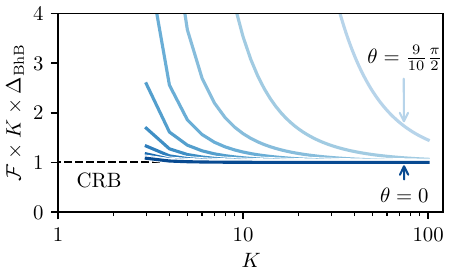}
   \caption{Bhattacharyya bounds, $\Delta_{\rm BhB}$, on the rescaled MSE for single-parameter estimation with the GHZ-state interferometer plotted against the number of measurement repetitions $K$. Solid lines depict BhB of order $\kappa=3$ for various measurement basis phases $\theta$ (see Sec.~\ref{subsubsec:unambiguity}) in shades of blue, ranging from dark (for $\theta=0$) to light blue (for $\theta=0.9\pi/2$) with increments of $0.1\pi/2$. The black dashed line indicates the CRB.}
   \label{fig:BhB_scaling}
\end{figure}
We evaluate the BhB of order $\kappa=3$ for the single-parameter GHZ-state interferometer. The rescaled BhB as a function of the number of measurements $K$ is shown in Fig.~\ref{fig:BhB_scaling} for different measurement basis phases $\theta$ (see Sec.~\ref{subsubsec:unambiguity}) in shades of blue, ranging from dark (for $\theta=0$) to light blue (for $\theta=0.9\pi/2$) in steps of $0.1\pi/2$. As the BhB is tighter than the CRB, the blue curves are positioned above the black dashed line that signifies the CRB. The figure shows that BhB highlights the sensor with $\theta=0$ measurement basis as optimal. It requires the fewest number of measurements~$K$ for its estimation error to approach the CRB, indicating that the $\theta=0$ sensor maximizes information acquired about the phase in each measurement repetition. Conversely, sensors with larger~$\theta$ values, approaching $\pi/2$, which position the working point ($\phi=0$) closer to the fringe boundary, may require an arbitrary number of measurements before nearing the CRB. This corresponds to a lower information acquisition per individual measurement.

This example underscores the relation between maximizing the information gained by the sensor per individual measurement and expanding its domain of unambiguous estimation [criterion (ii)]. Moreover, all the optimal sensor solutions (OQSs) obtained in Sec.~\ref{sec:two-parameter_QC} and Sec.~\ref{sec:three-parameter_QC} reinforce this relation by inherently maximizing the single-shot information gain while demonstrating an extended domain~$\Omega^*$. While the BhB can act as a cost function for defining optimal sensors meeting criteria (i) and (ii), the BhB optimization is challenging. Unlike the FIM and CRB, it is not additive and requires to consider the complete likelihood~$p(\bm\mu | \phi)$ rather than the single-shot conditional probability~$p(\mu | \phi)$. The findings of this appendix lend support to the discussions in Secs.~\ref{subsec:Bayesian_framework} and~\ref{subsec:Domain_HL_relation} of the main text.

\section{Simulated maximum likelihood estimation}
\label{app:mle}

In this Appendix, we summarize the procedure used to numerically simulate maximum likelihood estimation using the various quantum sensors discussed in this work. This method is utilized to generate the data shown in Fig.~\ref{fig:QC_vs_SLD}, and the orange curves in Figs.~\ref{fig:2D_MSE} and~\ref{fig:3D_MSE}.

We describe the procedure in the general context of estimating a $d$-dimensional parameter vector using a given sensor, comprising of input state and a POVM with $L$ elements. A total of $M$ true phase vectors are randomly chosen in a $d$-dimensional cube of side $2\varphi$ using the Latin Hypercube method. For each sample point $\bm\phi_0$, a maximum likelihood estimation experiment is performed following these steps:

\begin{enumerate}
    \item Determine the conditional probabilities $\{p(\mu_j|\bm\phi_0)\}$ of the $L$ single-shot measurement outcomes $\{\mu_j\},j=1,\ldots,L$, according to Eq.~(\ref{eq:cond_prob}).
    \item Simulate the frequencies $\{f_j\},j=1\ldots,L$ of the $L$ outcomes in a total of $K$ independent measurements using random numbers drawn from an $L$-dimensional multinomial distribution with probabilities given by $\{p(\mu_j|\bm\phi_0)\}$. 
    \item Define the log-likelihood function, parameterized by the frequencies $\{f_j\}$ and a parameter vector $\bm\phi$, as 
    \begin{equation}
        \mathcal{L}(\bm\phi,\{f_j\}) = \sum_{j=1}^L f_j \log p(\mu_j|\bm\phi).
    \end{equation}
    \item Start with an initial guess vector $\td{\bm\phi}$ and find the nearest local maximum $\bm{\xi}_{\bm{\mu}}^{\mathrm{ML}}$ of $\mathcal{L}(\bm\phi,\{f_j\})$ as per Eq.~\eqref{eq:MLE}. In practice, we achieve this by finding the local minimum of the function $-\mathcal{L}^2$ using the \texttt{trust-region} method of the \texttt{fminunc} routine in MATLAB. Empirically, we find that minimizing $-\mathcal{L}^2$ instead of $-\mathcal{L}$ leads to more robust convergence to the nearest local minimum. We supplement the minimizer with analytical formulas for the gradients $\partial_{\phi_i}\mathcal{L}$ , which can be evaluated using expressions for $\ket{\partial_{\phi_i}\psi_{\bm\phi}}$ derived in Appendix~\ref{app:QCRB_saturability}.
    \item Compute the squared error~(SE) in estimation as $\mathrm{SE} = (\bm\phi_0 - \bm{\xi}_{\bm{\mu}}^{\mathrm{ML}})\cdot (\bm\phi_0 - \bm{\xi}_{\bm{\mu}}^{\mathrm{ML}})$.
    \item Repeat Steps $2-4$ $R$ times to approximate the mean squared error as $\mathrm{MSE}=\sum_{r=1}^R \mathrm{SE}_r/R$. The averaging over $R$ independent simulations serves as a numerical approximation to the average over all possible measurement records that appears in Eq.~(\ref{eq:mse}).
\end{enumerate}

Table~\ref{tab:mle_params} shows the parameters used in generating the data shown in Figs.~\ref{fig:QC_vs_SLD}, \ref{fig:2D_MSE} and ~\ref{fig:3D_MSE}. 

\begin{table}[t]
\centering
\renewcommand{\arraystretch}{1.5}
\begin{tabular}{c|c|c|c|c|c}
Figure & $M$ & $\varphi$ & $K$ & $\td{\bm\phi}$ & $R$\\
\hline
Figs.~\ref{fig:QC_vs_SLD} and~\ref{fig:2D_MSE} &   $5000$ & $0.75$  & $10^5$    & $\{0.01,0.01\}$ &  $100$\\\hline
Fig.~\ref{fig:3D_MSE} & $20000$ &   $0.5$ & $10^5$ & $\{0,0,0\}$ & $100$ \\\hline
\end{tabular}
\caption{Parameters used for the maximum likelihood estimation simulations. The top (bottom) row corresponds to data generated for two-parameter (three-parameter) sensing. The symbols denote number of sampled true vectors ($M$), linear size of sampling region (2$\varphi$), number of measurement repetitions in a single simulated experiment ($K$), initial guess vector ($\td{\bm\phi}$) and number of experiments simulated at each true vector ($R$).}
\label{tab:mle_params}
\end{table}

\section{Derivation of the asymptotic cost function}
\label{app:Cost_derivation}

In this Appendix we provide derivation of the single-shot Bayesian cost Eq.~\eqref{eq:cost_Bayes_single_shot} as an asymptotic limit of the $K$-measurement Bayesian cost Eq.~\eqref{eq:cost_Bayes_K}. To this end, we consider an additional measurement repetition with an outcome $\mu$ on top of the $K$ measurements such that the posterior $p(\bm\phi|{\bm\mu})$, Eq.~\eqref{eq:posterior_K}, becomes the new prior for the additional single measurement. The posterior of the combined $K+1$ measurements reads:
\begin{equation}
p(\bm\phi|{\bm\mu},\mu) = \frac{p(\mu | \bm\phi) \, p(\bm\phi|{\bm\mu})}{p({\bm\mu,\mu})},
\label{eq:posterior_K+1}
\end{equation}
where the marginal probability reads $p({\bm\mu,\mu})=\int d\bm\phi\, p(\mu | \bm\phi)p(\bm\phi|{\bm\mu})$.

The corresponding Bayesian cost function for the $K+1$ measurements reads:
\begin{equation}
\mathcal{C} = \sum_{\bm\mu,\mu} \int d\bm\phi\, (\bm\phi-\bm\xi_{\bm\mu,\mu})^2\,p(\bm\phi|{\bm\mu},\mu)p(\bm\mu | \bm\phi_0) p(\mu | \bm\phi_0),
\label{eq:cost_Bayes_K+1}
\end{equation}
where we use shorthand notation for the scalar product, $(\bm\phi-\bm\xi_{\bm\mu,\mu})^2 \equiv (\bm\phi-\bm\xi_{\bm\mu,\mu})\cdot (\bm\phi-\bm\xi_{\bm\mu,\mu})$.

Making use of the asymptotic posterior $p(\bm\phi|{\bm\mu}) \to\mathcal{P}_K(\bm\phi-\bm\phi_0)$, Eq.~\eqref{eq:BvM_K}, we write the cost $\mathcal{C}$, Eq.~\eqref{eq:cost_Bayes_K+1}, as:
\begin{align*}
\mathcal{C} &= \sum_{\bm{\mu},\mu}\int \!d\bm{\phi}\,\frac{(\bm{\phi}-\bm{\xi}_{\bm{\mu},\mu})^{2}p(\mu | \bm\phi)\mathcal{P}_{K}(\bm{\phi}-\bm{\phi}_{0})}{\int d\bm{\phi}\,p(\mu | \bm\phi)\mathcal{P}_{K}(\bm{\phi}-\bm{\phi}_{0})}\\
&\qquad\quad\times p(\bm\mu | \bm\phi_0) p(\mu | \bm\phi_0)
\end{align*}
Given that now the asymptotic posterior distribution~$\mathcal{P}_{K}$ does not depend on outcomes $\bm{\mu}$ the estimator is also independent of $\bm{\mu}$, i.e. $\bm{\xi}_{\bm{\mu},\mu}\to\bm{\zeta}_{\mu}$. This allows us to take the sum $\sum_{\bm{\mu}}p(\bm\mu | \bm\phi_0)=1$, and the cost is calculated accordingly:
\begin{align}
\mathcal{C}	&=\sum_{\mu} \frac{p(\mu | \bm\phi_0)}{\int d\bm{\phi}\,p(\mu | \bm\phi)\mathcal{P}_{K}(\bm{\phi}-\bm{\phi}_{0})}\notag\\
&\qquad\times\int d\bm{\phi}\,(\bm{\phi}-\bm{\zeta}_{\mu})^{2}{p(\mu | \bm\phi)\mathcal{P}_{K}(\bm{\phi}-\bm{\phi}_{0})}\notag\\
	&\approx\sum_{\mu}\int d\bm{\phi}\,(\bm{\phi}-\bm{\zeta}_{\mu})^{2}p(\mu | \bm\phi)\mathcal{P}_{K}(\bm{\phi}-\bm{\phi}_{0})\equiv \Xi_K.
	\label{eq:derive_XI_K}
\end{align}
In the last equality we used the $\delta$-like property of the asymptotic prior in the integral, ${\int d\bm{\phi}\,p(\mu | \bm\phi)\mathcal{P}_{K}(\bm{\phi}-\bm{\phi}_{0})}\approx p(\mu | \bm\phi_0)$. The resulting equation~\eqref{eq:derive_XI_K} defines the new cost function $\Xi_K$ introduced in Sec.~\ref{subsec:new_cost} of the main text.

Note that the approximation step in Eq.~\eqref{eq:derive_XI_K} corresponds to dropping all higher order term in $K^{-1}$ expansion of $\varrho_\mu$, Eq.~\eqref{eq:scalar_vector_terms}. These terms, given by the even derivatives of $p(\mu | \bm\phi)$ (see Appendix~\ref{app:asymptotic_cost}), are generally unhelpful for sensing as they are symmetric around the optimization point $\bm\phi_0$, i.e. they are given by even powers of $(\bm\phi-\bm\phi_0)$. Hence, the optimization of asymptotic Bayesian cost effectively suppresses these terms. In QC solutions with symmetric sensitivity, these terms evaluate to zero, rendering the approximation an exact equality in Eq.~\eqref{eq:derive_XI_K}. The impact of these terms on QC solutions with asymmetric sensitivity remains a topic for future research.

\section{Numerical identification of quantum compass solutions}
\label{app:num_id}

In this Appendix, we introduce a numerical method to find the quantum compass solutions. Our approach is based on the numerical method demonstrated in Ref.~\cite{Kaubruegger2023}, which enables the minimization of the asymptotic Bayesian cost $\Xi_K$, Eq.~\eqref{eq:cost_Bayes_single_shot}, with respect to the state, measurement, and estimators. The numerical technique exploits the multi-convexity of the cost $\Xi_K$ with respect to the degrees of freedom of the state, measurement, and estimators. Using a block coordinate descent approach, it efficiently identifies optimal sensor solutions for a given prior distribution~$\mathcal{P}_K$.

Here, we employ this numerical algorithm to iteratively find sensor solutions in the limit $K\gg1$ corresponding to a very narrow asymptotic distribution $\mathcal{P}_K$ centered at $\bm\phi_0$ with a width of $\delta=\sqrt{\mathrm{var}(\mathcal{P}_K)}\sim1/\sqrt{K}\ll1$. The process begins with the optimization of the cost $\Xi_K$, defined by a normal prior distribution with a relatively large width of $\delta\sim1/N$. Under these conditions, the convergence to a solution is typically rapid. This allows for the testing of numerous random initial states and measurements to pinpoint the best candidate. Subsequently, we gradually reduce the width $\delta$ step by step using the previously found solution as an initial guess until reaching a solution that closely approaches the QCRB at $\bm\phi_0$ and remains unchanged with further reduction of the prior width. The input state and measurement of the limiting sensor approximate the quantum compass solution we are interested in.

However, identifying the optimal sensor when the asymptotic prior width limit is small, $\delta\to0$ ($K\to\infty$), is not straightforward if we directly consider the minimum of the asymptotic Bayesian cost $\Xi_K$. This is because it corresponds to the variance of the posterior phase distribution, $\smash{\Delta^2\phi=\min\Xi_K}$, and thus includes both the prior information and the intrinsic sensor noise. As the prior distribution narrows (the prior information increases), the relative contribution of the intrinsic sensor noise to the Bayesian cost decreases. Consequently, the cost becomes purely defined by the prior variance, $\Xi_K\to\mathrm{var}(\mathcal{P}_K)$ for $K\to\infty$. This complicates the numerical analysis of sensor properties.

To isolate the contribution of the sensor noise that we are interested in, we define an effective measurement variance $\Delta^2\phi_M$ using the following equation~\cite{Leroux2017,Kaubruegger2021}:
\begin{equation}
\frac1{{\Delta^2\phi}} = \frac1{\Delta^2\phi_M} + \frac1{\mathrm{var}(\mathcal{P}_K)}.
\label{eq:DeltaEffective}
\end{equation}
Equation~\eqref{eq:DeltaEffective} describes the Bayesian update of the Gaussian prior distribution, characterized by a variance~$\mathrm{var}(\mathcal{P}_K)$, to a Gaussian posterior distribution with a variance of $\Delta^2\phi$. The variance $\Delta^2\phi_M$ corresponds to the conditional Gaussian probability of a hypothetical measurement that would reduce our uncertainty on phase values to the same extent as the actual measurement. Importantly, this variance can be interpreted in a manner that assumes no prior knowledge, and thus, is subject to the CRB limitation, $\Delta^2\phi_M\geq\mathrm{Tr}\mathcal{F}^{-1}$. As we minimize the cost $\Xi_K$ for increasingly smaller prior width $\delta$, we study the convergence of $\Delta^2\phi_M$ to the CRB to identify the QC solution, as demonstrated below for a specific sensor example.

\begin{figure}[t] 
   \centering
   \includegraphics[]{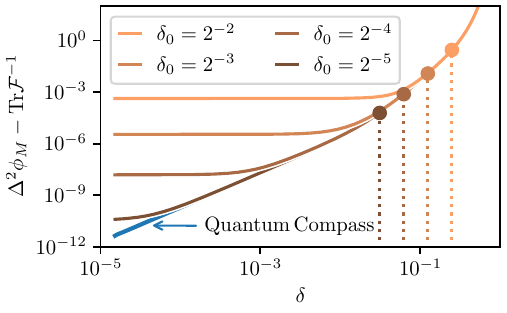} 
   \caption{Identifying 3D quantum compass for $N=8$ atoms}
   \label{fig:identify_compass}
\end{figure}

\emph{Example of identifying a 3D QC solution}. Figure~\ref{fig:identify_compass} illustrates the use of the iterative algorithm to find the quantum compass solution for a three-parameter SU(2) sensor comprising $N=8$ two-level atoms. We optimize the sensor at $\bm{\phi}_{0}=\{0,0,0\}$, where its sensitivity is at its maximum and isotropic. As such, we use an isotropic Gaussian with a width of $\delta$ as our prior, $\mathcal{P}_K(\bm{\phi})\propto\exp\left[-\bm{\phi}\cdot\bm{\phi}/(2\delta^{2})\right]$. The plot displays the difference between the effective measurement variance ($\Delta^2\phi_M$) and the QCRB, denoted as $\Delta^2\phi_M - \mathrm{Tr}\mathcal{F}^{-1}$, as a function of the prior width~$\delta$. For this particular case, the three-parameter QCRB for $N=8$ is calculated as $\mathrm{Tr}\mathcal{F}^{-1} = 9/[N(N+2)] = 0.1125$.

The four colored circles (ranging from orange to brown) represent the performance of quantum sensors that minimize the asymptotic Bayesian cost $\Xi_K$ at various widths of the normal prior distribution ($\delta_0 = 2^{-2},2^{-3},2^{-4},2^{-5}$) indicated by the vertical dotted lines. These sensors were obtained through iterative optimization, starting with the largest $\delta_0$, as described earlier. The curves passing through the circles of matching colors depict the performance of the corresponding sensors at different prior widths. In this analysis, we utilize estimators optimal for the prior width $\delta$, as given by Eq.~\eqref{eq:MMSEE}, while keeping the input state and measurement fixed and optimized at $\delta_0$.

By examining the plot, it becomes apparent that the sensor, when optimized with a larger width of $\delta_0 = 2^{-2}$ (orange line), reaches its performance limit around $\delta \sim 10^{-2}$ and fails to attain the QCRB as the prior width approaches zero. However, as we iteratively optimize the sensor with smaller and smaller prior widths $\delta_0$, its performance at smaller $\delta$ improves dramatically. In fact, the sensor optimized with $\delta_0 = 2^{-5}$ (brown line) achieves a virtually optimal sensing performance, exceeding the QCRB by only about $10^{-10}$ when $\delta \sim 10^{-5}$. Clearly, the orange-brown curves converge towards a unique quantum compass solution, represented by the blue line, which exactly meets the QCRB as $\delta$ approaches zero while maintaining its performance at finite widths, matching the performance of sensors specifically optimized for these widths. When we convert this back to the number of measurement repetitions, $K\sim\delta^{-2}$, we observe that the performance of the QC solution converges to the QCRB in the quickest possible manner as $K$ increases. This effectively illustrates the discussion presented in Section~\ref{subsec:Bayesian_framework} of the main text. 

Figure~\ref{fig:identify_compass} illustrates a typical behavior observed during the iterative sensor optimization process employed to identify a multiparameter quantum compass. Notably, in certain cases such as the two-parameter SU(2) sensor (Sec.~\ref{sec:two-parameter_QC}) or a three-parameter sensor (Sec.~\ref{sec:three-parameter_QC}) with a specific number of atoms~$N$, we observe a significantly faster convergence towards the limiting solution. This rapid convergence signifies the emergence of a highly symmetric solution.

In conclusion, the iterative approach presented in this appendix serves two primary purposes. Firstly, it provides a reliable starting point for finding an exact POVM sensor solution that establishes fundamental bounds on the sensor's performance. Secondly, it allows for the identification of the optimal sensor while considering relevant practical constraints, such as the restriction of measurements to the projective class. Moreover, this technique can be easily employed for the numerical exploration of QC solutions with anisotropic sensitivity by utilizing an anisotropic normal distribution $\mathcal{P}_K$. Every QC solution presented in Sections~\ref{sec:two-parameter_QC} and \ref{sec:three-parameter_QC} is derived by executing the outlined iterative process, which produces a plot similar to the Fig.~\ref{fig:identify_compass}.

\section{Asymptotic Bayesian cost Taylor expansion}
\label{app:asymptotic_cost}
In this Appendix, we discuss the asymptotic Taylor expansion of the cost function $\Xi_K$ in the form given by Eq.~\eqref{eq:Cost} in the limit of an infinitesimally narrow asymptotic prior distribution $\mathcal{P}_K$, $K\to\infty$. We consider isotropic prior distributions,
$\mathcal{P}(|\bm{\phi}|)$, which imply no preference for any particular direction in the parameter space, treating all parameters equally. This can be naturally
extended to anisotropic priors, which would allow us to focus more
on certain parameters of interest and address the sensors with anisotropic sensitivity. We use an isotropic Gaussian centered at~$\bm{\phi}_{0}$
as our asymptotic prior, $\mathcal{P}_K(|\bm{\phi}-\bm{\phi}_{0}|)\propto\exp\left[-(\bm{\phi}-\bm{\phi}_{0})\cdot(\bm{\phi}-\bm{\phi}_{0})K/2\right]$.

Given our interest in infinitesimally narrow prior distributions, we can expand the quantities $\varrho_{\mu}$ and $\bm{\varrho}_{\mu}$, Eq.~\eqref{eq:scalar_vector_terms},
in terms of the distribution parameter $K$:
\begin{align*}
\varrho_{\mu} & =p_{\mu}^{(0)}+p_{\mu}^{(2)}K^{-1}+\ldots\\
\bm{\varrho}_{\mu} & =\bm{p}_{\mu}^{(1)}K^{-1}+\bm{p}_{\mu}^{(3)}K^{-2}+\ldots
\end{align*}
Here we have introduced terms denoting certain derivatives of the
conditional probability $p(\mu | \bm\phi)$ evaluated at~$\bm{\phi}=\bm{\phi}_{0}$
\begin{align*}
p_{\mu}^{(0)} & =p(\mu | \bm\phi),\\
p_{\mu}^{(2)} & =\frac12\sum_{k=1}^{d}\frac{\partial^{2}}{\partial^{2}\phi_{k}}p(\mu | \bm\phi),\\
\bm{p}_{\mu}^{(1)} & =\bm{\nabla}p(\mu | \bm\phi),\\
p_{\mu,j}^{(3)} & =\frac16\sum_{k=1}^{d}\Big(\frac{\partial}{\partial\phi_{j}}\frac{\partial^{2}}{\partial^{2}\phi_{k}}\\
 & \quad+\frac{\partial}{\partial\phi_{k}}\frac{\partial}{\partial\phi_{j}}\frac{\partial}{\partial\phi_{k}}+\frac{\partial^{2}}{\partial^{2}\phi_{k}}\frac{\partial}{\partial\phi_{j}}\Big)p(\mu | \bm\phi),
\end{align*}
where $j=1,\ldots,d$ with $d$ the number of parameters.

The corresponding expansion of the asymptotic Bayesian cost Eq.~\eqref{eq:Cost}
reads
\begin{equation}
\Xi_K = \frac{d}{K} -\frac1{K^2}\left\{{C}^{(1)} + \frac{{C}^{(2)}}{K} + \frac{{C}^{(3)}}{K^{2}}+\ldots\right\},
\end{equation}

The first term is the variance of the $d$-dimensional Gaussian prior
which does not play a role in the cost minimization. The leading relevant
term is given by the trace of the Fisher matrix $\mathcal{F}=\sum_{\mu}[\bm{\nabla}p(\mu | \bm\phi)\bm{\nabla}^{T}p(\mu | \bm\phi)]/p_{\mu}^{(0)}$
\begin{equation}
C^{(1)}={\rm Tr}\,\mathcal{F}=\sum_{\mu}\frac{\bm{p}_{\mu}^{(1)}\cdot\bm{p}_{\mu}^{(1)}}{p_{\mu}^{(0)}}.
\label{eq:C1_cost}
\end{equation}
Thus, minimizing the Bayesian cost in the limit of infinitesimally
small prior width, $K\to\infty$, corresponds to maximizing the trace
of the Fisher information matrix. In the case where the Fisher matrix
is proportional to the identity, this is equivalent to minimizing
the mean squared error (MSE) for an unbiased estimator according to
the CRB, ${\rm MSE}(\bm{\phi}_{0})\geq{\rm Tr}\left\{ \mathcal{F}^{-1}\right\}$. Note, however, that the maximization of the term $C^{(1)}$ cannot be considered in isolation from the higher order terms that contribute to the cost $\Xi_K$ for a finite $K$.

Let us briefly consider these higher-order terms. The second-order term reads 
\begin{equation}
C^{(2)}={\rm Tr}\,\mathcal{I}^{(2)}=\sum_{\mu}\frac{1}{p_{\mu}^{(0)}}\left[2\bm{p}_{\mu}^{(1)}\cdot\bm{p}_{\mu}^{(3)}-\bm{p}_{\mu}^{(1)}\cdot\bm{p}_{\mu}^{(1)}\frac{p_{\mu}^{(2)}}{p_{\mu}^{(0)}}\right].
\label{eq:C2}
\end{equation}
This term includes the first and third-order derivatives of the conditional
probability $p(\mu | \bm\phi)$, and its structure resembles
that of the Fisher matrix. Hence, we refer to it as the trace of the
second-order information matrix,
\begin{equation}
\mathcal{I}^{(2)}=\frac1{p_{\mu}^{(0)}}\sum_{\mu}\left[\bm{p}_{\mu}^{(1)}\bm{p}_{\mu}^{(3)T}+\bm{p}_{\mu}^{(3)}\bm{p}_{\mu}^{(1)T}-\bm{p}_{\mu}^{(1)}\bm{p}_{\mu}^{(1)T}\frac{p_{\mu}^{(2)}}{p_{\mu}^{(0)}}\right].
\label{eq:2nd_inf_matrix}
\end{equation}

We observe that QC solutions employing unconstrained
POVM tend to be very symmetric, in particular $p_{\mu}^{(s>1)}=0$ for the QC solution with isotropic sensitivity considered in this work.
This simplifies the expressions for $C^{(s>1)}$, e.g. in
Eq.~\eqref{eq:C2} the second term in square brackets vanishes, thus
facilitating analytical optimization. Using this simplification the
next order term in the Bayesian cost expansion reads
\begin{equation}
C^{(3)}={\rm Tr}\,\mathcal{I}^{(3)}=\sum_{\mu}\frac{1}{p_{\mu}^{(0)}}\left[2\bm{p}_{\mu}^{(1)}\cdot\bm{p}_{\mu}^{(5)}+\bm{p}_{\mu}^{(3)}\cdot\bm{p}_{\mu}^{(3)}\right].
\label{eq:C3}
\end{equation}
Here, $\bm{p}_{\mu}^{(5)}$ represents the 5th-order derivative
of the conditional probability and is defined in a similar way as~$\bm{p}_{\mu}^{(3)}$. Note that in Eq.~\eqref{eq:C3},
the second term in square brackets can be omitted during optimization as~$\bm{p}_{\mu}^{(3)}$ is fixed by
the optimization of the lower order term~$C^{(2)}$. In
our analytical examples of SU(2) interferometry, we were able to
identify unique compass solutions without expanding the Bayesian cost
beyond~$C^{(3)}$.

Interestingly, higher-order information matrices, akin to those presented in Eqs.~\eqref{eq:C2} and \eqref{eq:C3}, were introduced already in works of Bhattacharyya~\cite{Bhattacharyya1947} which are contemporary to the CRB literature~\cite{Rao1945,Cramer1945}. Bhattacharyya presented a family of bounds that are tighter than the CRB and account for the unbiasedness of the estimator over a progressively wider region surrounding the true phase value by adding constraints on the higher-order derivatives of the conditional probability $p(\mu | \bm\phi)$. This suggests a possibility to interpret our quantum compass solutions as optimal in the sense of saturating not only the quantum Fisher information matrix but also the higher-order quantum information matrices. Notably, the Bhattacharyya bounds have recently been quantized for the case of single-parameter quantum metrology~\cite{gessner2023}.

It is important to note that analytical optimization of the asymptotic cost $\Xi_K$ (i.e., sequential optimization of the $C^{(\ell)}$ terms) can only be achieved by employing an effective ansatz for the state and measurements, which can be formulated using insights from numerically derived QC solutions (refer to Appendix~\ref{app:num_id}). The ansatz serves two primary functions. Firstly, it reduces the number of degrees of freedom. Secondly, and more importantly, it constrains the solution space by eliminating irrelevant ones. For instance, it rules out solutions with anisotropic sensitivity that might have the same $C^{(1)}={\rm Tr}\,\mathcal{F}$ as the isotropically sensitive solutions we are seeking. The analytical QC solution presented in Section~\ref{sec:three-parameter_QC} and Appendix~\ref{app:3D_QC_solutions} are derived using the Taylor expansion introduced in the appendix. Equations~\eqref{eq:C1_cost} and \eqref{eq:C2} are used to calculate $C^{(1)}$ and $C^{(2)}$ for the GHZ-state interferometer in Sec.~\ref{subsec:analytic_quantum_compass} of the main text.

\section{Evaluation of domains of unambiguous estimation}
\label{app:domains}
In this Appendix, we elaborate on the calculation of the exact shapes and radii of domains of unambiguous estimation. As discussed in Sec.~\ref{subsec:Domain_2D} of the main text, in order to assess domain boundaries we can inspect the roots of individual probabilities $p(\mu|\bm\phi)$. 

The analysis becomes simpler if the input state and measurement basis display antiunitary symmetry. This is true for the 2D QC solutions for even $N$ and some of the 3D QC solutions considered in this paper.
This antiunitary symmetry results in purely real amplitudes~$\braket{\mu|\psi_{\bm\phi}}$, which, in turn, determine the probabilities $p(\mu|\bm\phi)$. In the context of 2D QC, these amplitudes~$\braket{\mu|\psi_{\bm\phi}}$ can be expressed as real $N$th order polynomials with respect to the radial component $|\bm\phi|$ of the phase vector in polar coordinates. Hence, the primary domain of unambiguous estimation centered around $\bm\phi_0=\{0,0\}$ is defined by the roots of these amplitude polynomials that are closest to $\bm\phi_0$ for each direction of $\bm\phi$.

Numerically finding the roots of a polynomial is a straightforward process. In Fig.~\ref{fig:2D_QC_Likelihood}, the orange line indicates the set of roots of $p(\mu=0|\bm\phi)$ that are closest to $\bm\phi_0$. The radius of the domain, denoted as $\phi_{\rm max}$, is determined by the root that is closest to~$\{0,0\}$ across all directions of $\bm\phi$.

This procedure is straightforward to generalize to the case of 3D QC solutions with antiunitary symmetry. However, the general case of multiparameter estimation is more complex. Generally, the conditional probabilities can have isolated roots, making the identification of a domain of unambiguous estimation more complicated as it involves analyzing the convexity properties of the likelihood.

The method outlined in this Appendix is employed to calculate domain shapes and radii in Sections~\ref{sec:two-parameter_QC} and \ref{sec:three-parameter_QC} of the main text.

\section{Heisenberg Limit in Three-Parameter SU(2) Sensing}
\label{app:HL_in_3D}

In this Appendix, we determine the Heisenberg limit (HL) in three-parameter SU(2) sensing. Following the approach discussed in Sec.~\ref{subsec:HL_in_2D} we search for the HL in three-parameter sensing by minimizing the CRB at some bias phase value $\bm\phi_0$ over all states and measurements. 

Specifically, we consider a sensor composed of even number of atoms $N$ and optimize the asymptotic Bayesian cost $\Xi_K$ using anisotropic infinitesimally narrow prior distributions $\mathcal{P}_K$ centered at $\bm\phi_0=\{0,0,\phi_0\}$ (we choose coordinates with $z$-axis along $\bm\phi_0$), for various $N$ values. This allows us to identify the following ansatz for the input states that maximizes the QFIM that defines the covariance matrix of the anisotropic prior $\mathcal{P}_K$. In the $\ket{J,m}$ basis, the input state can be expressed as follows:
\begin{equation}
\label{eq:psi_HL_3D}
\ket{\psi_{\rm in}^{\rm 3D}}_{\bm\phi_0} = \frac{1}{\sqrt{2+\lambda^2}}\left(\ket{J,-J} + \ket{J,J} - \lambda\ket{J,0}\right),
\end{equation}
where $\lambda$ is a free parameter. It is important to note, that the state Eq.~\eqref{eq:psi_HL_3D} is \emph{not} the 3D QC state for a general~$N$, but rather it is a state that yields the same QFIM as the 3D QC state for the same~$N$. This has been verified for $N=6,8,12$ and we conjecture that the state~\eqref{eq:psi_HL_3D} maximizes the QFIM for larger even $N$. More details about the anisotropic 3D QC solutions and their relationship to the state Eq.~\eqref{eq:psi_HL_3D} will be discussed in an upcoming Ref.~\cite{Shankar202x}.

Since the state $\ket{\psi_{\rm in}^{\rm 3D}}_{\bm\phi_0}$ is quasiclassical, the QCRB provides an achievable lower bound. We calculate the QFIM, Eq.~\eqref{eq:QFIM}, and minimize the QCRB for the state, Eq.~\eqref{eq:psi_HL_3D}, with respect to the parameter $\lambda$ for a given $\bm\phi_0$. The minimum is achieved at
\[
\lambda({\bm\phi_0})=\sqrt{\frac{2}{J+1}\left[\frac{2 J}{\left|\mathrm{sinc}({|\bm\phi_0|}/{2})\right| }-1\right]}
\]
and the corresponding minimal QCRB defining the HL is given by Eq.~\eqref{eq:3D_HL} in the main text. The HL surpasses the QCRB, Eq.~\eqref{eq:QCRB_3D}, found in \cite{Baumgratz2016} for all values of $|\bm\phi_0|$, supporting the conjecture that the state~\eqref{eq:psi_HL_3D} maximizes the QFIM for general even $N$.

\section{3D Quantum Compass Solutions}
\label{app:3D_QC_solutions}

This appendix outlines the 3D QC solutions discussed in Sec.~\ref{subsec:3D_QC_gallery} of the main text. First, we delve into the QC solution involving $N=4$ qubits, explicitly demonstrating the resolution of the degeneracy in solutions that saturate the quantum Fisher matrix through the optimization of higher-order information terms. Subsequently, we provide POVM solutions for~$N=6,8,12$ qubits, along with the optimal state for~$N=32$.

\subsection{$N=4$ QCS}
The POVM, $M_j=\ket{\mu_j}\bra{\mu_j}$, $j=1,\ldots,6$, as defined in Eq.~\eqref{eq:POVM_n4_3D} in the main text, relies on the fiducial state $\ket\eta$. To optimize the fiducial state, we parameterize it using a real-valued vector $\vec x=\{x_1,x_2,\ldots,x_9\}$ and express it in the $\ket{J=2,m}$ basis as (the phase is selected to render the third term imaginary):
\begin{equation*}
\ket\eta=\{x_1 + ix_2, x_3 + ix_4, i x_5, x_6 + ix_7, x_8 + i x_9\}.
\end{equation*}

Our objective is to minimize the metrological cost Eq.~\eqref{eq:Cost_expansion} under the POVM constraint $\mathcal{L}\equiv\sum_j M_j=\mathbb{1}_5$. The POVM ansatz Eq.~\eqref{eq:POVM_n4_3D} results in 7 nontrivial and independent constraint equations. This includes constraints on 5 diagonal elements, $\bra{2,m}\mathcal{L}\ket{2,m}=1$, and constraints on real and imaginary parts of an off-diagonal element, $\bra{2,2}\mathcal{L}\ket{2,-1}=0$.

First, we maximize the Fisher information term $C^{(1)}$ from the cost expansion Eq.~\eqref{eq:Cost_expansion}, explicitly given by Eq.~\eqref{eq:C1_cost} in Appendix~\ref{app:asymptotic_cost}. The 7 constraints are incorporated using Lagrange multipliers, resulting in a system of polynomial equations for the vector $\vec x$ and 7 Lagrange multipliers. In general, solving this kind of equations may be feasible with Gr\"obner bases. Alternatively, exact solutions can be constructed by converting high-precision numerical solutions into algebraic numbers, leveraging the fact that QC solutions tend to be ``simple'', i.e. they are expressible in radicals (all identified solutions conform to this pattern). This technique parallels the approach used to determine exact symmetric informationally complete quantum measurements (SIC POVMs)~\cite{Appleby_2018}.

The solution that maximizes $C^{(1)}$ is given by:
\begin{multline}
\ket\eta=\Big\{\frac{-x_6}{\sqrt2} - \frac{i(1+\sqrt2 x_7)}{2}, \frac{\sqrt2+i}{3\sqrt2}, \\
\frac{-i}{\sqrt6}, x_6 + ix_7, \frac{1-i\sqrt2}{3\sqrt2}\Big\},
\label{eq:C1_solution}
\end{multline}
with one remaining independent constraint:
\begin{equation}
\bra{2,2}\mathcal{L}\ket{2,2}=\frac56+3x_6^2+x_7(\sqrt2+3x_7)=1.
\label{eq:constraint1}
\end{equation}
Thus, the solution has one degree of freedom, and through direct substitution, it can be verified that the Fisher matrix is maximized, remaining independent of this degree of freedom, yielding $\mathcal{F}=\mathcal{F}_Q=8\,\mathbb{1}_3$. The corresponding maximized cost term is $C^{(1)}=24$.

To identify the unique solution defining the QC, we maximize the next term $C^{(2)}$ in the cost expansion Eq.~\eqref{eq:C2} provided in Appendix~\ref{app:asymptotic_cost}. This term reads:
\begin{equation}
C^{(2)} = -8(47+56x_6).
\end{equation}
The extrema of $C^{(2)}$ subject to the constraint given by Eq.~\eqref{eq:constraint1} occur at $x_7=-\frac1{3\sqrt2}$ and $x_6=\pm\frac13$. The choice of~``$+$'' leads to the minimum $C^{(2)}=-472$, while~``$-$'' results in the maximum $C^{(2)}=-280$, representing the QC solution. The associated second-order information matrix Eq.~\eqref{eq:2nd_inf_matrix} is isotropic, $\mathcal{I}^{(2)}=-\frac{280}3\mathbb1_3$. Substituting $x_7=-\frac1{3\sqrt2}$ and $x_6=-\frac13$ into Eq.~\eqref{eq:C1_solution} and selecting the phase such that the first element of the state vector is real gives the fiducial state presented in Eq.~\eqref{eq:3D_eta_n4} of the main text.

\subsection{$N=6$ QCS}
The QC POVM consists of 8 antiunitary invariant elements $M_j=\ket{\mu_j}\bra{\mu_j}$, $j=1,\ldots,8$:
\begin{align*}
\ket{\mu_\ell} &= e^{-i\frac{\pi}{2}(\ell-1)J_z}\ket\eta, &\ell&=1,\ldots,4,\\
\ket{\mu_{4+\ell}} &= e^{-i\frac{\pi}{2}(\ell-1)J_z}e^{-i\pi J_y}\ket\eta, &\ell&=1,\ldots,4.
\end{align*}
The antiunitary invariant fiducial state reads:
\begin{multline}
\ket\eta = \frac1{\sqrt8}
\Big\{\frac{i \sqrt{3}+\sqrt{5}}{2
   \sqrt{2}},i^{\frac32},\frac{\sqrt{3}+i \sqrt{5}}{2
   \sqrt{2}},1,\\
   \frac{-\sqrt{3}+i \sqrt{5}}{2
   \sqrt{2}},(-i)^{\frac32},\frac{i \sqrt{3}-\sqrt{5}}{2
   \sqrt{2}}\Big\}.
\end{multline}
The resulting Fisher information matrix saturates the QFIM, $\mathcal{F}=\mathcal{F}_Q=16\,\mathbb{1}_3$, and maximizes the domain of unambiguous estimation, see Table~\ref{tab:gt} ($J=3$).

\subsection{$N=8$ QC}
The QC POVM comprises 10 antiunitary invariant elements $M_j=\ket{\mu_j}\bra{\mu_j}$, $j=1,\ldots,10$, organized into two groups. The first group, corresponding to the two red estimators in Table~\ref{tab:gt} ($J=4$), is expressed in the $\ket{J=4,m}$ basis as follows:
\begin{align*}
&\begin{multlined}
\ket{\mu_1} = \textstyle\Big\{\frac{1}{12}
   \big(\sqrt{\frac{77}{5}}-2\big)+\frac{i}{\sqrt{5
   }},0,0,0,\frac{1}{6} \sqrt{\frac{83}{10}+2
   \sqrt{\frac{77}{5}}},\\
   \textstyle 0,0,0,\frac{1}{12}
   \big(\sqrt{\frac{77}{5}}-2\big)-\frac{i}{\sqrt{5
   }}\Big\},
\end{multlined} \\
&\ket{\mu_2} = e^{-i\pi J_y}\ket{\mu_1}.
\end{align*}

The second group, related to the green estimators in Table~\ref{tab:gt} ($J=4$), reads:
\begin{align*}
\ket{\mu_{2+\ell}} &= e^{-i\frac{\pi}{2}(\ell-1)J_z}\ket\eta, &\ell&=1,\ldots,4,\\
\ket{\mu_{6+\ell}} &= e^{-i\frac{\pi}{2}(\ell-1)J_z}e^{-i\pi J_y}\ket\eta, &\ell&=1,\ldots,4.
\end{align*}
The antiunitary invariant fiducial state is parametrized as follows:
\begin{multline*}
\ket\eta = \{x_1+i x_2,x_3+i x_4,x_5+i x_6,x_7+i
   x_8,x_9,\\
   -x_7+i x_8,x_5-i x_6,-x_3+i x_4,x_1-i
   x_2\},
\end{multline*}
where $x_2=\frac1{4\sqrt5}$, $x_5=x_6=\frac14$. There are two solutions for the remaining parameters that maximize the metrological cost up to the $C^{(2)}$ term. The first solution is:
\begin{equation}
\begin{aligned}
x_3 &= -x_8,\quad x_4 = -x_7,\\
x_7 &= \textstyle-\frac{1}{8} \sqrt{4+2
   \sqrt{\frac{7}{5}}+\frac{3}{\sqrt{5}}},\\
x_8 &= \textstyle-\frac{1}{8}
   \sqrt{4-2
   \sqrt{\frac{7}{5}}-\frac{3}{\sqrt{5}}},\\
x_1 &= \textstyle-\frac{1}{12}
   \left(\sqrt{\frac{7}{5}}+\frac{\sqrt{11}}{2}\right
   ),\\
x_9 &= \textstyle\frac{1}{12} \sqrt{\frac{97}{10}-2
   \sqrt{\frac{77}{5}}},
\end{aligned}
\end{equation}
It corresponds to the estimators configuration and the shape of the domain of unambiguous estimation presented in Table~\ref{tab:gt} ($J=4$). The second solution reads:
\begin{equation}
\begin{aligned}
x_3 &= x_8,\quad x_4 = x_7,\\
x_7 &= \textstyle\frac{1}{8} \sqrt{4+2
   \sqrt{\frac{7}{5}}-\frac{3}{\sqrt{5}}},\\
x_8 &= \textstyle-\frac{1}{8}
   \sqrt{4-2
   \sqrt{\frac{7}{5}}+\frac{3}{\sqrt{5}}},\\
x_1 &= \textstyle\frac{1}{12}
   \left(\sqrt{\frac{7}{5}}+\frac{\sqrt{11}}{2}\right
   ),\\
x_9 &= \textstyle-\frac{1}{12} \sqrt{\frac{97}{10}-2
   \sqrt{\frac{77}{5}}}.
\end{aligned}
\end{equation}
It corresponds to the domain and estimators configuration twisted in the opposite direction compared to the ones shown in Table~\ref{tab:gt} ($J=4$). The Fisher information matrix corresponding to the solutions saturates the QFIM, $\mathcal{F}=\mathcal{F}_Q=\frac{80}{3}\,\mathbb{1}_3$. The maximized metrological cost expansion coefficients read $C^{(1)}=80$ and $C^{(2)}=4240-{320 \sqrt{385}}/{3}$.

\subsection{$N=12$ QCS}
The QC POVM comprises 20 antiunitary invariant elements $M_j=\ket{\mu_j}\bra{\mu_j}$, $j=1,\ldots,20$. To define the optimal measurement, we change the basis by performing a rotation around the $y$-axis, orienting a face of the Majorana constellation of the input state $\ket{\psi_{\rm in}}$ [defined in Table~\ref{tab:gt} ($J=6$)] upward, aligning the face normal with the $z$-axis. The rotated state in the $\ket{J=6,m}$ basis is given by:
\begin{align*}
\ket{\tilde\psi_{\rm in}} &= e^{-i\arctan(3-\sqrt5)J_y}\ket{\psi_{\rm in}}\\
&\begin{multlined}
 = \frac1{9\sqrt3}\Big\{2
   \sqrt{7},0,0,-\sqrt{77},0,0,-\sqrt{33},\\
   0,0,\sqrt{77},0,0,2
   \sqrt{7}\Big\}.
\end{multlined}
\end{align*}

The input state is antiunitary invariant and, in this basis, it also displays symmetry under rotations around the $z$-axis by angles of $\pm2\pi/3$, together with the estimators configuration. Consequently, the POVM element corresponding to the estimator parallel to the $z$-axis shares these symmetries. We select it as our fiducial state, $\ket{\mu_1}=\ket\eta$, due to its economical parametrization as follows:
\begin{multline*}
\ket\eta = \Big\{x_1+i y_1,0,0,x_2+i y_2,0,0,x_3,\\
0,0,-x_2+i y_2,0,0,x_1-i y_1\Big\}.
\end{multline*}

The next nine POVM elements are obtained by rotating the fiducial state around a vertex of the top face of the Majorana constellation, followed by a rotation around the $z$-axis:\begin{equation}
\ket{\mu_{1+\ell+3(k-1)}} = e^{-i\frac{2\pi}{3}(k-1)J_z}e^{-i\frac{2\pi}{5}\ell H}\ket\eta,
\end{equation}
where $H = \sqrt{\frac{2}{3}(1-1/\sqrt5)} J_x + \sqrt{\frac{1}{3}(5+2/\sqrt5)} J_z$ and $\ell,k=1,2,3$. Finally, the remaining 10 elements are obtained by flipping the first 10 around the $y$-axis: $\ket{\mu_{10+\ell}} = e^{-i\pi J_y}\ket{\mu_{\ell}}$, with $\ell=1,2,\ldots,10$.

The optimal parameters defining the QC can be found be optimizing the metrological cost expansion terms up to the $C^{(3)}$ term. The solution reads:
\begin{align*}
x_1 &= \frac{1}{45} \sqrt{195-\frac{10
   \sqrt{770}}{3}},\\
x_2 &= \frac{1}{18}
   \sqrt{\frac{93}{10}+\frac{\sqrt{770}}{3}},\\
x_3 &= \frac
   {1}{9}
   \sqrt{\frac{81}{20}+\sqrt{\frac{77}{10}}},\\
y_1 &= \frac
   {\sqrt{33}-6}{9 \sqrt{10}},\\
y_2 &= \frac{1}{6} \left(4
   \sqrt{\frac{2}{15}}+\sqrt{\frac{11}{10}}\right).
\end{align*}
The resulting Fisher information matrix saturates the QFIM, $\mathcal{F}=\mathcal{F}_Q=56\,\mathbb{1}_3$, and maximizes the domain of unambiguous estimation, see Table~\ref{tab:gt} ($J=6$). Interestingly, in this case, the domain has a shape resembling icosahedron of the input state Majorana constellation.

\subsection{$N=32$ QCS input state}
This system size allows for a remarkably symmetric Majorana constellation, taking the form of a pentakis dodecahedron. We conjecture that the associated state is the QC input state, substantiated by preliminary numerical optimizations. The non-zero \emph{unnormalized} state amplitudes~$\Psi_m$ in the $\ket{J=16,m}$ basis read:
\begin{align*}
\Psi_{\pm15} &= \pm\frac1{4\sqrt2},\\
\Psi_{\pm10} &= -\frac{1}{12}\sqrt{\frac{217}{29}},\\
\Psi_{\pm5} &= \mp\frac{1}{4}\sqrt{\frac{2015}{4002}},\\
\Psi_{0} &= \frac{1}{3}\sqrt{\frac{5890}{11339}}.
\end{align*}
The state is discussed in Sec.~\ref{subsec:3D_QC_gallery} of the main text.

\subsection{$N=3$ QCS}
\label{subapp:3D_QCS_N3}

For this system size, the QC solution exhibits anisotropic sensitivity. Hence, we present two solutions: one derived using an isotropic prior for comparison with the circuit-based sensor discussed in Sec.~\ref{sec:CircuitModel} of the main text, and the other utilizing the prior defined by the FIM. The resulting performance of these two solutions, in terms of asymptotic Bayesian cost, is very similar.

Both solutions share the same optimal state, which is a GHZ state:
\begin{equation}
\ket{\psi_{\rm in}^{\rm 3D}}_{N=3} = \frac1{\sqrt{2}}\left(\ket{\tfrac32,\tfrac32} + \ket{\tfrac32,-\tfrac32}\right).
\label{eq:psi_3D_N3}
\end{equation}
The state possesses a QFIM $\mathcal{F}_Q=\mathrm{diag}\{3,3,9\}$, which is maximal among all states in the symmetric subspace of the 3-qubit Hilbert space.

The optimal measurement for an isotropic prior $\mathcal{P}_K$ is projective, with the measurement basis in the $\ket{J=\frac32,m}$ basis reading:
\begin{equation}
\begin{aligned}
\ket{\mu_1} &= \frac12\{1,1,i,-i\},\\
\ket{\mu_2} &= \frac12\{1,-1,-i,-i\},\\
\ket{\mu_3} &= \frac12\{1,-1,i,i\},\\
\ket{\mu_4} &= \frac12\{1,1,-i,i\}.
\end{aligned}
\label{eq:PM_3D_N3}
\end{equation}

The optimal measurement for an anisotropic prior $\mathcal{P}_K$, with the covariance matrix given by the QFIM, is a POVM $M_j=\ket{\mu_j}\bra{\mu_j}$, $j=1,\ldots,5$. The POVM elements are expressed in the $\ket{J=\frac32,m}$ basis as follows:
\begin{equation}
\begin{aligned}
\ket{\mu_{1}} &= \Big\{x,\frac{1}{\sqrt{3}},-\frac{1}{\sqrt{3}},x\Big\},\\
\ket{\mu_{2}} &= e^{-i\frac{2\pi}{3}J_z}\ket{\mu_{1}},\\
\ket{\mu_{3}} &= e^{i\frac{2\pi}{3}J_z}\ket{\mu_{1}},\\
\ket{\mu_{4}} &= \Big\{y+\frac{i}2,0,0,y-\frac{i}2\Big\},\\
\ket{\mu_{5}} &= e^{-i\pi J_x}\ket{\mu_{4}}.
\end{aligned}
\end{equation}
The parameters are $x=-\sqrt{(6-\sqrt{3})/33}$ and $y = -\frac{1}{2} \sqrt{(2\sqrt{3}-1)/11}$.

\begin{figure}[t] 
   \centering
   (a)
   \includegraphics[width=1.35in]{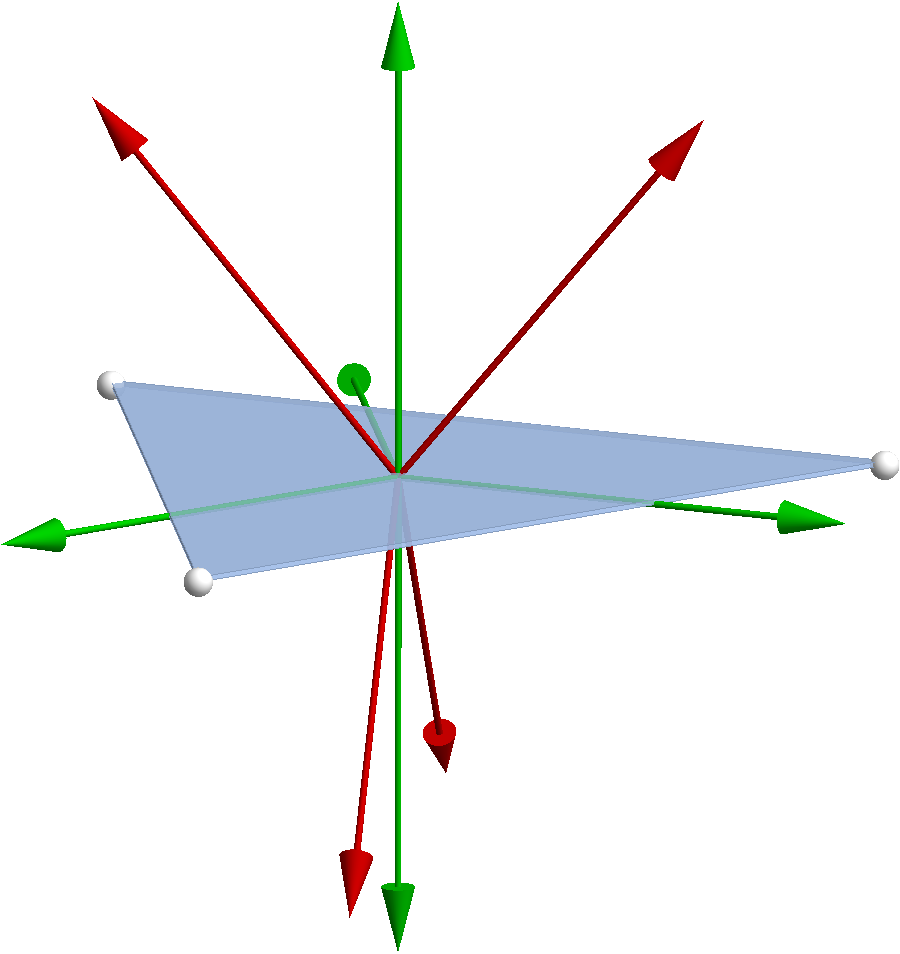}
   \hspace{0.3cm}
   \includegraphics[width=1.35in]{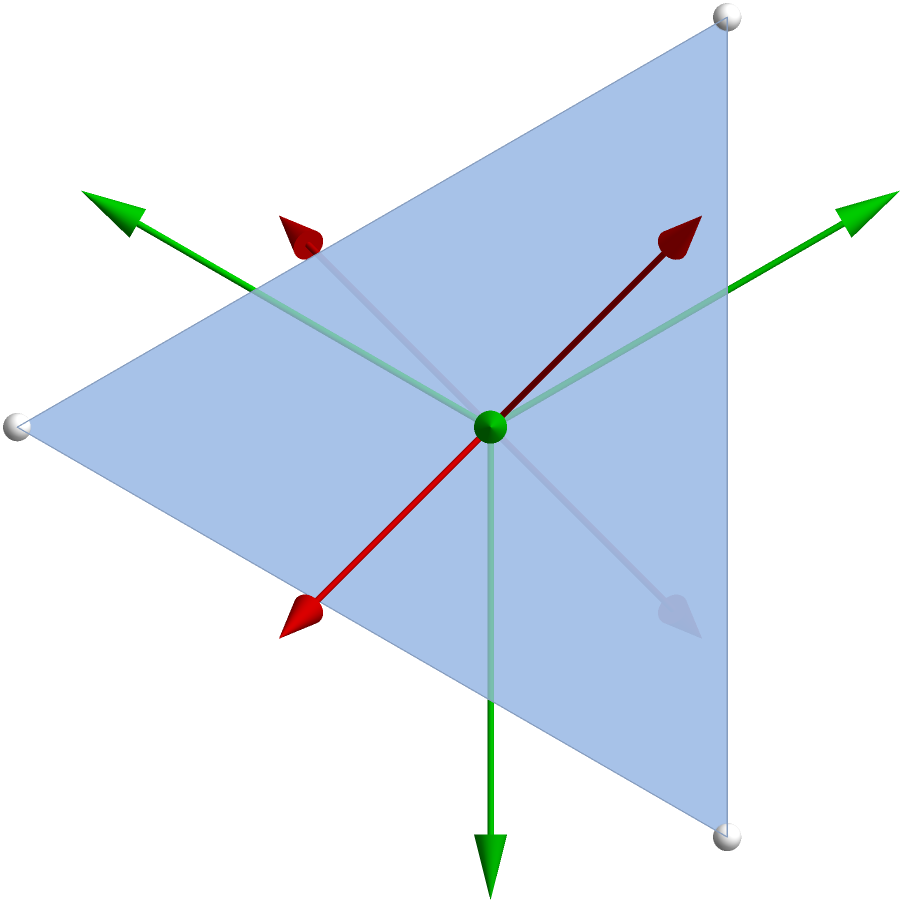}
   (b)
   \caption{Three-parameter QC solutions for $N=3$. The (a) panel depicts a side view, while the (b) panel shows a top view. Majorana constellation illustrating the GHZ input state (blue polygon) is shown alongside the directions of single-shot estimators for PM-based QC (red arrows) and POVM-based QC (green arrows).
   }
   \label{fig:Majorana_N3_QC_PM_POVM}
\end{figure}

Both sensors exhibit an anisotropic FIM saturating the QCRB, $\mathcal{F}=\mathcal{F}_Q=\mathrm{diag}\{3,3,9\}$. Figure~\ref{fig:Majorana_N3_QC_PM_POVM} displays the configurations of single-shot estimators for the two QC solutions. The performance of the sensors, in terms of metrological cost expansion coefficients [see Eq.~\eqref{eq:Cost_expansion} of the main text], is very similar. When evaluated using an isotropic prior, the first-order coefficient for both sensors is $C^{(1)}=15$, determined by the trace of the FIM. The second-order term is $C^{(2)}=-157$ for the PM-based QC and $C^{(2)}=-2 (65+21 \sqrt{3})\approx-202.7$ for the POVM-based QC. For an anisotropic prior, we rescale the prior width such that the first-order coefficient matches that of the isotropic case, $C^{(1)}=15$. Then, the second-order term is $C^{(2)}=-475/3\approx-158.3$ for the PM-based QC and $C^{(2)}=-\frac{25}{21} (115+9 \sqrt{3})\approx-155.5$ for the POVM-based QC. Further analysis of estimating performance within the domain $\Omega^*$ will be explored in future work.

The state Eq.~\eqref{eq:psi_3D_N3} and measurement Eq.~\eqref{eq:PM_3D_N3} are discussed in Sec.~\ref{subsec:circuit-based_sensors} of the main text.

\clearpage
\bibliography{quantum_compass}

\end{document}